\documentclass[12pt]{article}
\usepackage[english]{babel}
\usepackage[utf8x]{inputenc}
\usepackage[T1]{fontenc}
\usepackage{authblk, algorithm}
\usepackage{amsfonts}
\usepackage{amsmath}
\usepackage{amsthm}
\usepackage{mathtools}
\usepackage{graphicx}
\usepackage{float}
\usepackage[colorinlistoftodos]{todonotes}
\usepackage[colorlinks=true, allcolors=black]{hyperref}
\usepackage{subcaption}
\usepackage{cite}    
\usepackage{natbib}  
\usepackage{tgheros}
\usepackage{siunitx}
\usepackage{textcomp}
\usepackage{bm}
\usepackage[a4paper, textwidth= 15cm]{geometry}

\newcommand{\blind}{1}

\def\T{{ \mathrm{\scriptscriptstyle T} }}
\newtheorem{theorem}{Theorem}[section]
\newtheorem{lemma}[theorem]{Lemma}

\newtheorem{defn}[theorem]{Definition}
\newtheorem{proposition}[theorem]{Proposition}

\def\mr{\mathrm}
\newcommand\norm[1]{\left\lVert#1\right\rVert}

\newcommand{\bgamma} {\mbox{\boldmath $\gamma$}}
\newcommand{\btheta}{\bm{\theta}}
\newcommand{\bw}{\mbox{\boldmath $w$}}
\newcommand{\balpha} {\mbox{\boldmath $\alpha$}}

\newcommand{\bk} {{\bf k}}

\newcommand{\bt} {{\bf t}}

\newcommand{\by} {{\bf y}}

\newcommand{\bepsilon}{\boldsymbol{\epsilon}}

\title{Neuronized Priors for  Bayesian Sparse Linear Regression}
\author[1]{Minsuk Shin}
\author[2]{Jun S Liu}
\affil[1]{Department of Statistics, University of South Carolina}
\affil[2]{Department of Statistics, Harvard University}
\date{}

\begin{document}

\def\spacingset#1{\renewcommand{\baselinestretch}%
{#1}\small\normalsize} \spacingset{2}

\if1\blind
{\centering
  

    
    
    
  \maketitle
} \fi

\if0\blind
{
  \bigskip
  \bigskip
  \bigskip
  \begin{center}
    {\LARGE\bf Neuronized Priors for  Bayesian Sparse Linear Regression}
\end{center}
  \medskip
} \fi

\def\spacingset#1{\renewcommand{\baselinestretch}%
{#1}\small\normalsize} \spacingset{1}
\begin{abstract}
{Although Bayesian variable selection methods have been intensively studied, their routine use in practice has not caught up with their non-Bayesian counterparts such as Lasso,  likely due to difficulties in both computations and  flexibilities of prior choices.
To ease these challenges, we propose the neuronized priors to unify and extend some popular shrinkage priors, such as Laplace, Cauchy, horseshoe, and  spike-and-slab priors.
A neuronized prior can be written as the product of a Gaussian weight variable and a  scale variable transformed from Gaussian via an activation function.
Compared with classic spike-and-slab priors, the neuronized priors achieve the same explicit variable selection without employing any latent indicator variables, which results in both more efficient and flexible posterior sampling and  more effective posterior modal estimation.
Theoretically, we provide specific conditions on the neuronized formulation to achieve the optimal posterior  contraction rate, and show that a  broadly applicable MCMC algorithm achieves an exponentially fast convergence rate under the  neuronized formulation.
We also examine various simulated and real data examples and  demonstrate that using the neuronization representation is computationally more or comparably efficient than its standard counterpart in all well-known cases. An \texttt{R} package \texttt{NPrior} is provided in the CRAN for using neuronized priors in Bayesian linear regression.}%
\end{abstract}
\noindent%

\section{Introduction}
\def\spacingset#1{\renewcommand{\baselinestretch}%
{#1}\small\normalsize} \spacingset{1.8}

We consider the standard linear regression model of the form
\begin{eqnarray}\label{eq:linearM}
\by = X\btheta + \bepsilon,
\end{eqnarray}
where $\by=\{y_1,\dots,y_n\}^\T$ is the vector of responses,  $X$ is the $n\times p$ covariate matrix,   $\btheta=\{\theta_1,\dots,\theta_p\}^\T\in\mathbb{R}^p$ is the coefficient vector, and $\bepsilon\sim N(0,\sigma^2 \mr{I})$. 
To model the sparsity of $\btheta$
when $p$ is large, one often imposes a shrinkage prior on the $\theta_j$'s. 
A popular choice is  the one-group {\it continuous shrinkage prior}, which can be represented as a hierarchical scale-mixture of Gaussian distributions: 
\def\spacingset#1{\renewcommand{\baselinestretch}%
{#1}\small\normalsize}\spacingset{1.8}
\begin{eqnarray} \label{eq:shrink}
	\theta_j \mid \nu_w^2, \tau_j^2 \sim N(0, \nu_w^2 \tau_j^2)\\ \nonumber
	\tau_j^2 \sim \pi_\tau\:\mbox{(or $\tau_j\sim\pi_\tau'$) \:\:and}  \:\:\:\nu_w \sim \pi_g, 
\end{eqnarray}
for $j=1,\dots,p$, where $\pi_\tau$ and $\pi_g$ are some distributions chosen by the user. The {\it local} shrinkage parameter $\tau_j^2$   governs the shrinkage level of each individual parameter, whereas the {\it global} shrinkage parameter $\nu_w^2$ controls the overall shrinkage effect  \citep{polson2010shrink}. It is common that the variance of the Gaussian prior in \eqref{eq:shrink} contains the unknown error variance $\sigma^2$ of the model. However, as shown in \cite{moran2018variance}, the inclusion of $\sigma^2$ in (\ref{eq:shrink}) can result in inconsistency of $\sigma^2$ under high-dimensional settings.  We thus offer a choice to not mix $\sigma^2$ in the prior of $\btheta$.  
  
A few choices of $\pi_\tau$ have been shown to induce desirable shrinkage on the regression parameters, including  the Strawderman-Berger prior with $\pi_\tau$ being a mixture of gamma distributions \citep{berger1996choice}, the Bayesian Lasso \citep{park2008bayesian} with 
   $\pi_\tau$  being  an exponential distribution, 
   the horseshoe prior \citep{carvalho2010horseshoe} with $\pi_\tau'$  being a half-Cauchy distribution, the generalized double Pareto \citep{armagan2011generalized} with $\pi_\tau$ being  a mixture of Laplace distributions, 
   and the Dirichlet-Laplace prior \citep{bhattacharya2015dirichlet}  with   $\pi_\tau$ being the product of a Dirichlet and a Laplace random variables. 
   Some recent theoretical investigations 
   show that the marginal prior density of $\theta_j$ with a heavy tail and a sufficient mass around zero achieves the minimax optimal rate of posterior contraction \citep{ghosh2017asymptotic,van2016conditions,song2017nearly}.

Another popular class of shrinkage priors is the class of  {\it spike-and-slab} (SpSL) priors  \citep{mitchell1988bayesian, george1993variable}, also known as two-group mixture priors, which can be  written as:
\begin{eqnarray}\label{eq:spike}
\theta_j\mid \gamma_j &\sim& (1-\gamma_j)\pi_0(\theta_j) + \gamma_j\pi_1(\theta_j)\\ \nonumber
\gamma_j &\sim& Bernoulli(\eta),
\end{eqnarray}
for  $j=1,\dots,p$. Distribution $\pi_0$ is typically chosen to be highly concentrated around zero, i.e., the ``spike'', whereas $\pi_1$ is relatively disperse, i.e., the ``slab''. Thus, when $\gamma_j=0$, coefficient $\theta_j$ is strongly shrunk towards zero, whereas when $\gamma_j=1$, the slab part allows  $\theta_j$ to be nearly unshrunk.   Parameter $\eta$ controls the sparsity of the model  \citep{scott2010bayes}.  
When a point-mass  at zero is used for $\pi_0$, we call the resulting prior   a  {\it discrete SpSL} prior; otherwise we  call it a  {\it continuous SpSL} prior.
 Common choices of $\pi_0$ and $\pi_1$ for a continuous SpSL prior are  Gaussian distributions with a small and a large variance, respectively \citep{george1993variable}.
 Under some regularity conditions, 
  it has been shown that an appropriate choice of $\eta$ leads to model selection consistency \citep{Narisetty2014} and the optimal posterior contraction \citep{castillo2012needles,castillo2015bayesian} for high-dimensional linear regression and the normal means model. 
 
 
 With continuous shrinkage priors, MCMC sampling of $\theta_j$  given the local and global shrinkage parameters can be efficiently implemented by taking advantage of the conjugacy. However, while  continuous shrinkage priors have computational advantages over discrete SpSL priors, 
  the resulting posterior inference does not automatically provide sparse estimates of the coefficients, so that  extra and {\it ad hoc} steps are needed for variable selection  \citep{hahn2015decoupling}. Computational implementations of SpSL priors often employ a binary latent vector indicating which of the two components each coefficient comes from.
 When a discrete SpSL prior is employed, the 
 posterior inference of $\btheta$ is notoriously challenging. MCMC sampling strategies \citep{dellaportas2002bayesian,guan2011bayesian} and stochastic search strategies \citep{hans2007shotgun, berger2005posterior, zhang2007look} have been proposed to counter the computational difficulty, mostly relying on the conjugacy of each component of the prior. An MCMC strategy for non-conjugate discrete SpSL priors, such as the one that uses reversible jump proposals \citep{green1995reversible}, is rarely practical especially under high-dimensional settings.      
  
  As a computationally scalable strategy, 
  \cite{rovckova2014emvs} proposed the Expectation Maximization Variable Selection (EMVS), which is an EM algorithm to obtain the {\it maximum a posteriori} (MAP) estimator of the regression coefficients under continuous SpSL priors with Gaussian components. \cite{rovckova2016spike} further extended their idea to cases with a SpSL Lasso (SSLasso) prior by adopting  Laplace distributions for $\pi_0$ and $\pi_1$. These procedures, however, provide only point estimates, and are insufficient for quantifying  uncertainties in model selection and estimation.

To address these  practical issues in using shrinkage priors, we propose  {\it neuronized priors},  which provide a unified form for popular shrinkage priors such as the horseshoe, Cauchy, SpSL, and more.
In the form of neuronized priors, each regression coefficient is reparameterized as a product of a weight parameter and  a transformed scale parameter via an activation function, as follows:

\def\spacingset#1{\renewcommand{\baselinestretch}%
{#1}\small\normalsize} \spacingset{1.8}
\begin{defn}{\bf (Neuronized prior)}\label{def_np}
For a non-decreasing activation function $T$ and  hyper-parameters $\alpha_0$ and $\tau_w$, a neuronized prior for $\theta_j$ is defined as:
\begin{eqnarray}\label{eq:neuro}
\theta_j \coloneqq  T(\alpha_j - \alpha_0)w_j,  
\end{eqnarray}
where the scale parameter $\alpha_j$ follows $N(0, 1)$ and the weight parameter $w_j$ follows $N(0, \tau_w^2)$, all independently for  $j=1,\dots,p$. 
\end{defn}
\def\spacingset#1{\renewcommand{\baselinestretch}%
{#1}\small\normalsize} \spacingset{1.8}

As the name implies, this formulation is inspired by the use of activation functions in neural network models \citep{rosenblatt1958perceptron,rumelhart1986learning}. Under this setting, we can write down the joint distribution as:
\begin{eqnarray}\label{eq:linear_post}
\pi(\balpha, \bw \mid \by, \alpha_0, \sigma^2) \propto {1\over \sigma^n} \exp\left\{   -\frac{\norm{\by - X\btheta(\balpha,\bw,\alpha_0)}^2_2}{2\sigma^2} -\frac{\balpha^\T\balpha}{2} - \frac{\bw^\T \bw}{2\tau_w^2} \right\},
\end{eqnarray}
where  $\balpha = \{\alpha_1,\dots,\alpha_p\}^\T$,  $\bw = \{w_1,\dots,w_p\}^\T$, $\pi(\sigma^2)$ is  the prior on  $\sigma^2$, and  
\begin{equation} \label{def_theta}
\btheta(\balpha,\bw,\alpha_0) = \{T(\alpha_1-\alpha_0)w_1,\dots, T(\alpha_p-\alpha_0)w_p\}^\T \stackrel{def}{=} D_\alpha \bw,
\end{equation}
where $D_\alpha$ is the diagonal matrix with diagonal elements the $T(\alpha_j - \alpha_0)$'s.
We  show that for most existing shrinkage priors we can find specific activation functions such that the resulting neuronized priors approximate the existing ones. Therefore, existing theoretical properties of various shrinkage priors can be directly applied to posterior behaviors based on the neuronized priors.  This theoretical equivalence will be discussed in Section  \ref{sec:neuro}.  
We also show that variable selection procedures based on neuronized priors offer following advantages:

\def\spacingset#1{\renewcommand{\baselinestretch}%
{#1}\small\normalsize} \spacingset{1.8}

\begin{itemize}{
\item{\it Unification.} 
Various classes of shrinkage priors can be practically implemented by just changing the activation function. This characteristic significantly reduces practical hurdles for the user to test out different priors simultaneously, which can be a valuable option. For example,  we may find that horseshoe prior is appropriate for analyzing GWAS data on bipolar disorders, whereas SpSL priors work much better for  Type-1 diabetes \citep{song2020polygenic}.

\item{\it Flexibility and efficient computation.}
Without having to rely on prior-likelihood conjugacy, neuronized  priors still attain comparable or better efficiency for MCMC-based posterior inference  compared with the standard procedures, thus can easily accommodate non-conjugate priors. In addition, neuronized priors also enable a scalable coordinate descent optimization algorithm for posterior modal estimation,  even with discrete SpSL priors. 

\item{\it Desirable theoretical properties.} 
 We give explicit conditions on the activation function and hyperpriors so that the resulting neuronized Bayesian regression achieves  the optimal posterior  contraction rate (Section~\ref{sec:theory}), and show that
a random-walk \emph{Metropolis-Hastings} (RWMH)  algorithm converges to the target  
distribution at an exponential rate, even for non-conjugate priors. 
}
\end{itemize}
\def\spacingset#1{\renewcommand{\baselinestretch}%
{#1}\small\normalsize} \spacingset{1.8}

The rest of the article is organized as follows. Section~\ref{sec:neuro} details the neuronized counterparts of a few popular shrinkage priors for Bayesian linear  regression: the  discrete SpSL, the Bayesian Lasso, and the horseshoe and Cauchy priors.  Section 3 shows how to manage  neuronized priors to achieve one's intended goals, such as matching a target prior or controlling the sparsity level. Section 4 details main computational strategies and advantages of neuronized priors. Section 5 studies theoretical properties of the neuronized priors, including sufficient conditions for achieving an optimal posterior contraction rate and geometric ergodicity of  MCMC algorithms under a simple setting. 
Section 6 reports  simulation studies to compare the effects of different priors and their neuronized counterparts.
Two real data examples are analyzed in Section 7, and a short conclusion is given in Section 8. Proofs of the main results, efficiency comparisons of some MCMC algorithms, and additional simulation studies are provided in the Supplementary Materials. 


\section{Neuronization of Standard Sparse Priors}
\label{sec:neuro}
\subsection{Discrete and continuous SpSL priors}

\begin{figure}[H]
    \centering
    \begin{subfigure}[b]{0.3\textwidth}
        \includegraphics[width=\textwidth]{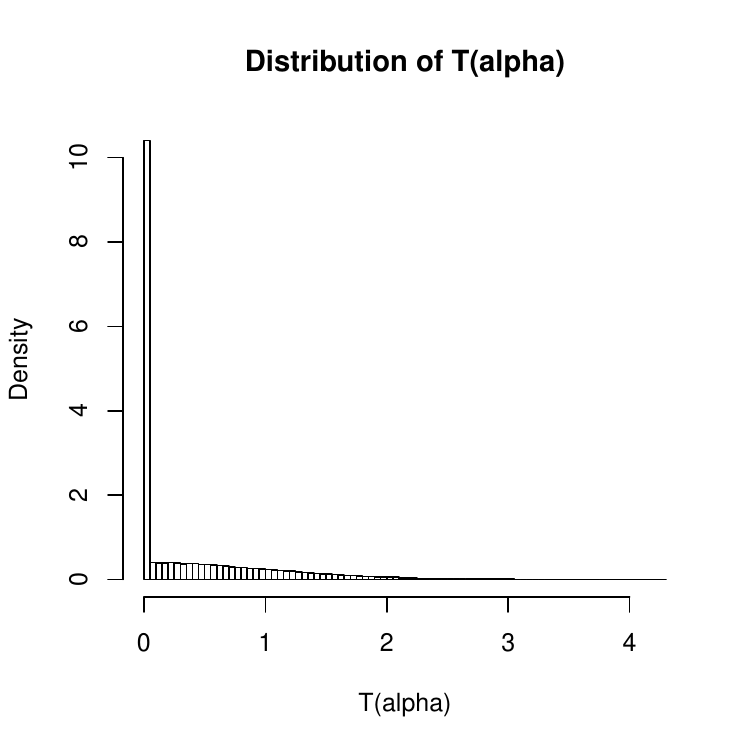}
        \caption{}
        \label{fig:T}
    \end{subfigure}
    ~ 
    \begin{subfigure}[b]{0.3\textwidth}
        \includegraphics[width=\textwidth]{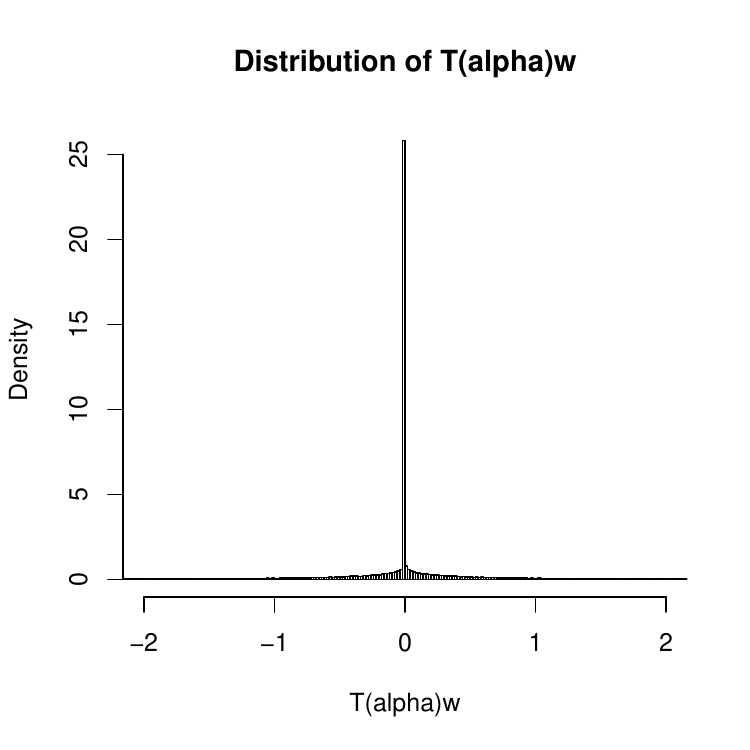}
        \caption{}
        \label{fig:Tw}
    \end{subfigure}
        \begin{subfigure}[b]{0.3\textwidth}
        \includegraphics[width=\textwidth]{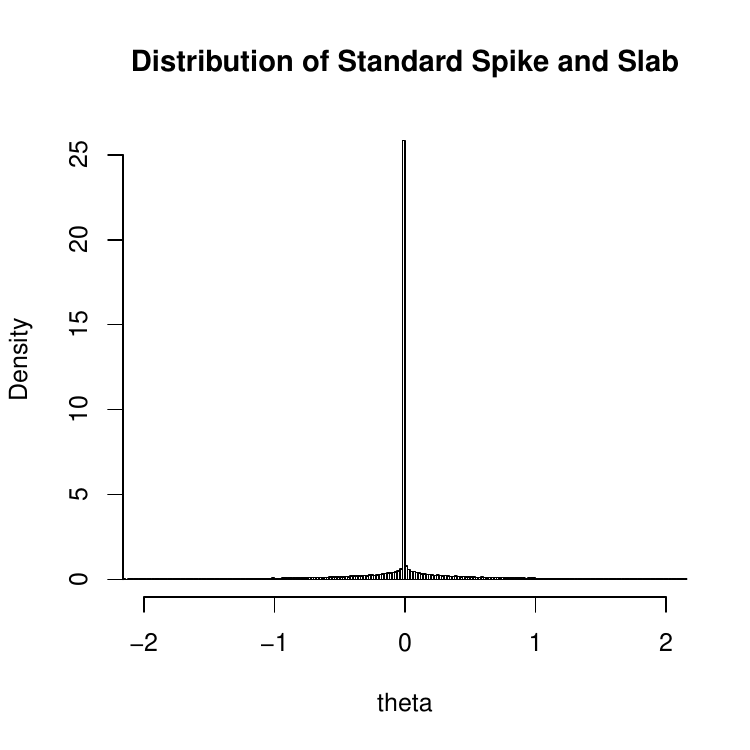}
        \caption{}
        \label{fig:theta}
    \end{subfigure}
    \caption{(a) histogram of  $T(\alpha)$; (b) histogram of $T(\alpha)w$; (c) histogram of the standard SpSL prior in   \eqref{eq:disc_spike}.}\label{fig:hist}
\end{figure}

Let the activation function in \eqref{eq:neuro} be the  
Rectifier Linear Unit (ReLU) function, 
$T(t) = \max\{0,t\}$.
When $\alpha_0=0$,  
$T(\alpha_j-\alpha_0)$ follows an equal mixture of the point-mass at zero and the half standard Gaussian, as shown in Figure~\ref{fig:hist}(a). This implies that the marginal density of $T(\alpha_j)w_j$ 
is a SpSL distribution
of the form
 \begin{eqnarray}\label{eq:disc_spike} \nonumber
\theta\mid\gamma &\sim& (1-\gamma)\delta_0(\theta) + \gamma \pi(\theta), \\
\gamma &\sim& Bernoulli (1/2),
\end{eqnarray}
where $\pi$ is the marginal density of the product of two independent standard Gaussians,  which is shown to have an exponential tail in Proposition~\ref{prop:BL}.  This tail behavior  is desirable, since \cite{castillo2012needles} and \cite{castillo2015bayesian} showed that the optimal minimax rate of  posterior contraction can be achieved when the tails of the slab part  of \eqref{eq:disc_spike} are  exponential or heavier. 
 We note that continuous SpSL priors  can be obtained from formulation \eqref{eq:neuro}  by adopting a ``leaky'' ReLU activity function \citep{maas2013rectifier}, i.e. $T(t) = \max\{ct, t\}$ for some $c<1$. 

\centerline{\framebox[1.1\width]{Figure \ref{fig:hist} Here}} 
	
More generally, hyper-parameter $\alpha_0$ controls the prior probability of sparsity: 
 $P(T(\alpha_j - \alpha_0) = 0\mid \alpha_0) = P( \alpha_j < \alpha_0\mid \alpha_0) = \Phi(\alpha_0)$, with $\Phi$ being the standard Gaussian CDF. Thus,
setting $\gamma \sim Bernoulli(\Phi(-\alpha_0))$ in \eqref{eq:disc_spike} leads to  the same distribution as that implied by (\ref{eq:neuro}).
 Conversely,  $\forall \eta\in(0,1)$, we choose $\alpha_0 = -\Phi^{-1}(\eta)$ to achieve the desired sparsity. 
\cite{scott2010bayes} showed that, 
for the sparsity parameter $\eta$ in model (\ref{eq:spike}), the Beta hyper-prior 
 \begin{eqnarray}\label{eq:model_prior}
 \eta \sim Beta(a_0,b_0),
 \end{eqnarray}
with $a_0=b_0=1$ results in a strong effect on multiplicity correction.  \cite{castillo2012needles} and \cite{castillo2015bayesian} found that the resulting SpSL procedure achieves model selection consistency and the optimal posterior contraction rate
if one chooses
$(a_0,b_0)=(1,p^a)$ for $a>1$, under an asymptotic regime where the number of predictors $p$ increases at a sub-exponential rate of $n$, i.e., $p \asymp \exp\{n^c\}$ for $c<1$. The neuronized priors can accommodate this Bernoulli-beta hyper-prior by adopting a hyper-prior on $\alpha_0$ as below: 
 

\begin{proposition}\label{prop:a0_prior}
Consider  
\eqref{eq:neuro} with $T(\cdot)$ being ReLU and a hyper-prior on $\alpha_0$,
\begin{eqnarray}\label{eq:a0_prior}
\pi(\alpha_0)\propto\Phi(-\alpha_0)^{a_0-1}(1-\Phi(-\alpha_0))^{b_0-1}\phi(\alpha_0),
\end{eqnarray}
 where $\phi$ and $\Phi$ are the pdf and cdf of $N(0,1)$, respectively. Then, the resulting prior distribution is identical to the form of (\ref{eq:spike}) with the Beta prior  \eqref{eq:model_prior} on $\eta$.
\end{proposition}



 { Since $\alpha_0$ is highly correlated with other parameters such as $\balpha$, an MCMC algorithm equipped with naive random-walk proposals would result in low sampling efficiency and poor mixing quality. Instead, we consider an efficient group-move update via a generalized Gibbs sampler \citep{liu2000generalised}. 
 The details of this computational strategy is provided in Section \ref{sec:MTM}.  }
 
\begin{figure}[H]
    \centering
        \includegraphics[width=5cm,height=4.5cm]{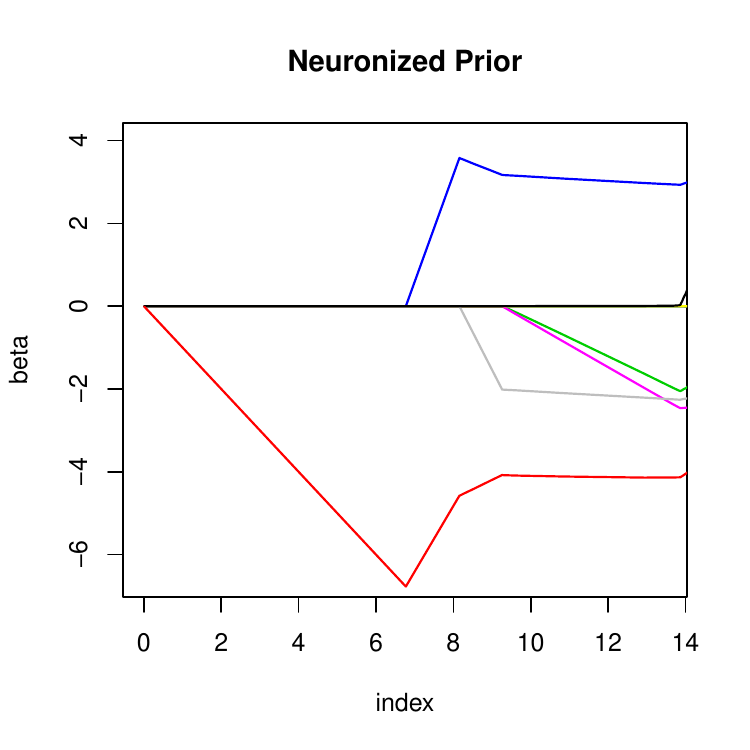}
        \includegraphics[width=5cm,height=4.5cm]{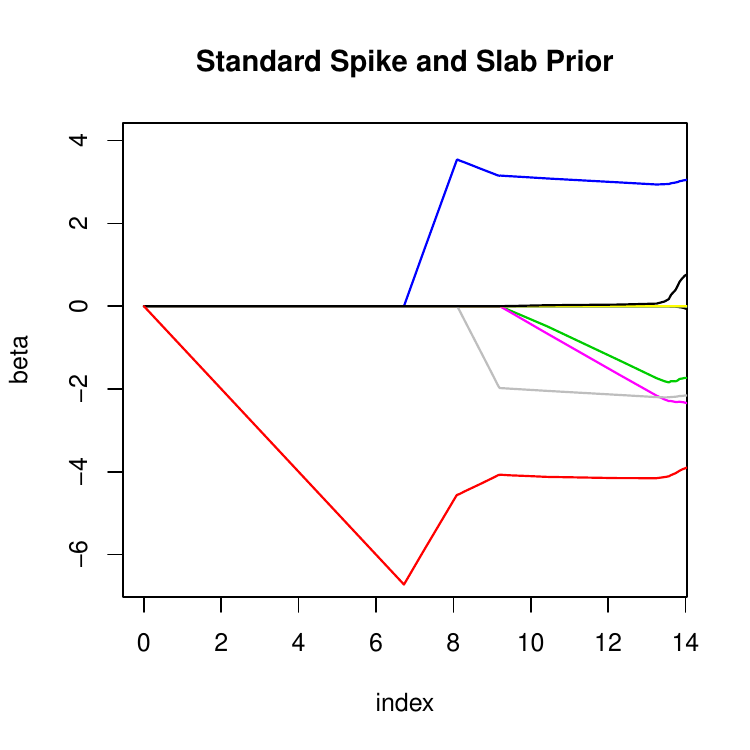}
\caption{Solution paths of the neuronized prior and the discrete SpSL prior.}\label{fig:Solution_Spike}
\end{figure}

 As a demonstration, we analyze  the Boston housing price data with linear regression.
 The dataset contains $n=506$ median housing prices of owner-occupied homes  in the Boston area, together with 10 variables that might be associated with the median prices.  Under the Jeffreys prior on $\sigma^2$, which is $1/\sigma^2$,  we consider  the independent neuronized prior: $\theta_j = T(\alpha_j - \alpha_0)w_j$, where $\alpha_j\sim N(0,1)$ and $w_j\sim N(0,\tau_w^2)$ for $j=1,\dots,p$.  
 As shown in Figure \ref{fig:Solution_Spike},  the solution path resulting from the neuronized prior with the ReLU activation function is almost identical to that resulting from the standard  discrete SpSL prior. 
 
 \centerline{\framebox[1.1\width]{ Figure \ref{fig:Solution_Spike} Here}}
 
\subsection{The Bayesian Lasso}

The Bayesian Lasso imposes a Laplace prior on $\theta_j$ and uses a Gaussian mixture representation to facilitate efficient MCMC computations \citep{park2008bayesian}.  
We shall show that the neuronized prior with $T(t) = t$ approximates the Bayesian Lasso.

  \begin{lemma}\label{lem:BL}
  With the activation function $T(t) = t$, the marginal density of $\theta$ resulting from the neuronized prior  is proportional to $ \int_0^\infty z^{-1}\exp\{ - \theta^2/(2\tau_w^2z^2) - z^2/2 \} dz$. 
 \end{lemma}
  Since 
  
  $\exp\{-|\theta|/\tau_w\} \propto \int_0^\infty \exp\{  -\theta^2/(2\tau_w^2 z^2) -  z^2/2\}dz$,
  the Laplace density differs from the form in Lemma~\ref{lem:BL}  only by a  term $z^{-1}$ in the integrand. Furthermore, the following proposition shows that the tail of this neuronized prior decays at an exponential rate like the Bayesian Lasso prior.
 
\begin{proposition}  \label{prop:BL}
Let $\pi_{L}$ be the marginal density function of $\theta$ defined in \eqref{eq:neuro} with $T(t) = t$ and $\alpha_0=0$. Then, $\forall \epsilon\in(0,1)$,  $\exists \theta_0$ and constants $c_1, c_2>0$, such that $c_1 \exp\{  - (1+\epsilon)^{1/2} |\theta|/ \tau_w \} \leq \pi_L(\theta) \leq c_2\exp\{ -(1-\epsilon)^{1/2}|\theta|/ \tau_w \} \:\:\mbox{when $\theta>\theta_0$.}$

\end{proposition}

 \cite{hoff2017lasso} also pointed out the similarity between the Bayesian Lasso and the product representation of the parameter (i.e., the neuronized prior with an identity activation function). He  showed that the MAP estimator based on the product representation of the parameter is identical to the standard Lasso.
 
\begin{figure}[H]
    \centering
    \begin{subfigure}[b]{0.32\textwidth}
        \includegraphics[width=5cm,height=4cm]{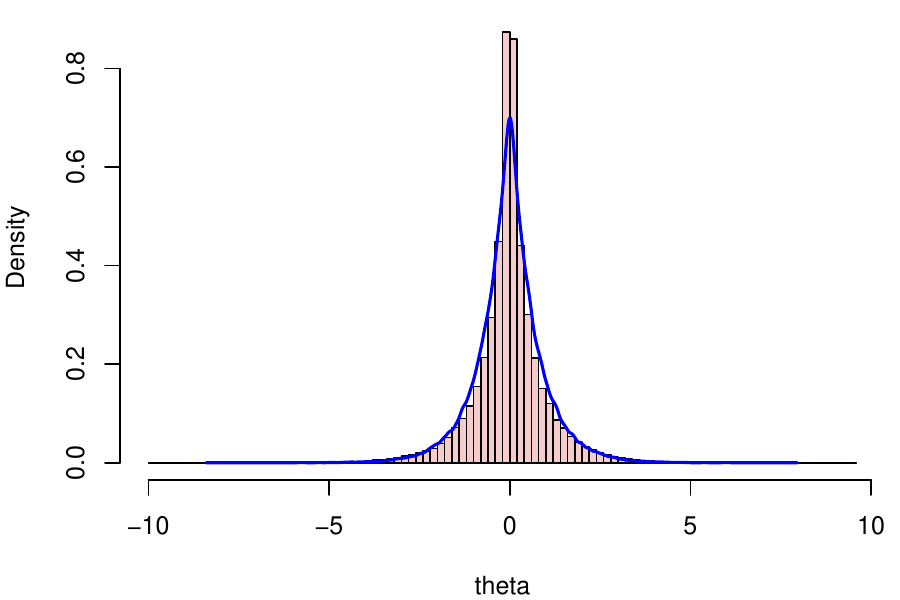}
        \caption{Bayesian Lasso}
        \label{fig:gull}
    \end{subfigure}
    \begin{subfigure}[b]{0.32\textwidth}
        \includegraphics[width=5cm,height=4cm]{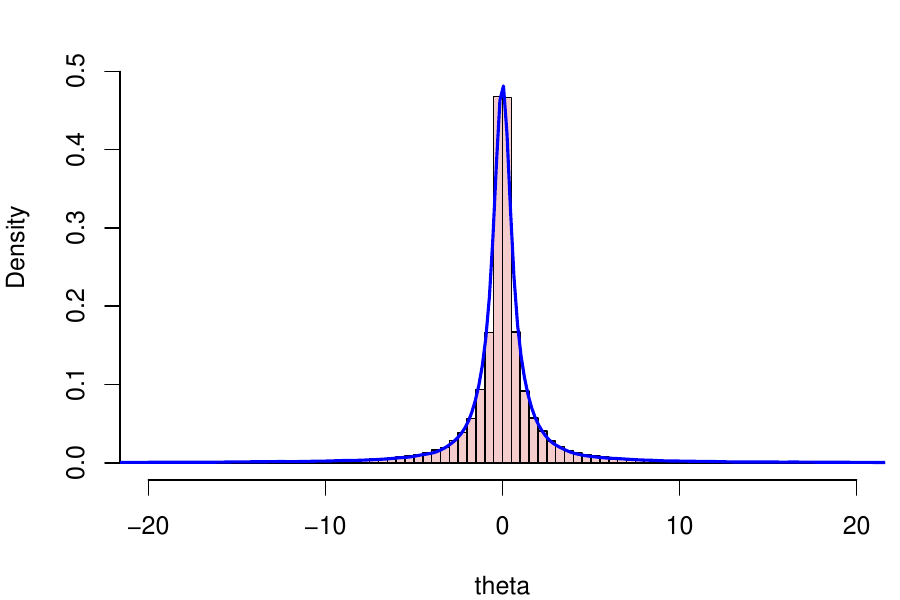}
        \caption{Horseshoe prior}
        \label{fig:mouse}
    \end{subfigure}
    \begin{subfigure}[b]{0.32\textwidth}
        \includegraphics[width=5cm,height=4cm]{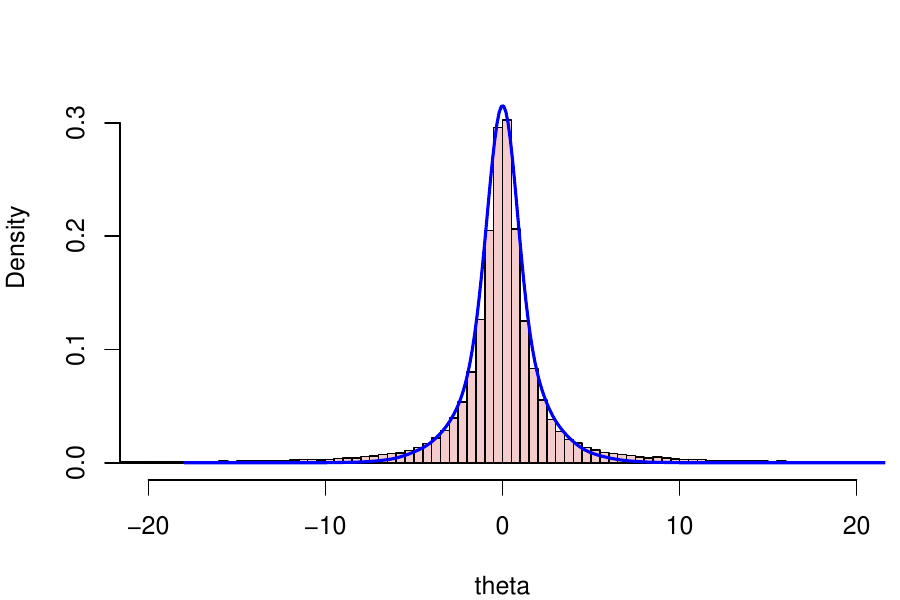}
        \caption{Cauchy prior}
        \label{fig:cuachy}
    \end{subfigure}
    
    \caption{ Histograms of some prior distributions. The blue lines indicate the density functions of their neuronized counterpart. }\label{fig:QQplots}
\end{figure}
\begin{figure}[H]
    \centering
        \includegraphics[width=14cm,height=4.5cm]{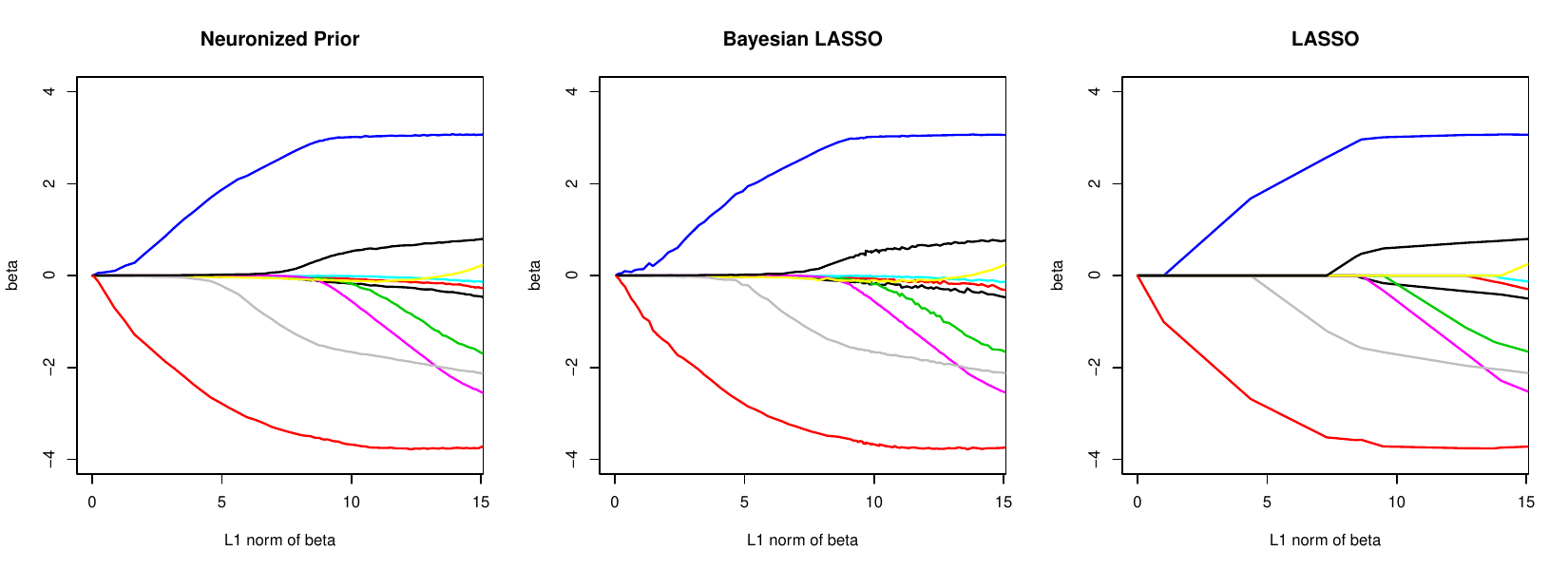}
    \caption{Solution paths of the neuronized prior, the Bayesian Lasso and the Lasso.}\label{fig:Solution_path}
\end{figure}

The histogram in Figure \ref{fig:QQplots}(a) compares the Bayesian Lasso prior with its neuronized version, verifying that the two distributions are indeed very similar. However, the Laplace prior has slightly more density around zero than the neuronized counterpart.  Figure \ref{fig:Solution_path} shows the solution paths of the Bayesian Lasso, the neuronized Bayesian Lasso, and the standard Lasso for the analysis of the Boston housing price data set, which are almost identical.  

\centerline{\framebox[1.1\width]{Figure \ref{fig:QQplots} Here}} 

\centerline{\framebox[1.1\width]{Figure \ref{fig:Solution_path} Here}} 

\subsection{Horseshoe,   Cauchy and their generalizations}\label{sec:HS}

We start with a simple result for transforming Normal to a heavy tail distribution. Then, we show some activation functions that can  make the corresponding neuronized priors approximate  the horseshoe and Cauchy priors.

\begin{lemma}\label{lem:T}
Let $T(t)=\exp(\lambda_1 sign(t) t^2)$ with $\lambda_1\in(0,1)$, and let $Z\sim N(0,1)$ and $U=T(Z)$. Then, the density function of $U$ is $f_U(u)\propto u^{-1-{{sign}(\log u)\over 2\lambda_1}} |\log (u)|^{-{1\over 2}}$ for $u>0$. If $\lambda_1 <(1+k)^{-1}$, we have $E(U^k)={1\over 2}\left( {1\over \sqrt{1-2 k\lambda_1}}+{1\over \sqrt{1+2k\lambda_1}}\right)$. 
\end{lemma}

 The proof is straightforward, and thus omitted.  This lemma implies that any polynomial tails of the local shrinkage prior 
can be constructed by ``neuronizing'' a Normal random variable through an exponential function, up to a logarithmic factor. 
We further show that the adoption of this exponential activation function induces a marginally polynomial-tailed prior on $\theta=T(\alpha)w$, as the following result: 
 \begin{proposition} \label{prop:HS} 
Let $\pi_{E}$ be the marginal density  of $\theta$ defined in \eqref{eq:neuro} with $T(t) = \exp(\lambda_1 sign(t) t^2)$ for $0<\lambda_1 \leq 1/2$. Then, for any $\kappa>0$, there exists $\theta_0$ such that $
c_1 (\log |\theta|)^{-{1\over 2}}
|\theta|^{(-1-\frac{1}{2\lambda_1})(1+\kappa)}\leq \pi_E(\theta) \leq c_2(\log |\theta|)^{-{1\over 2}}|\theta|^{(-1-\frac{1}{2\lambda_1})(1-\kappa)}$ if  $\theta>\theta_0$,
where $c_1$ and $c_2$ are some positive constants.
\end{proposition}

 As $\lambda_1$ dictates the tail behavior of  a neuronized prior with an exponential activating function, we  consider the following class of activating functions:
 \begin{equation}\label{eq:ghs}
     T(t)=\exp\{\lambda_1 \mbox{sign}(t)t^2 + \lambda_2 t+\lambda_3\},
 \end{equation}
 with $\lambda_1\geq 0$,
 which results in a class of {\it  generalized horseshoe priors}. {We recommend to choose $\lambda_3$ so that the resulting distribution for $\theta_j/\tau_w$ in \eqref{eq:neuro} has a similar interquartile range as that for the standard horseshoe distribution, i.e.,  1.1$\sim$1.5.
  We numerically found that, 
  with $T(t)=\exp\{0.5\text{sign}(t)t^2+0.733t\}$, the neuronized prior for $\theta_j/\tau_w$ 
  approximates the horseshoe prior well (the details of the numerical evaluation is deferred to Section \ref{sec:numT}). In the same sense, 
  the neuronized prior for $\theta_j/\tau_w$ approximates the
  standard Cauchy distribution if
  $T(t)=\exp\{0.5t^2-1.27t+0.29\}$.
  We may therefore regard the neuronized priors induced by $T(t)=\exp\{\lambda_1 t^2+\lambda_2 t+\lambda_3\}$ as {\it }generalized Cauchy priors}, which differ from those induced by \eqref{eq:ghs}  in having weaker shrinkage effects for weak signals because $T(t)$
  is bounded below by $\exp\left\{\lambda_3+\min\left(0,{\lambda_2 |\lambda_2|  \over 4\lambda_1}\right)\right\}$.
\color{black}

\begin{figure}[H]
    \centering              \includegraphics[width=5cm,height=4.5cm]{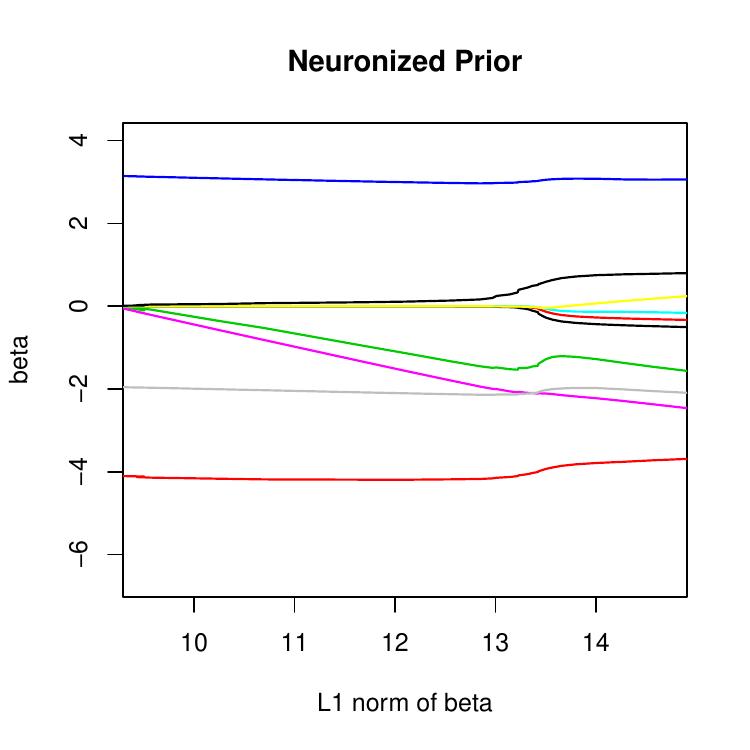}
        \includegraphics[width=5cm,height=4.5cm]{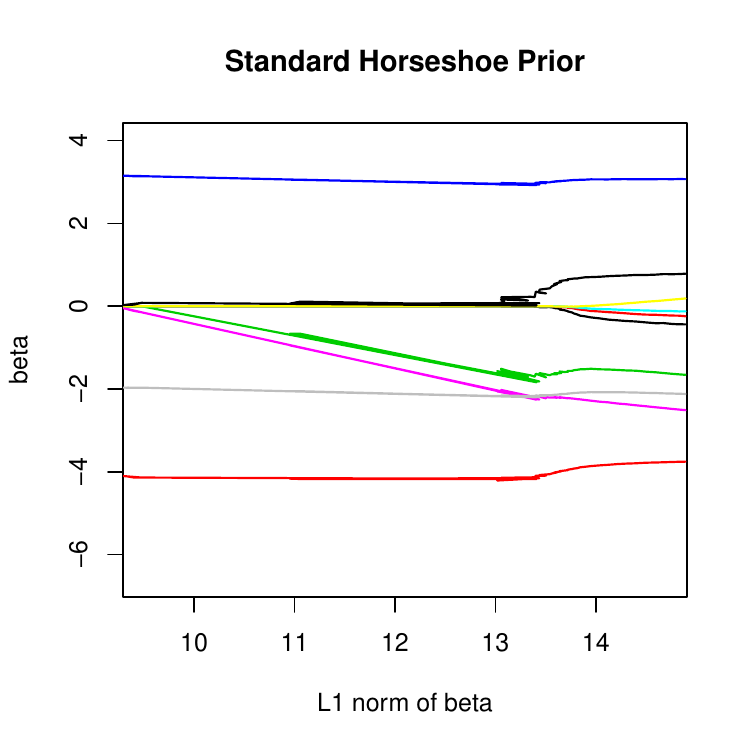}
    \caption{Solution paths of the neuronized prior  and the horseshoe prior.}\label{fig:Solution_HS}
\end{figure}

Figure \ref{fig:QQplots}(b) and (c) show  histograms contrasting the horseshoe and  Cauchy priors with their corresponding neuronized versions, respectively.
 Figure \ref{fig:Solution_HS} compares the solution paths under the neuronized and  standard horseshoe priors  for the same Boston housing price data,  demonstrating their nearly identical behaviors.
Table \ref{tab:neuro} summarizes the results in this Section. 

\centerline{\framebox[1.1\width]{Figure \ref{fig:Solution_HS} Here}}

\centerline{\framebox[1.1\width]{Table \ref{tab:neuro} Here}}

Although it covers a  large class of prior densities as demonstrated, the  current neuronization formulation as in \eqref{eq:neuro}  still has difficulties emulating some  distributions. For example,  nonlocal priors  \citep{johnson2010use,johnson2012bayesian,rossell2017non}, which   are bimodal  and symmetric around zero, cannot be easily constructed using \eqref{eq:neuro}. However, one may  still capture the bimodality of a desired prior by changing the distribution of $w$ and $\alpha$ in \eqref{eq:neuro} to be bimodal. Also, dependent prior densities cannot be represented by  a product of neuronized priors. These examples include the Zellner's $g$-prior \citep{zellner1986} and the Dirichlet-Laplace prior \citep{bhattacharya2015dirichlet}. But an extension of the neuronized prior to a multivariate version may overcome this limitation.


\def\spacingset#1{\renewcommand{\baselinestretch}%
{#1}\small\normalsize} \spacingset{1}
\section{Managing Neuronized Priors}
\subsection{Find the activation function to match a given prior}\label{sec:numT}
\def\spacingset#1{\renewcommand{\baselinestretch}%
{#1}\small\normalsize} \spacingset{1.8}
 Section~\ref{sec:neuro} presents neuronized formulations for some  popular existing priors.
More generally, if we want to find an activation function $T$  so that the resulting neuronized prior matches a  desired target  distribution $\pi(\theta)$ symmetric about zero, 
we may consider a family of activation functions $\{T_\phi\}$ parameterized by $\phi$, and then numerically find $\widehat\phi$ so that $T_{\widehat\phi}$ minimizes a certain discrepancy measure between the resulting neuronized prior and the target $\pi(\theta)$. 
For example, we can consider a family of exponential functions as in \eqref{eq:ghs} by setting $\phi=\{\lambda_1,\lambda_2,\lambda_3\}$ to construct a generalized horseshoe prior. More flexibly, the function space spanned by a class of B-spline basis functions can be a reasonable choice, i.e.,  $T_\phi(t)=\mathbf{B}(t)\boldsymbol{\phi}$, where $\mathbf{B}$ is a vector of $K$ B-spline basis functions  and  $\boldsymbol{\phi}\in \mathbb{R}^K$.

 {If we aim to match the polynomial tail of a general target prior, we can consider an additive mixture of an exponential function as in Proposition \ref{prop:HS} and a basis expansion.
 More precisely, we define a class of activation functions parameterized by 
 $\boldsymbol{\zeta}=\{\lambda_1,\boldsymbol{\phi}\}$:  
 \[T_{\boldsymbol{\zeta}}(t) = \exp\{\lambda_1 \mbox{sign}(t)t^2\}  +\mathbf{B}(t)\boldsymbol{\phi}, \ \ \mbox{with }  \lambda_1>0.\]
 These activation functions naturally  lead to polynomial tails for the corresponding neuronized priors because the effect of B-spline bases are minimal as $|t|\to \infty$.             
 To find an appropriate $\boldsymbol\zeta$, we can first find $\lambda_1$ to match the tails of the target prior based on the results of Proposition~\ref{prop:HS}. For example, if the target prior decays at the rate of $|x|^{-b}$, we choose $\lambda_1={1\over 2(b-1)}$.

   Once $\lambda_1$ is fixed,} 
   we generate a large number $S$ of i.i.d. samples from the the neuronized prior: $\widetilde\theta_{\zeta,i} = T_{\lambda_1,\phi}(\alpha_i - \alpha_0)w_i$, where  $(\alpha_i,w_i) \sim N(0,1)\times N(0,1)$, for $i=1,\ldots, S$; 
and also generate $\theta_i \stackrel{iid}{\sim} \pi(\theta)$ for $i=1,\dots,S$, where $\pi(\cdot )$ is the target prior. We measure the discrepancy between these two samples, for example, by $D(\boldsymbol{\zeta})=\sum_{i=1}^S|\widetilde\theta^{(i)}_{\zeta} - \theta^{(i)} |$, where $\widetilde\theta^{(i)}_{\zeta}$ and $\theta^{(i)}$ are the $i$-th largest value of the generated samples $\{\widetilde\theta_{\zeta,i}\}_{i=1,\dots,S}$ and $\{\theta_i\}_{i=1,\dots,S}$, respectively. Some other attractive measures  are the \emph{$l_2$ distance} or the \emph{Wasserstein distance}.
Then, we can minimize $D(\boldsymbol{\zeta})$ with respect to $\boldsymbol{\zeta}$ by using a grid search algorithm or a simulated annealing algorithm \citep{kirkpatrick1983optimization}. This optimization is not computationally intensive as long as the dimension of $\boldsymbol{\zeta}$ is  moderate. 


\subsection{Choosing hyper-parameters}\label{sec:hyper}

 Neuronized priors have two hyper-parameters: the variance of the global shrinkage parameter $\tau_w^2$ and the bias parameter $\alpha_0$. The roles of these hyper-parameters are different according to the choice of the activation function. When we consider  neuronized  continuous shrinkage priors,  $\alpha_0$ is set at 0 by default. When we use neuronized discrete SpSL prior via the ReLU activation function,  
 the prior probability for each coefficient to be non-zero is $\Phi(-\alpha_0)$. 
{As shown in Proposition \ref{prop:a0_prior}, we can impose a hyper-prior on $\alpha_0$ so that the sparsity level is adaptively controlled by the data set.  However, sampling $\alpha_0$ conditional on other parameters in Gibbs sampling is not trivial and naive random-walk proposals for a MH algorithm is highly inefficient due to its high posterior correlation with other parameters.  We describe an efficient group-move in the next section.

The choice of $\tau_w^2$ is a bit complicated. When $\mathbb{E}(T^2(\alpha))$ is bounded,  the prior expected signal-to-noise ratio for the regression model is 
$\mathbb{E}\norm{ \btheta}^2/\sigma^2$ 
$=p\tau_w^2 \mathbb{E}[T^2(\alpha_j-\alpha_0)]/\sigma^2$. Thus, the choice of $\tau_w$ needs to reflect our prior knowledge about the signal strength in the data. 
Although some theoretical analysis has been attempted on the normal means model \citep{van2014horseshoe} under a fixed $\tau_w^2$,
a theoretically justified selection of the hyper-prior for $\tau_w^2$ has not been found.}

{When $\mathbb{E}(T^2(\alpha))$ does not exist as in the horseshoe and Cauchy cases, the signal strength interpretation is not valid. As noted by \cite{carvalho2010horseshoe}, for horseshoe priors the shrinkage factor $\kappa_j=(1+T^2(\alpha_j)\tau_w^2)^{-1}$ 
determines the shrinkage level of $\theta_j$ and can be interpreted as an approximation of $\mathbb{E}(1-\gamma_j)$ in \eqref{eq:spike}. We thus numerically search  $\tau_w^2$ so that $1-\mathbb{E}(\kappa_j) = \pi_0$ for some prior belief on the proportion of non-zero parameters $\pi_0\in(0,1)$, which is set at 
$\min\{0.01,0.1\times n/p\}$ by default.}
We subsequently use this setting and show that the empirical performance of the resulting  procedure is promising in various simulation and real data examples. 

As shown in \cite{moran2018variance}, the traditional conjugate prior for linear models, i.e., $\btheta\mid \sigma^2\sim N(\bm{0}, \lambda_0{\sigma^2} I_p)$ and  $\sigma^2\sim\mbox{Inv-Chisq}(\nu_0)$ {\it a priori},
can lead to  inconsistency in high-dimensional problems. 
To avoid this undesirable situation, we assume that $w_j\sim N(0,\tau^2_w)$ and $\sigma^2\sim$ Inv-Gam$(a,b)$ are  independent {\it a priori}. 

\def\spacingset#1{\renewcommand{\baselinestretch}%
{#1}\small\normalsize} \spacingset{1}
\section{Sampling and Optimization with Neuronized Priors}\label{sec:comp}
\subsection{MCMC sampling with neuronized priors}
\def\spacingset#1{\renewcommand{\baselinestretch}%
{#1}\small\normalsize} \spacingset{1.8}

Consider the linear regression model in \eqref{eq:linearM} and the 
unnormalized  joint  distribution of $\balpha$, $\bw$, and $\sigma^2$
as in (\ref{eq:linear_post}).
  The conditional posterior distribution of $\bw$ given $\balpha$ and other hyper-parameters is Gaussian: 
  \begin{eqnarray}\label{eq:w_Gibbs}
  \bw \mid \by, \balpha, \sigma^2, \tau_w^2 \sim N(\widetilde \mu, \sigma^2 \widetilde \Sigma ), 
  \end{eqnarray}
where $\widetilde\Sigma = (D_\alpha X^\T X D_\alpha + \sigma^2\tau_w^{-2} \mr{I})^{-1}$ and $\widetilde\mu = \widetilde\Sigma D_\alpha X^\T\by$, with
$D_\alpha$ as defined in (\ref{def_theta}).
When an Inv-Gam($a, b$)  
 is imposed on $\sigma^2$, the conditional distribution of $\sigma^2$ given other parameters is
 Inv-Gam$(n/2 + a, \norm{\by - X\btheta}^2_2/2  + b).$ When $p$ is large relative to $n$, the numerical calculation of  $(D_\alpha X^\T X D_\alpha + \sigma^2 \tau_w^{-2}I)^{-1}$ is highly expensive.  \cite{bhattacharya2016fast} proposed a fast sampling procedure that reduces the computational complexity from $O(p^3)$ to $O(n^2p)$, which is employed here.  Conditional on $\bw$ and $\balpha_{(-j)}$, each $\alpha_j$ can be sampled by a naive RWMH algorithm, for $j=1,\dots,p$. Since  $w_j$ and $\alpha_j$ tend to be highly correlated {\it a posteriori}, a better strategy is to integrate out $w_j$ so as to draw $\alpha_j^\ast$ from $\pi(\alpha_j\mid \by, \bw_{(-j)}, \balpha_{(-j)})$, and then draw $w_j$ from $\pi(w_j\mid \by, \bw_{(-j)}, \balpha_{(-j)},\alpha_j^\ast)$.
 
  \centerline{\framebox[1.1\width]{Algorithm \ref{alg:MCMClin} Here}}
 
   The RWMH step in Algorithm \ref{alg:MCMClin} is  local and cheap, and is thus iterated $M$ times for sampling each $\alpha_j$. We set $M = 10$ in all our numerical examples and find the resulting  algorithm to perform well. We use $N(\alpha_j^{(t)}, 2^2)$ as the proposal distribution,
which enables $\alpha_j$ to propose efficiently between the regions $\{\alpha_j: \alpha_j<\alpha_0\}$ and $\{\alpha_j: \alpha_j\geq\alpha_0\}$. We subsequently use Algorithm \ref{alg:MCMClin} as the default to implement the posterior inference based on the  neuronized prior. Parameter $\alpha_0$ is set at 0  for neuronized continuous shrinkage priors, but will follow a hyper-prior distribution as in (\ref{eq:a0_prior}) for neuronized SpSL priors, whose MCMC update is detailed next.

\subsection{Sampling $\alpha_0$ efficiently}
\label{sec:MTM}
For neuronized discrete SpSL priors, we may want to impose a prior distribution on $\alpha_0$ to accommodate some vague prior knowledge of the sparsity level as in Proposition \ref{prop:a0_prior}. 
Due to  high correlation between $\alpha_0$ and the $\alpha_j$'s, however, a naive MH approach in which $\alpha_0$ is updated by a MH step conditioned on $\balpha$ is highly inefficient. 
To overcome this difficulty, we consider a group-move via the generalized Gibbs sampling formulation \citep{liu2000generalised}: update $\balpha$ and $\alpha_0$ simultaneously by a common shift $\delta\in\mathbb{R}$. More precisely, $(\balpha, \alpha_0)$ is updated as 
\begin{eqnarray*}
(\balpha,\alpha_0)\to (\balpha + \delta\mathbf{1}, \alpha_0 + \delta),
\end{eqnarray*}
where $\delta$ is drawn from the distribution  $g(\delta)\propto \pi^*(\balpha+\delta\mathbf{1}, \alpha_0 + \delta)$, where $\pi^*(\balpha, \alpha_0)$ is the conditional posterior density of $\balpha$ and $\alpha_0$. After this group-move, it is necessary to update each $\alpha_j$ conditionally to distinguish the individual posterior behavior, but we do not need to  consider an extra step to update  $\alpha_0$ individually.  

When the  prior is of the form \eqref{eq:a0_prior} with $a_0=b_0=1$,   $g(\delta)$ is simply Gaussian:
\begin{eqnarray}\label{eq:update_a0}
N((\balpha^\T\balpha+\alpha_0 )/(p+1), (p+1)^{-1}). 
\end{eqnarray}
However, when $a_0\neq1$ or $b_0 \neq1$, the distribution $g(\delta)$ is non-standard, and an extra approximation step is needed for updating $\delta$. To this end, we propose a multiple-try MH independence sampler (MTM-IS) to sample $\delta$, following the ideas in \cite{liu2000multiple}. This algorithm proposes multiple candidates $\delta_{1},\dots,\delta_{m}$   drawn independently from a proposal distribution (such as the Gaussian distribution in \eqref{eq:update_a0}), and then chooses one from them with  probability proportional to their importance weights. The acceptance-rejection ratio is adjusted to account for this selection effect.
The detailed algorithm is as follow.

  \centerline{\framebox[1.1\width]{Algorithm \ref{alg:MTM-IS} Here}}


A proof of the correctness of this algorithm follows immediately the approach in \cite{liu2000multiple} and thus omitted. 

\subsection{MCMC strategies for discrete SpSL priors}\label{sec:relu}
A most direct and effective approach for conducting sparse Bayesian linear regression is to employ a discrete SpSL prior for the coefficients. When the continuous component of this prior is conjugate to the Gaussian likelihood,  a well-known computational strategy  is the {\it collapsed Gibbs sampler} \citep{liu1994collapsed}, which integrates out all the continuous parameters (e.g., regression coefficients) and samples, via MCMC, the binary indicator vector $\bgamma$ defined in \eqref{eq:spike} from the posterior distribution $\pi(\bgamma\mid \by, \sigma^2) = \frac{m_\gamma(\by\mid \sigma^2)h(\bgamma)}{\sum_{\gamma'}m_{\gamma'}(\by\mid \sigma^2 )h(\bgamma')}$, 
 where $m_\gamma(\by)$ is the marginal likelihood of $\bgamma$ and $h(\cdot)$ is the model prior mass function.
 Note that $\sigma^2$ is still present in the marginal likelihood because our prior is not fully conjugate with respect to the error variance.
 This collapsed sampler can become highly inefficient 
 {if one calculates the marginal likelihood by brute force at every iteration. A more efficient strategy  is to update the  required matrix inversion and determinant incrementally. For example, to add or remove a variable from the current model, we need to modify the sample covariance matrix by adding or deleting one row and one column. The corresponding inverse and determinant can be updated using the formulas in Section~\ref{appendix:algebra} of Supplementary Materials. However, even with this efficient implementation, the fully collapsed sampler is still rather slow.}
 

 {Alternatively, we can consider a half-collapsed sampling  strategy, which appears to be computationally more efficient. Instead of integrating out all the $\theta$'s, at each iteration we sample $\gamma_j$ from the conditional distribution  $[\gamma_j \mid \btheta_{(-j)},\by]$, with $\theta_j$ integrated out, and then update $\theta_j$ conditional on $\gamma_j$. Although each iteration step of this half-collapsed sampler is less efficient than the fully-collapsed one, a major advantage of this approach is that every step is much faster to compute. A comparison between the fully-collapsed and the half-collapsed Gibbs sampler is provided in the Supplementary Materials, suggesting that the half-collapsed Gibbs sampler is ten times or  more efficient than the fully-collapsed one for the examined examples.
 
 However, both collapsing approaches become unavailable if one cannot analytically integrate out the continuous parameters. In such cases, either a crude and/or time-consuming approximation strategy, or a cleverly designed, yet case-specific, data augmentation strategy \citep{polson2013Polya}, or a much less efficient reversible-jump scheme \citep{green1995reversible}, has to be employed.}
 In contrast, the neuronized priors 
 can achieve the same effect as standard discrete SpSL priors while  permitting  more efficient computation even if one cannot marginalize out  continuous components in the joint posterior distribution.    
 When a ReLU activation function is adopted, the result below further shows  that conditional distribution $\pi(\alpha_j\mid \by, \bw_{(-j)}, \balpha_{(-j)})$ is a mixture of two truncated Gaussians and can be sampled exactly. 
 
\begin{proposition}\label{prop:cond_alpha}
Let $r_j = \by - \sum_{k\neq j}X_k\theta_k$ and $\balpha=(\alpha_1,\dots,\alpha_p)$,  and 
let $N_{tr}(a,b;c,d)$  denote the truncated Gaussian with mean $a$ and variance $b$ on $(c,d)$. 
The conditional distribution $ [\alpha_j\mid \balpha_{-j},\bw,\by,\sigma^2]$  based on the posterior distribution  \eqref{eq:linear_post} with  the ReLU
activation function is
$\kappa N_{tr}(0,1;-\infty, \alpha_0 ) + (1 - \kappa) N_{tr}(\widetilde\alpha_j, \widetilde \sigma_j^2; \alpha_0, \infty),$ 
where 
$\widetilde\alpha_j = \frac{(r_j+X_j\alpha_0w_j)^\T X_jw_j}{X_j^\T X_jw_j^2 + \sigma^2 }$, $ \widetilde\sigma^2_j  = \sigma^2\left(X_j^\T X_jw_j^2 + \sigma^2\right)^{-1}$, and
\begin{eqnarray*}
\kappa &=&  \frac{\Phi(\alpha_0)\exp\left\{ -\frac{\norm{r_j}^2_2}{2\sigma^2} \right\}}{\Phi(\alpha_0)\exp\left\{ -\frac{\norm{r_j}^2_2}{2\sigma^2} \right\} + \left\{  1- \Phi\left(\frac{\alpha_0 - \widetilde\alpha_j}{\widetilde\sigma_j}\right)\right\} \widetilde\sigma_j \exp\left\{\frac{ \widetilde\alpha_j^2}{2\widetilde\sigma_j^2} - \frac{\norm{r_j + X_j\alpha_0w_j}^2_2 }{2\sigma^2} \right\} }. 
\end{eqnarray*}
\end{proposition}

   There is another computational advantage of using the ReLU activation function. When sampling $\bw$ in a Gibbs step, the conditional posterior distribution can be decomposed as a product of  independent Gaussian densities so that the numerical inversion of the $p\times p$ matrix $\widetilde\Sigma$ in \eqref{eq:w_Gibbs} can be avoided. 
 We can rewrite that 
   \begin{equation*}
      \widetilde\Sigma = \begin{pmatrix} 
\widetilde\Sigma^* & 0 \\
0 &  \sigma^{-2}\tau_w^{2}\mr{I}
\end{pmatrix}, \ \ 
\quad
\widetilde\mu = \begin{pmatrix} 
\widetilde\mu^* \\
0
\end{pmatrix},
   \end{equation*}
in \eqref{eq:w_Gibbs},  where $\widetilde\Sigma^*=(D_\alpha^*X^{*\T}X^*D_\alpha^* + \sigma^2\tau_w^{-2} \mr{I})^{-1}$, $\widetilde\mu^* = \widetilde\Sigma^{*}D_\alpha^*X^{*\T}\by$,
and $D_\alpha^*$ and $X^*$ are the sub-matrices induced by the index of the nonzero regression coefficients. This expression means that for those $j$ with $\alpha_j<\alpha_0$,  coefficient $\theta_j$ is set to zero and the sampling of $w_j$ follows  $N(0,\sigma^2\tau_w^2)$ independently.  The conditional distribution of the sub-vector $\bw^*=\{w_j: \alpha_j>\alpha_0\}$ is  $N(\widetilde\mu^*, \sigma^2\widetilde\Sigma^*)$. To sample $\bw^*$, we only need to compute $\widetilde\Sigma^*$,
which has a much smaller size than the $p\times p$ matrix $\widetilde\Sigma$, reducing computational complexity from $O(p^3\wedge np) $ to $O\left(|\bw^*|^3\wedge p \wedge n |\bw^*|\right)$, where $a\wedge b$ is the minimum operator between $a$ and $b$.


\subsection{A scalable  algorithm for finding posterior modes}\label{sec:opt}

  For massive-sized data sets, MCMC algorithms may not be practical and one needs to consider optimization-based algorithms. 
  We here propose the Coordinate-Ascent Algorithm for Neuronized priors (CAAN)  to find the MAP estimator. 
CAAN   adopts a warm start strategy as in  \cite{rovckova2016spike}  by initiating with a hyper-parameter that results in a weak shrinkage and increasing gradually the strength of the shrinkage. 
  While this warm start strategy requires multiple implementations of the optimization  with various hyper-parameters,
  it reduces the chance of being trapped in a local optimum. 
  Although it cannot be guaranteed to converge to a global optimum,  empirical results 
  in Sections \ref{sec:sim} and \ref{sec:real} show that CAAN  performs  similarly as SSLasso and significantly better than other considered methods. 
 
  \centerline{\framebox[1.1\width]{Algorithm \ref{alg:Opt} Here}}
  
A key to the success of CAAN is the optimization with respect to $\alpha_j$ while fixing other parameters, $\balpha_{(-j)}$ and $\bw$.  
Because the function of $\alpha_j$ in ($\Diamond$)  of  Algorithm \ref{alg:Opt} is a linear combination of a quadratic function and a function of $T(\alpha_j -\alpha_0)$,
we divide the optimization space into two parts: $\{\alpha_j: \alpha_j>\alpha_0\}$ and $\{\alpha_j: \alpha_j\leq \alpha_0\}$,
and find a local maximum from each part. Then, we  update $\alpha_j$ to the  best of the two local maxima. 
This one-dimensional optimization problem can be easily solved by  
many existing algorithms
and we adopt the secant algorithm of \cite{brent1973algorithms}.
Vector $\bw$ is updated jointly conditioning on $\balpha$ by taking advantage of the Gaussian conjugacy. 

The algorithm employs a temperature scheme to help with the optimization task. With $t$ taking values in an $(2L+1)$-level schedule,  $t_0\geq \cdots \geq t_{2L}=1$, CAAN maximizes the objective function 
\[-{1\over 2t\sigma^2} \norm{\by - X\btheta(\balpha,\bw)} - \frac{\balpha^\T\balpha}{2t} - \frac{\bw^\T \bw}{2t\tau_w^2} - {n\over 2}\log\sigma^2\] with respect to $\balpha$, $\bw$, and $\sigma^2$. At each temperature $t_k$, we conduct coordinate ascent iterations  $M$ times. This approach is different from simulated annealing  \citep{kirkpatrick1983optimization} in that (a) the term ${n\over 2} \log \sigma^2$ is free of the temperature, (b) temperature $t$ is bounded below by one, and (c) we do coordinate-ascending  instead of MCMC sampling at each iteration. 
Consequently, at a warm temperature the  solution tends to select a large-sized model, and irrelevant features get eliminated as the temperature decreases. At default, we set $M = 20$, $L=10$, $N = 20$,  $t_k=\left(3-{2k\over L}\right)^2$ for $k=0,\ldots,L$, and $t_{L+1}=\cdots=t_{2L}$=1.  
To reduce the chance of getting trapped in a local optima, in the first ${L }$ levels of schedule,  we add a random noise $\xi\sim \mbox{Exp}(1)$ to $\sigma^2$
after every $N$ iterations. 

For the ReLU activation function, $\alpha_0$ affects the sparsity level since it sets  the prior probability  for each coefficient to be non-zero as $\psi_0=\Phi(-\alpha_0)$. 
By using Proposition \ref{prop:a0_prior}, we deploy a hyper-prior on $\alpha_0$ so that the induced prior on $\psi_0$ is  $Beta(a_0,b_0)$. 
As a default in all SpSL procedures, we set $(a_0, b_0) = (1,1)$, and this  beta-binomial prior on the sparsity has been shown to have a strong effect on multiplicity control \citep{scott2010bayes}.  

\subsection{Comparisons with other posterior optimization procedures} \label{comp_optim}

We consider four optimization procedures for the Bayesian SpSL variable selection problem: a Majorization-Minimization (MM) algorithm \citep{yen2011majorization}, EMVS \citep{rovckova2014emvs}, SSLasso \citep{rovckova2016spike}, and our CAAN. To compare the algorithms and track their solution paths, we adopt as goodness measures the mean-squared error (MSE) and the \emph{Extended Bayesian information criterion} (EBIC; \cite{chen2008}), i.e., 
\begin{equation}
    \mbox{EBIC}({\bf k}) = \mbox{BIC} + \zeta |{\bf k}|\log p,
\end{equation} 
where ${\bf k}$ denote the set of selected variables
$\zeta$ is a tuning parameter, and BIC is the \emph{Bayesian information criterion} \citep{schwarz1978estimating}.  We set $\zeta = 1$ as suggested   by \cite{chen2008}.

MM finds the MAP estimator of our problem
by approximating the $l_0$-norm by a continuous function:
$\norm{\btheta}_0 = \lim_{\tau_3 \to 0}\sum_{j=1}^{p}\log(1+\tau_3^{-1}|\theta_j|)/(\log(1+\tau_3^{-1}))$. 
 In practice,  we need to choose  $\tau_3$ in advance, which strongly affects the performance of the approximation. While a smaller $\tau_3$ leads to a better approximation to the original posterior distribution, the resulting target function becomes highly non-concave and is much more difficult to optimize.
 
 EMVS  and SSLasso  were proposed to  evaluate the MAP estimator   based on an EM formulation when using a continuous SpSL prior as in \eqref{eq:spike}. The prior for EMVS  is a mixture of  $\pi_0 = N(0,\nu_0)$ and $\pi_1 = N(0,\nu_1)$, and that for SSLasso is a mixture of  $\pi_0 = {Laplace}(\lambda_0)$ and $\pi_1 = {Laplace}(\lambda_1)$, where $\nu_0 \ll \nu_1$ and $\lambda_0 \gg \lambda_1$. 
Since the spike prior part is not a point mass,   $\nu_0$ (or $\lambda_0$) needs to be carefully chosen to control how much the spike prior density is concentrated around zero.  
We impose a uniform prior on $\eta$ in \eqref{eq:spike}. We choose $\nu_1=100$ and $\nu_0^{-1} \in (1,1000)$ for EMVS; and choose $\lambda_1 = 1$ and  let $\lambda_0$ vary in (5,50) for SSLasso. To implement EMVS, we use the \texttt{EMVS} library in \texttt{R}. For SSLasso, we follow the recommendations  in \cite{rovckova2016spike}. At the beginning, we fix $\lambda_0=\lambda_1  (=1)$; then, we increase the value of $\lambda_0$ by 1 after the convergence of the optimization step and use the solution of the previous evaluation as the initial point for the following optimization. At the end, we track the solutions of SSLasso with varying $\lambda_0$.

We generate synthetic data based on the Bardet-Biedel data set \citep{scheetz2006regulation} to be detailed in Section \ref{sec:real}. Specifically, we retain the original predictors,  set the  error variance  $\sigma^2=1$, and let  the first ten elements of the coefficient vector be $\pm 2$ with random signs and the rest  zero.  
EBIC and log-MSE paths for each procedure are examined as iteration increases. For each procedure, we consider ten initial points randomly generated from i.i.d. standard Gaussian.

\begin{figure}[H]
    \centering
        \begin{subfigure}[b]{0.24\textwidth}
        \includegraphics[width=\textwidth, height=3cm]{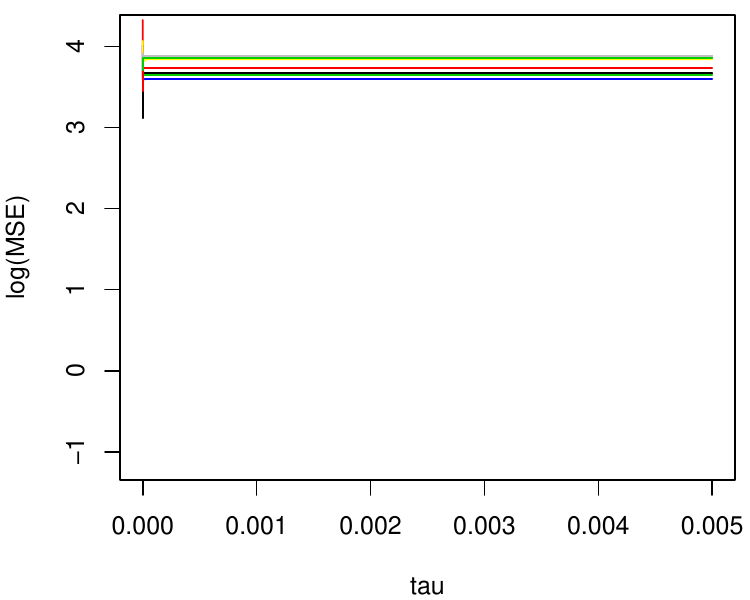}
\end{subfigure}                
\begin{subfigure}[b]{0.24\linewidth}
        \includegraphics[width=\textwidth, height=3cm]{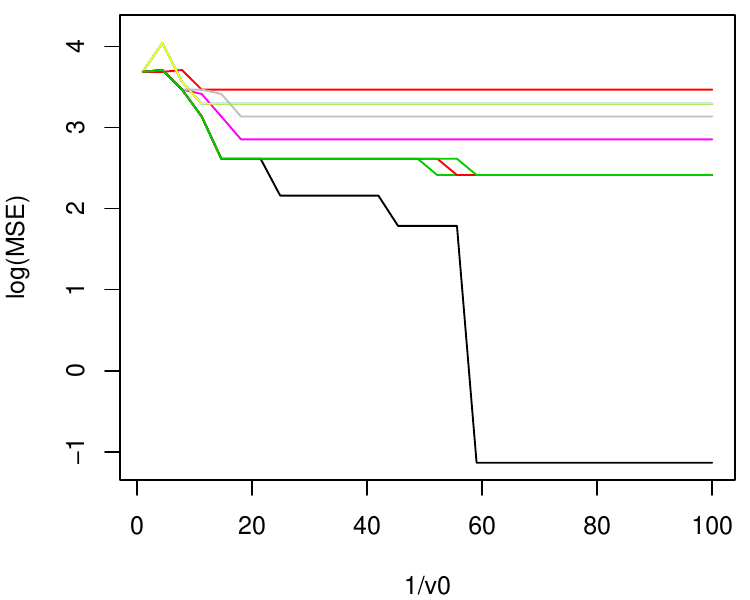}
\end{subfigure}
 \begin{subfigure}[b]{0.24\linewidth}
        \includegraphics[width=\textwidth, height=3cm]{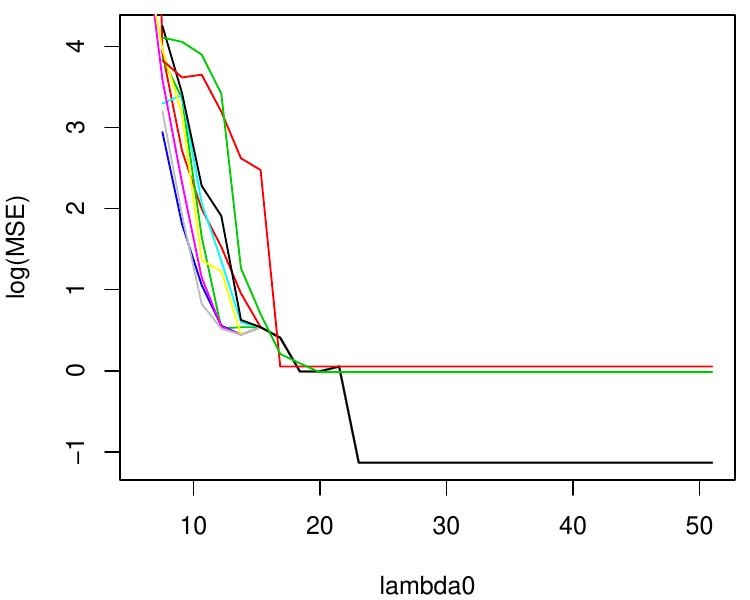}
\end{subfigure} 
 \begin{subfigure}[b]{0.24\linewidth}
        \includegraphics[width=\textwidth, height=3cm]{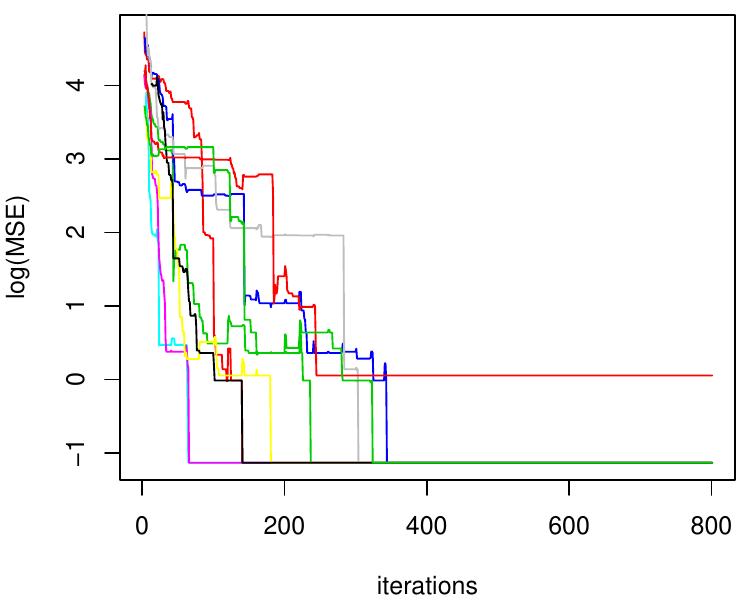}
\end{subfigure}
        \begin{subfigure}[b]{0.24\textwidth}
        \includegraphics[width=\textwidth, height=3cm]{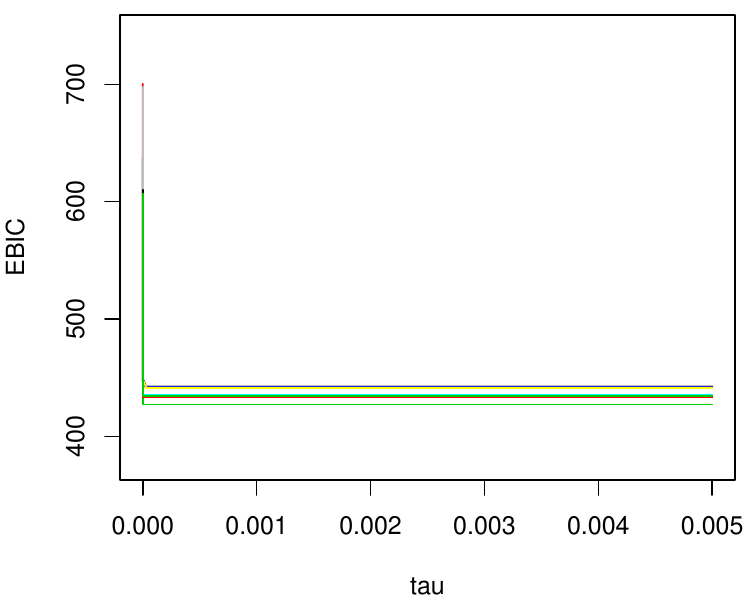}\caption{MM}
\end{subfigure}                
\begin{subfigure}[b]{0.24\linewidth}
        \includegraphics[width=\textwidth, height=3cm]{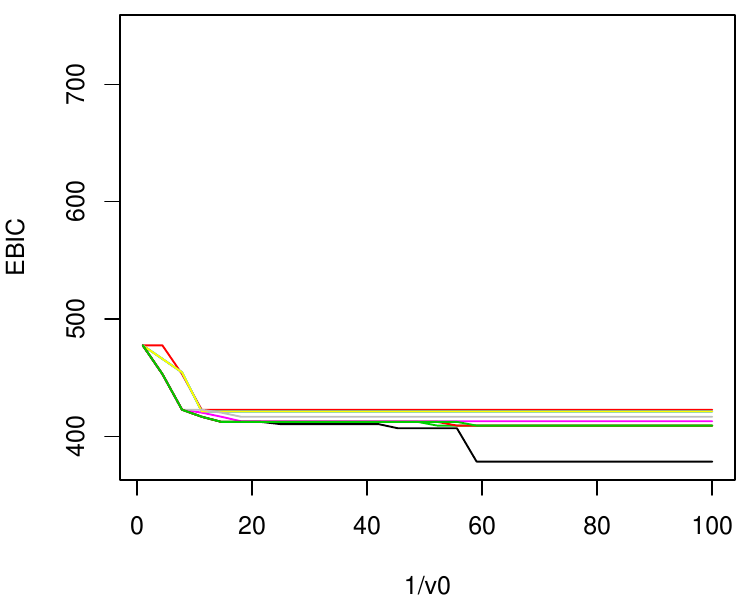}\caption{EMVS}
\end{subfigure}
 \begin{subfigure}[b]{0.24\linewidth}
        \includegraphics[width=\textwidth, height=3cm]{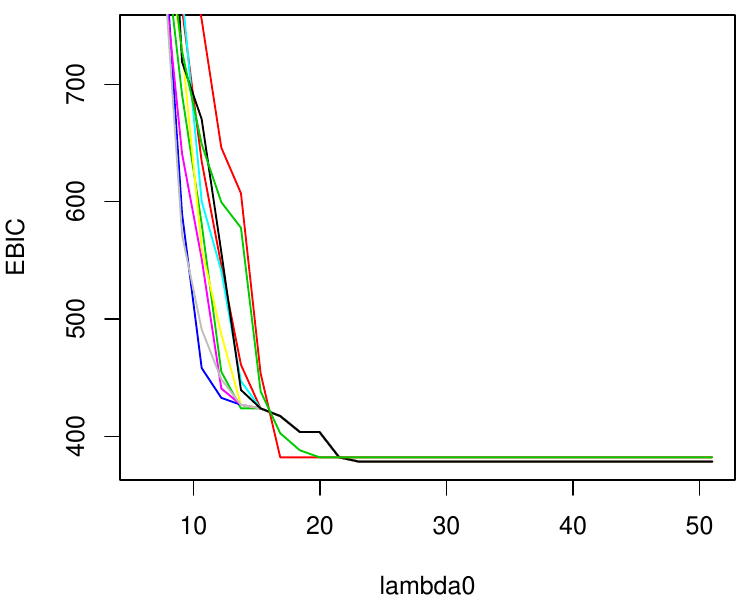}\caption{SSLasso}
\end{subfigure} 
 \begin{subfigure}[b]{0.24\linewidth}
        \includegraphics[width=\textwidth, height=3cm]{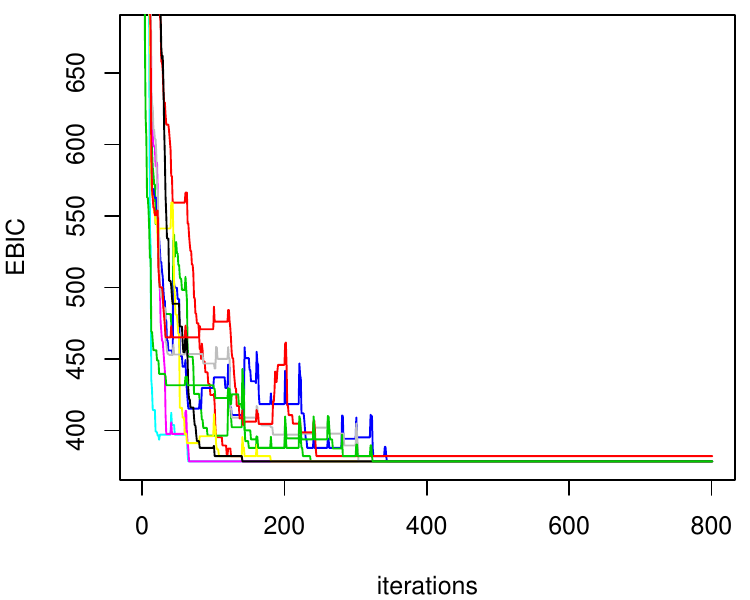}\caption{CAAN}
\end{subfigure}
\caption{Trace plots of the log-MSE (top row) and EBIC (bottom row) paths from 10 different initial points for the four optimization algorithms, based on a synthetic data set generated from the Bardet-Biedl dataset ($n= 120$ and $p=  200$) with the true model
10.  The MM procedure used $\tau_3=10^{-2}$.
}\label{fig:p10}
\end{figure}

 Figures  \ref{fig:p10} displays the optimization paths of  MM, EMVS, SSLasso, and CAAN. 
 We observe that the optimization paths of MM and EMVS quickly converged to some sub-optimal models,
corresponding to different solutions when started with different random initializations. 
 Although all procedures failed to provide consistent results when initialized with different starting configurations, CAAN and SSLasso showed  similar behaviors and were more stable than EMVS  and MM in that the searched models of CAAN and SSLasso tend to have smaller EBIC values. In the Supplementary Materials, we also provide an additional example where the true model size is five, and all methods performed better. In particular, CAAN and SSLasso consistently chose the same model with ten different initializations.
 

  \centerline{\framebox[1.1\width]{Figure \ref{fig:p10} Here}}

\def\spacingset#1{\renewcommand{\baselinestretch}%
{#1}\small\normalsize} \spacingset{1}
\section{Theoretical Properties of Neuronized Priors }\label{sec:theory}
\subsection{Posterior contraction rates}\label{sec:theory1}
\def\spacingset#1{\renewcommand{\baselinestretch}%
{#1}\small\normalsize} \spacingset{1.8}

Because  neuronized priors are naturally related to standard ones as demonstrated in Section \ref{sec:neuro},   existing theoretical results for standard frameworks can also be applied to their neuronized counterparts. 
In this section, we formalize more specific conditions on neuronized priors to achieve optimal theoretical properties as with standard Bayesian sparse regression procedures in high-dimensions.  

We first introduce some notations. For two sequences $a_n$ and $b_n$, $a_n \succ b_n$ means that $a_n/b_n \to \infty$ as $n\rightarrow\infty$,  $a_n\succeq b_n$ indicates that $b_n=O(a_n)$, and $a_n \asymp b_n$ denotes that the asymptotic rates of $a_n$ and $b_n$ are the same. For a symmetric matrix $A$, $\lambda_{min}(A)$ and $\lambda_{max}(A)$ denote the minimum and maximum eigenvalues of $A$, respectively. 
We assume that the true regression coefficient vector  $\btheta_0\in\mathbb{R}^p$ is indeed sparse, and we denote the corresponding  set of relevant variables as $\bt=\{j:\theta_{0,j}\neq 0\}$. The size of a finite set $\bk$ is denoted by $|\bk|$, the  sub-matrix of $A$ implicated by the index set $\bk$ is $A_\bk$, and the corresponding sub-vector of $\btheta$ is $\btheta_\bk$.  

We say that the posterior contraction rate of a parameter $\btheta\in\mathbb{R}^p$
is  $\epsilon_n$, if for any constant $M$,   $\sup_{{{\btheta}_0}}\mathbb{E}_{{\btheta_0}}\left\{\pi\left[d(\btheta,\btheta_0)>M\epsilon_n\mid \by, X\right]\right\} \to 0$, where $\mathbb{E}_{\btheta_0}$ is the expectation with respect to the sampling distribution of the data under the true parameter $\btheta_0$, and $d$ is a discrepancy measure, such as the $l_1$ or $l_2$ distance.
It has been shown that the minimax optimal contraction rate can be achieved
for linear regression coefficients under 
discrete SpSL priors \citep{castillo2015bayesian},  continuous SpSL priors \citep{rovckova2016spike, rovckova2018bayesian, Narisetty2014}, and  continuous shrinkage priors \citep{song2017nearly, bhattacharya2015dirichlet, ghosh2017asymptotic}. 
To obtain sufficient conditions for neuronized priors to achieve  desirable theoretical properties, we consider the following conditions.

\vspace{0.1in}

\noindent{\bf Regularity conditions:}
There exist constants $C_1, C_2, C_3, C_4>0$ such that

\noindent (A1) Sparsity: $|\bt|^2 (\log p) /n = o(1)$. \\
\noindent (A2) Feature magnitudes: $C_1\sqrt n\leq\min_{1\leq j \leq p}\norm{X_j}_2\leq\max_{1\leq j \leq p}\norm{X_j}_2 \leq C_2\sqrt n$.  \\
\noindent (A3) Eigenvalues of the design matrix: $\inf_{\bk:|\bk| \leq |\bt|\log n} \lambda_{min}(X_{\bk}^\T X_{\bk}) > C_3 n$. \\
\noindent (A4) Signal strength: $\min_{j\in\bt}\theta_{0,j}^2 \succ |\bt|\log p /n$ and $\max_{j\in\bt}\theta_{0,j}^2 <C_4$. 

\vspace{0.15in}

Condition (A3) is commonly considered in recovering the true model \citep{buhlmann2011statistics,  song2017nearly, shin2018scalable, Kim2012, Narisetty2014} when $p$  increases much faster than $n$.
Condition (A4) is imposed to prevent degenerating situations where the true coefficients decay or diverge at an extremely fast rate.
\begin{theorem}\label{theo:post_SpSL}
  Assume that \emph{(A1) -- (A4)} hold and $\sigma^2$ is known. Suppose,  for the neuronized prior defined 
  in Definition \ref{def_np} with $T$ be the ReLU function, 
  $(n\log p )^{-1}/16\leq\tau_w^2\leq n^{-1}p^2$ and $\alpha_0$ follows the distribution in (\ref{eq:a0_prior}) with $(a_0,b_0) = (1, p^u)$ for some constant $u>1$. Then, the  posterior distribution based on  this neuronized prior achieves the optimal posterior contraction rate $\epsilon_n$, i.e.,
 \begin{eqnarray}\label{eq:post_rate0}
  \epsilon_n =\begin{cases}
    |\bt|\sqrt{\log p /n}, \:\: \mbox{under $l_1$ norm}, \\
    \sqrt{|\bt|\log p/n}, \:\: \mbox{under $l_2$ norm}.
  \end{cases}
  \end{eqnarray}
\end{theorem}

\cite{song2017nearly} investigated  a similar posterior contraction problem 
under standard continuous shrinkage priors. They showed that when the tails of a prior decay at a polynomial rate and the prior possesses enough density around the true regression coefficients, the resulting posterior distribution contracts to the true coefficient at the optimal minimax rate. Following their approach, we show that the same claim can be applied to the neuronized version of continuous shrinkage priors as follows:

\begin{theorem}\label{theo:post_conti}
Assume that \emph{ (A1) -- (A4)} hold and $\sigma^2$ is known. Suppose that $T(t) =  \exp\{ t^2/\{2(r-1)\} \}$ for $r\geq 2$, and let $ \tau_w \preceq p^{-(u+1)/(r-1)}|\bt|\log p/n$ and $-\log \tau_w = O(\log p)$ for some $u>0$, and  $\alpha_0 = 0$. Then, the posterior distribution of $\theta$ based on the corresponding neuronized prior achieves the optimal contraction rate in \eqref{eq:post_rate0}.
\end{theorem}

Two practical implications follow immediately from these theorems: for  discrete neuronized priors, cares are required for specifying a hyper-prior on  $\alpha_0$ (in particular, the choice of $b_0$) to control the asymptotic sparsity level; for continuous neuronized  priors, the choice of the activation function is important.      

\subsection{Convergence of naive MCMC algorithms}\label{sec:fast}

Convergence properties of MCMC algorithms have been of interest to many researchers. In particular,  \emph{geometric ergodicity} of the Markov chain underlying a practical MCMC algorithm has been deemed  necessary  \citep{johnson2013component, roberts1996geometric, jarner2000geometric,  roberts2004general}. 
A Markov chain with a transition kernel $P$ and the target distribution 
$\pi(\cdot\mid\by)$ is said to be geometrically ergodic if,  $\forall \ \btheta^{(0)}\in\Theta$ and $\forall t=1,2,\ldots,$   $\norm{P^t(\btheta^{(0)}, \cdot) - \pi(\cdot\mid\by) }_{TV} \leq C(\btheta^{(0)})\rho^t$, for some $\rho\in(0,1)$ and a finite function $C(\cdot)$, where  $\norm{W - G}_{TV} = \sup_{A \in \mathcal F}|W(A) - G(A)|$, with  $\mathcal F$  being a Borel $\sigma$-algebra of subsets of $\Theta$, is called the total variation  between distributions $W$ and $G$. A geometrically ergodic Markov chain is called \emph{uniformly ergodic} if $C$ is uniformly bounded on $\Theta$. 
Geometric ergodicity implies that a generalized central limit theorem is valid 
for estimates based on
MCMC samples
\citep{atchade2011kernel, flegal2011implementing, jones2006fixed}.



\cite{tan2013geometric} investigated convergence behaviors of MH algorithms with different proposal distributions and showed that   a Gibbs sampler and its MH-within-Gibbs algorithm either are both geometrically ergodic  or are both not. By using this fact, we show  geometric ergodicity of Algorithm \ref{alg:MCMClin} for a wide class of neuronized priors characterized by activation functions with {\it stable tables}, including all cases discussed previously.  



\begin{defn}\label{def:stable}
Function $T(x)$, $x\in \mathbb{R}$, is said to have {\it stable tails}
if 
there exist constants $ C_1, C_2, C_3 >0$ such that (a) when $ x<-C_3$, either $|T'(x)|\leq C_1$ or $|T'(x)|\geq C_2$ and the sign of $T'(x)$ does not change;
and (b) when $ x>C_3$, either $|T'(x)|\leq C_1$ or $|T'(x)|\geq C_2$ and the sign of $T'(x)$ does not change.
\end{defn}

\begin{theorem}\label{theo:ge_N}
Consider the case with $X$ being orthogonal,  $\sigma^2$  known, and $\alpha_0$  fixed. Suppose the activation function $T$ for a neuronized prior  has  stable tails. 
Then,  Algorithm \ref{alg:MCMClin}  is geometrically ergodic. 
\end{theorem}

\begin{theorem}\label{theo:ge_HS} Under the standard Bayesian linear regression setting, suppose we employ
a standard continuous shrinkage prior as in \eqref{eq:shrink} with a heavy-tailed distribution $\pi_\tau$ such that $\pi_\tau (x)\succeq \exp\{-cx^{\kappa}\}$, $x>0$, for some constants $c>0$ and $0<\kappa<1$. 
Then, the corresponding MCMC algorithm  cannot achieve geometric ergodicity if one updates $\tau_j$ conditional on other variables by a RWMH algorithm.
\end{theorem}

Theorem \ref{theo:ge_N} implies that  a naive MH algorithm can be practical for  neuronized priors provided that the activation function is  not too erratic. All activation functions in Table \ref{tab:neuro} attain stable tails, so the considered neuronized Bayesian shrinkage procedures achieve a fast convergence of their MCMC. In contrast, Theorem~\ref{theo:ge_HS} shows that, under the conventional setting, geometric ergodicity  cannot be achieved by a RWMH algorithm under a  heavy-tailed prior on $\tau_j$; e.g.,  the horseshoe prior. In this setting, the conditional posterior distribution of $\tau_j$, used in the Gibbs sampler, is also heavy-tailed (at least sub-exponential). As shown in \cite{mengersen1996rates}, when the target distribution of a RWMH algorithm is heavy-tailed, the resulting MCMC algorithm cannot be geometrically ergodic.  


To attain an optimal rate of posterior contraction, however, we need to choose a heavy-tailed prior on $\tau_j$'s  as discussed in Section~\ref{sec:theory1}. 
Thus, some clever, but case-specific, MCMC moves need to be designed. For example, using a slice sampler for updating $\tau_j$ in horseshoe priors can be shown to be geometrically ergodic \citep{roberts1999convergence}. 
Even so,  empirical results in Sections \ref{sec:sim} and \ref{sec:real} show that employing the neuronized horseshoe prior with Algorithm \ref{alg:MCMClin} is   computationally more efficient than an efficient MCMC algorithm using the slice sampling under the conventional framework, which may be due to high correlations between the $\tau_j$'s and $\theta_j$'s when using representation (\ref{eq:shrink}) for such a  prior.   
The form of the neuronized prior can be viewed as a transformed parameter expansion \citep{liu1999parameter} via an activation function, which improves the mixing property of Algorithm \ref{alg:MCMClin}. This advantage of parameter expansion is also discussed in \cite{scott2010parameter}.

\def\spacingset#1{\renewcommand{\baselinestretch}%
{#1}\small\normalsize} \spacingset{1}
\section{Simulation Studies}\label{sec:sim}
 \subsection{Simulation setups and evaluation criteria}

\def\spacingset#1{\renewcommand{\baselinestretch}%
{#1}\small\normalsize} \spacingset{1.8}

Under the Bayesian regression framework, we compare the effect of some standard priors, such as  Bayesian Lasso, the horseshoe, and the discrete SpSL as in \eqref{eq:spike}, with that of their  neuronized counterparts. We also include a scalable approximation algorithm called Skinny Gibbs (SkG; \cite{narisetty2019skinny}) for continuous SpSL priors. By ignoring the correlation between selected variables and the other variables,  SkG  improves computational efficiency.

Among the  optimization-based algorithms in  comparison, we include  two penalized  likelihood procedures, Lasso \citep{tibshirani1996regression} and
SCAD \citep{fan2001variable}. 
Cross-validations (CV) and  either BIC (when $n>p$)
or EBIC (when $n<p$) are used to select tuning parameters for both LASSO and SCAD.
As a calibration, we provide  the oracle estimate, i.e., the OLS estimate under the true model. 
Posterior mode-finding algorithms includes  MM, EMVS, SSLasso and CAAN under two neuronized  SpSL priors, in which the slab part matches either Laplace or Cauchy (denoted as  N-SpSL-L and N-SpSL-C, respectively; see Table \ref{tab:neuro}). 
We impose $a_0=b_0=1$ in \eqref{eq:model_prior},  $\nu_1=10$ for EMVS, and $\lambda_1 = 0.1$ for SSLasso. Then, we evaluate the MAP estimators based on different choices of  $\nu_0$ for EMVS and  $\lambda_0$ for SSLasso and select a value that minimizes BIC for low-dimensions and EBIC for high-dimensions.  \texttt{R} packages \texttt{EMVS} and \texttt{SSLASSO} (available on the CRAN) are used for the implementation.

 To evaluate the estimation performances, we report both the {\it Mean Squared Error} (MSE) and  the cosine of the angle  between the true  coefficient vector $\btheta_0$ and its estimate $\widehat\btheta$, i.e., ${ \btheta_0^\T \widehat\btheta\over \norm{\btheta_0}  \Vert\widehat\btheta\Vert}$, for each method.  The angle measure is more stringent as it cannot benefit from a simple shrinkage. To measure  model selection performances, we examine the {\it Matthews correlation coefficient} (MCC; \cite{matthews1975comparison}) defined as $\mbox{MCC} = \frac{\mbox{TP}\cdot \mbox{TN} - \mbox{FP} \cdot \mbox{FN}}{ \sqrt{(\mbox{TP}+\mbox{FP})(\mbox{TP}+\mbox{FN})(\mbox{TN}+\mbox{FP})(\mbox{TN}+\mbox{FN})} }$, where TP, TN, FP, and FN denote the numbers of true positives, true negatives, false positives, and false negatives, respectively. 
 The value of MCC is bounded by one, and the closer to one MCC is, the better a model selection procedure is. 
The {\it Effective Sample Size} (ESS) is adopted as an efficiency measure for a MCMC procedure, 
which is defined as $\mbox{ESS} = {N \over 1+2\sum_t^\infty \rho(t)}$,
where $N$ is number of MCMC samples and $\rho(t)$ is the lag-$t$  autocorrelation. We report the average of the ESS (per second) of the ten ``most significant'' coefficients, i.e., with the largest posterior variances. 


{We consider a Toeplitz  design (i.e., AR(1) dependence) to generate the covariates:   $X_i\sim N(0,\Sigma)$ for  $i=1,\ldots, n$, where  $\Sigma=(\sigma_{lk})$ with  
$\sigma_{lk}=0.7^{|l-k|}$  for $1\leq l,k \leq p$. Additional simulation settings,
such as  one with i.i.d. standard Gaussian covariates, can be found in  Supplementary Materials.}
Two ``low''-dimensional cases  are tested: (a) $n=100, p =50$; and (b) $n=400, p =100$. The number of nonzero $\beta_j$'s is $p_1=p/10$, with each taking $\pm 0.2$  randomly. 
Another two ``high''-dimensional cases (with $n<p$) are also tested: (c) $n=100, p=300$; and  (d) $n=150,p=1000$.
We let the  coefficient vector be  $\beta_0 = \{\pm 0.4,\pm 0.45,\pm 0.5,\pm 0.55,\pm 0.6,0,\dots,0\}$. The  error variance is set at $\sigma^2=1$ for all scenarios.

\def\spacingset#1{\renewcommand{\baselinestretch}%
{#1}\small\normalsize} \spacingset{1}
\subsection{Technicalities about computational strategies}\label{sec5.2}
\def\spacingset#1{\renewcommand{\baselinestretch}%
{#1}\small\normalsize} \spacingset{1.8}

For using regular SpSL priors in \eqref{eq:spike}, we impose a uniform distribution on $\eta$. For its neuronized version, we impose  a hyper-prior on $\alpha_0$ as in \eqref{eq:a0_prior}. We consider the Jeffrey's prior on $\sigma^2$  for all Bayesian procedures; i.e.  $\pi(\sigma^2)\propto1/\sigma^2$. For the horseshoe prior and its neuronized version, { we  numerically find a proper $\tau_w^2$ 
as discussed in Section \ref{sec:hyper}.}
For  Bayesian Lasso and its neuronized version, we choose the global shrinkage parameter that matches the tuning parameter value $\lambda_{CV}$ determined by cross-validations for the standard Lasso procedure. 

For standard discrete SpSL priors, we examine both the Gaussian  and  Cauchy distributions for the slab part. We employ the half-collapsed Gibbs sampler as discussed in Section~\ref{sec:relu}, denoted as SpSL-G(HCG) and SpSL-C(HCG) for Gaussian slabs and Cauchy slabs, respectively. 
Note that the use of a Gaussian  slab  does not match  the neuronized SpSL prior with a ReLU activation function since the product of two independent Gaussians in the neuronization formulation results in a Laplace-like slab distribution. Nevertheless, we use the standard Gaussian SpSL prior to sustain computational efficiency. 

We let ``N-SpSL-L(Exact)'' denote the neuronized SpSL prior 
implemented via the exact Gibbs sampler as in Proposition  \ref{prop:cond_alpha}, {and use ``N-SpSL-L(RW)'' and ``N-SpSL-C(RW)'',  corresponding to a Laplace-like and a Cauchy slab, respectively, to denote that implementation via   Algorithm \ref{alg:MCMClin}, which uses RWMH to update $\alpha_0$. Since  ``N-SpSL-L(RW)'' produces identical results as ``N-SpSL-L(Exact)'' but is 60\% - 80\% less efficient (see the Supplementary Materials for a detailed comparison), we omit its results from the comparison tables. }    
For the standard SpSL prior with a Cauchy slab (i.e., ``SpSL-C''), we lose the conjugacy   and need to use numerical integration (a trapezoidal rule) to  marginalize out each coefficient in a Gibbs sampler.
In contrast, its neuronized version N-SpSL-C(RW)  can be implemented by Algorithm \ref{alg:MCMClin} directly, only requiring one to choose an appropriate activation function 
as in Table \ref{tab:neuro}. 
{We note that, due to the existence of a location-shift by $\alpha_0$, the resulting neuronized prior  differs slightly from the standard SpSL prior with a Cauchy slab, although they share the same behavior at tails, i.e., decaying at the rate of $x^{-2}$.} 
The Bayesian Lasso is implemented by an efficient Gibbs sampler as in \cite{park2008bayesian}.
For the standard  horseshoe prior, we use a slice sampler to sample each local shrinkage parameter. For both procedures, since the posterior distribution does not provide a sparse solution, 
 we set a threshold of $0.1\times\widehat\sigma$,  where $\widehat\sigma^2$ is {the posterior mean of the regression error variance,}
and select only those predictors whose posterior mean estimates of the  coefficients have a magnitude higher than the threshold. 
 For all procedures, we generate  10,000  MCMC samples after 2,000 burn-in iterations,  replicate 100 data sets, and average the results over the replications. 

\centerline{\framebox[1.1\width]{Tables \ref{tab:sim_low_dep}, \ref{tab:sim_high_dep}  Here}}

\subsection{Results discussion}

Tables~\ref{tab:sim_low_dep} and \ref{tab:sim_high_dep} summarize  low-dimensional and high-dimensional simulation results, respectively. In general, we observe that (a) no procedure clearly dominate others in all situations for all criteria; (b) Bayesian averaging results in a better performance than the corresponding MAP estimator; (c) the Lasso-based procedures typically show the best estimation performance under the low-dimensional settings, but they tends to select more false positives; (d) the SpSL-based procedures attain  competitive model selection performances under high-dimensional settings.

SpSL-G(HCG) shows the most efficient performance in terms of ESS because it takes advantage of the conjugacy to marginalize continuous components, which, however, is also restrictive.
For example, with a Cauchy slab, SpSL-C(HCG) has a much reduced ESS because it has to employ a numerical integration method for marginalization.
In contrast, its neuronized counterpart N-SpSL-C(RW), which is implemented via a single unified algorithm that can accommodate any activation function, obtained an ESS 80\%  larger 
than that of SpSL-C(HCG). 

In general, neuronized priors performed 
robustly throughout all situations, with improved computational efficiency in comparison with their standard counterparts for most cases. In particular, the N-HS was at least two times more efficient than the HS in terms of ESS in all simulation scenarios, which might be due to the highly correlated latent structure between $\tau_j$'s and $\theta_j$'s in the standard horseshoe prior.  
We verified via very long MCMC iterations that our implementations of the horseshoe prior and its neuronized counterpart indeed produce identical posterior inference results (more details are given in Supplementary Materials). 
Their differences shown in the tables  are due to numerical approximation errors. 
The tables also list the performances of optimization-based SpSL procedures including the CAAN, the MM algorithm, the EMVS, and the SSLasso. The results show that,  overall, the CAAN and the SSLasso significantly outperformed the  MM and the EMVS algorithms in terms of estimation and model selection.

\section{Real Data Examples}\label{sec:real}

We analyze both the Boston housing data set introduced in Section \ref{sec:neuro} and the Bardet-Biedl data set available in the \texttt{R} package \texttt{flare}.
The Bardet-Biedl data set contains mRNA expression values of 31,042 probe sets in
eye tissues of 120 twelve-week old male rats,
normalized by the robust multi-chip averaging method \citep{irizarry2003exploration}. This data set has been analyzed  previously \citep{huang2008adaptive, kim2008smoothly, fan2011nonparametric}. As with those papers, our goal is to find a subset of probe sets  that are associated with the probe set {\it 1389163\_at}, corresponding to  gene {\it TRIM32}, which is linked to the Bardet-Biedl syndrome. 
All probe sets are ranked according to the magnitudes of their marginal correlations with {\it 1389163\_at}, and the top 200  are retained for the regression analysis ($n=120$ and $p=200$).

\begin{figure}[H]
    \centering 
    \begin{subfigure}[b]{0.45\textwidth}
        \includegraphics[width=7.3cm, height=4.1cm]{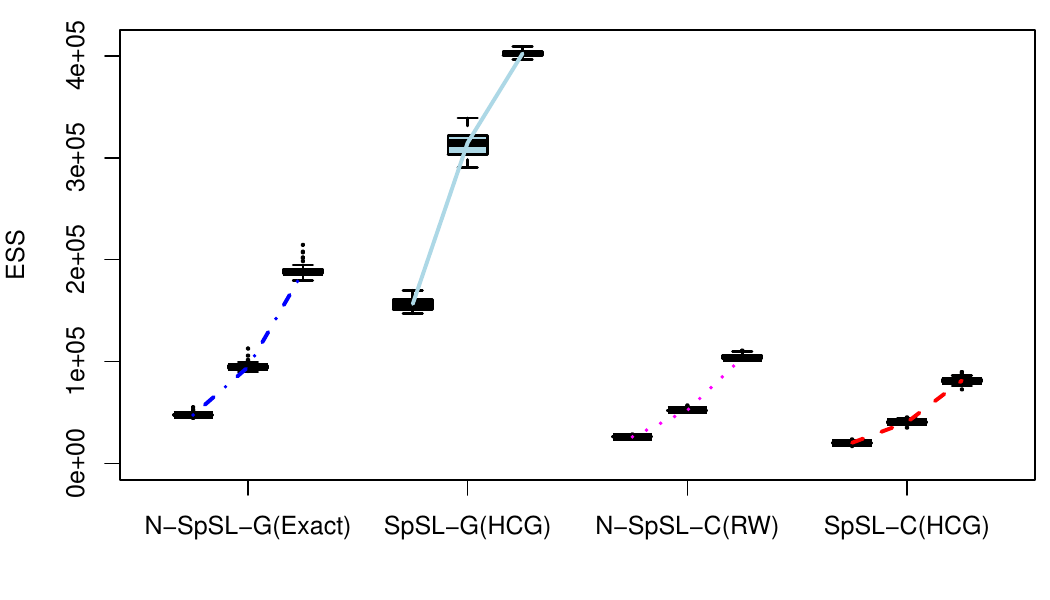}
        \caption{}
    \end{subfigure}
    \begin{subfigure}[b]{0.45\textwidth}
        \includegraphics[width=7.3cm, height=4.1cm]{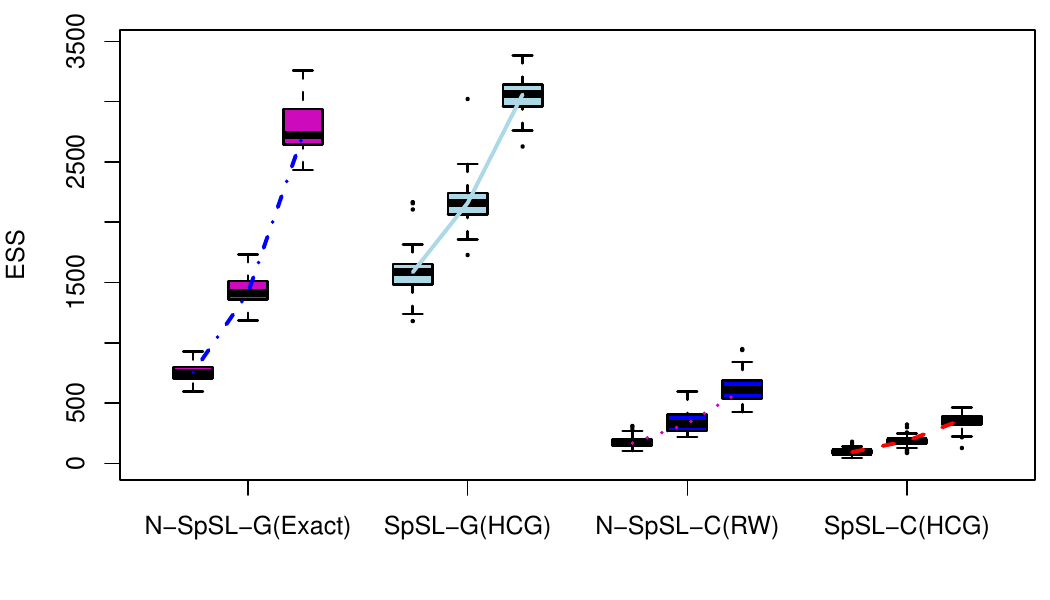}
        \caption{}
    \end{subfigure}
      \centering
    \begin{subfigure}[b]{0.45\textwidth}
        \includegraphics[width=7.3cm, height=4.1cm]{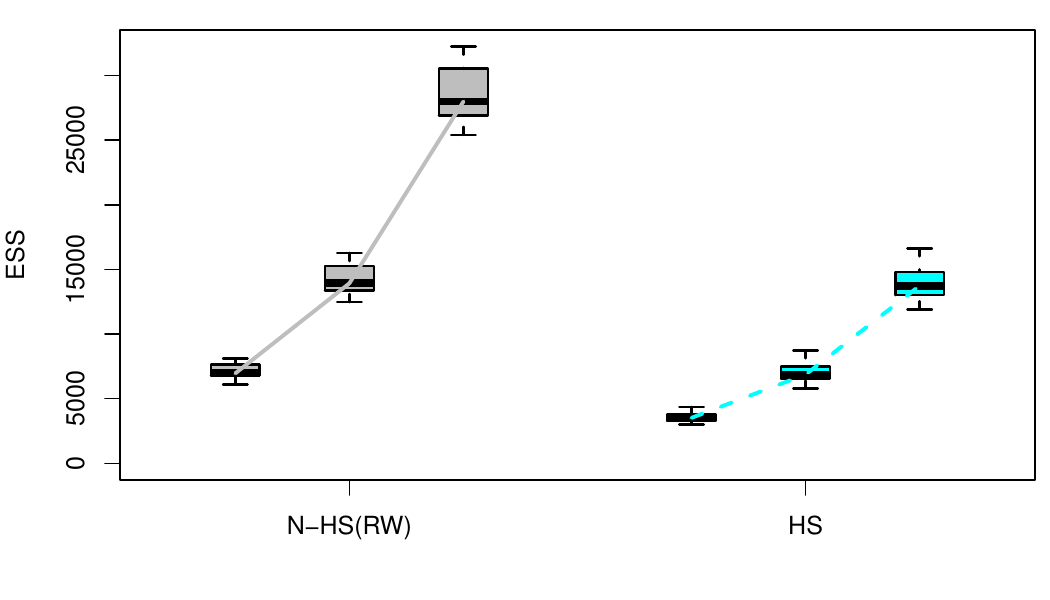}
        \caption{}
    \end{subfigure}
    \begin{subfigure}[b]{0.45\textwidth}
        \includegraphics[width=7.3cm, height=4.1cm]{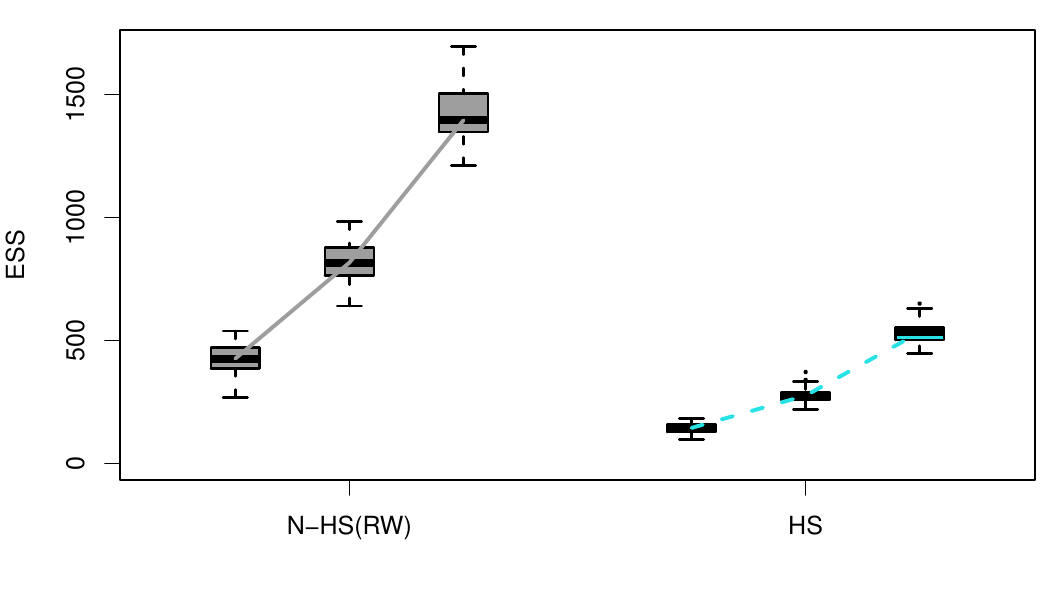}
        \caption{}
    \end{subfigure}
    \caption{Effective samples size versus actual computation time for the Boston housing data set (the first column) and the Bardet-Biedl data set (the second column). For each procedure, the first, second, and third boxplot indicates the ESS evaluated at 5 seconds, 10 seconds, and 20 seconds, respectively.   
    }\label{fig:ess_comp}
\end{figure}

Figure \ref{fig:ess_comp} 
shows the  ESS  obtained at 5 seconds, 10 seconds, and 20 seconds, respectively,  by the MCMC algorithms corresponding to different priors
for the Boston housing data set and the Bardet-Biedl data set. The efficiency comparison results are consistent with  those in the simulation study. In (a), we observe that SpSL-G(HCG) obtained the largest ESS, and N-SpSL-L(Exact) was about 50\% less efficient. With the Cauchy slab, SpSL-C(HCG) and N-SpSL-C(RW) performed similarly. For (b), the advantage of SpSL-G(HCG) over N-SpSL-L(Exact) appeared to have shrunk, and  N-SpSL-C(RW) attained 50\% more ESS than SpSL-C(HCG) does under the same time unit.
In  (c) and (d), we see clear evidences that
the neuronized horseshoe formulation results in significantly more efficient computation than the standard one. 
 
 \centerline{\framebox[1.1\width]{Figure \ref{fig:ess_comp} Here}}
 \centerline{\framebox[1.1\width]{Table \ref{tab:real1} Here}}

  We employ the out-of-sample mean squared prediction error (MSPE) to measure the prediction performance of each procedure by setting aside a randomly selected 10\% of the samples for testing.
  We also consider the cosine angle between the test responses and the corresponding predicted values; i.e., $\by_\text{ test}^\T\widehat\by/(\norm{\by_\text{ test}}_2\cdot\norm{\widehat\by}_2)$. This measure is useful  in cases where people care more about how correlated the prediction is with the observation, such as in financial market forecasting.   The process is replicated 100 times and the averages are reported in Table \ref{tab:real1},
which shows that the neuronized priors performed comparably  with their standard counterparts.   In particular, N-SpSL(MAP) achieve the smallest MSPE for both data sets.
For the Boston housing data set, 
the sizes of the models selected by different approaches are comparable.
  For the Bardet-Biedl data set,  
  however, the Bayesian Lasso and its neuronized version N-BL(RW) selected much larger models than other methods. We also noticed that both EMVS and SkG selected the null model  but had different prediction results, which is due to their adoption of different non-degenerate priors for the model parameters.

\section{Discussion}

Inspired by the idea of neuron activation, which is central to all neural network-based methods, we propose to use an activation function and a product representation to unify and extend  shrinkage priors employed in high-dimensional Bayesian regression analyses. 
 By simply changing the activation function, our unified framework (together with its companion software package) enables practitioners to easily test out effects of different classes of priors for a regression model. We show that the neuronization procedure can be efficiently implemented to emulate a wide class of distributions including many non-conjugate and  mixture priors, which is a clear advantage over existing Bayesian  regression frameworks. The neuronization formulation can also be easily extended to a broad class of nonlinear models (such as logistic regression), where the lack of prior conjugacy may hinder the applicability and scalability of conventional Bayesian regression procedures, especially when one wants to employ discrete SpSL priors.
 
 Furthermore, the neuronization idea  can be  applied to construct structured sparsity priors for more complicated models. For example,  some sparsity patterns may be spatially correlated, which is computationally challenging if one directly imposes  spatial correlations among the latent indicator variables that underlie either a discrete or a continuous multivariate SpSL prior. In contrast, a multivariate structure can be easily imposed on the $\alpha_j$'s in the neuronized prior setting (\ref{eq:neuro}). Because all parameters in such a setting are continuous and non-latent, a Hamiltonian Monte Carlo algorithm can be used to efficiently sample from the posterior distribution.

All introduced algorithms are coded in the \texttt{R} package \texttt{NPrior}  available on the CRAN. 
\subsection*{Acknowledgment}
This research is supported in part by the NSF grants 
DMS-1903139, DMS-2015528, and DMS-2015411. 

\newpage
\bibliography{myReference.bib}

\newpage

\def\spacingset#1{\renewcommand{\baselinestretch}%
{#1}\small\normalsize} \spacingset{1}
\noindent
{\bf\Large Algorithms}

\vspace{8mm}

\begin{algorithm}[H]
\caption{A general MCMC algorithm for neuronized priors}\label{alg:MCMClin}
\begin{tabbing}
Initialize the parameters $\balpha$, $\alpha_0$, $\bw$, $\sigma^2$.
\enspace {For  $i=1,\dots N$}\\
    \enspace\quad  $\bullet\:\:$Sample $\bw$ conditional on $\by, \balpha, \sigma^2$      from \eqref{eq:w_Gibbs}.          \\
   \enspace\quad $\bullet\:\:$Set $\mathbf{r} = \by - X\btheta(\balpha,\bw)$. \\
   \enspace\quad For $j = 1,\dots,p$ \\                                               
   \enspace\qquad $\bullet\:\:$Update $\mathbf{r} = \mathbf{r} + X_jT(\alpha_j-\alpha_0) w_j$.  \\
\enspace\qquad     Repeat $M$ times  \\
   \enspace\qquad\quad $\bullet\:\:$ Sample $\alpha_j$ from $[\alpha_j\mid \by,\balpha_{(-j)}, \bw_{(-j)}, \sigma^2, \tau_w^2]$ by using a RWMH step for \\
    \enspace\qquad\qquad  the log-target function  
   $   -\log (v_j )/2 - \alpha_j^2/2+ v_jm_j^2/(2\sigma^2)$,  \ \ \ \ ---$(*)$ \\
   \enspace\qquad\qquad  where $v_j = X_j^\T X_j T^2(\alpha_j - \alpha_0) + \sigma^2/\tau_w^2$ and $m_j = \mathbf{r}^\T X_j T(\alpha_j-\alpha_0)/v_j$. \\
   \enspace\qquad\quad $\bullet\:\:$Sample $w_j$ from $[w_j \mid \by, \balpha_{(-j)},\alpha_j, \bw_{(-j)},\sigma^2, \tau_w^2]$, which is    $N(m_j,\sigma^2 v_j^{-1})$. \\
   \enspace\enspace\enspace\enspace\enspace  End. \\
   \enspace\qquad $\bullet\:\:$Update $\mathbf{r} = \mathbf{r} - X_jT(\alpha_j-\alpha_0) w_j$.   \\
     \enspace\enspace\enspace End. \\
  \enspace\quad $\bullet\:\:$ Sample $\sigma^2$ from $[\sigma^2\mid \by ,\balpha,\bw,\tau_w^2]$, which is an inverse Gamma.\\
\enspace\quad $\bullet\:\:$ When $a_0 = b_0 = 1$, sample $\delta$ from \eqref{eq:update_a0}. In case where $a_0\neq  1$ or $b_0\neq 1$,\\ 
\enspace\quad draw $\delta$ via Algorithm \ref{alg:MTM-IS}. Then, update $\balpha = \balpha + \delta \mathbf{1}$ and $\alpha_0 = \alpha_0 + \delta$.\\
\enspace{End.}
\end{tabbing}
\end{algorithm}

\vspace{0.25in}

\noindent
\begin{algorithm}[H]
\caption{The Multiple-Try-Metropolized Independence Sampler (MTM-IS)}\label{alg:MTM-IS} 

\vspace{0.2in}

Let the target density be $g(\delta)$, and let the  trial/proposal density be $h(\delta)$ (a default is \eqref{eq:update_a0}). We define $w(\delta)={g(\delta)
/h(\delta)}$.
Let $\delta^{(t)}$ be the sample at step $t$.  Then, at step $t+1$,
\begin{itemize}
    \item  Draw $\delta_1,\ldots, \delta_m$ i.i.d. from the trial density $h( )$;
    \item 	Select $\delta'=\delta_j$ from  $\{\delta_1,…,\delta_m \}$ with probability $\propto w(\delta_j) $
    \item Compute $p^{(t)}=\min\left\{1,\frac{\sum_{k=1}^m w(\delta_k)}{w(\delta^{(t)})+\sum_{k\neq j} w(\delta_k)}\right\}$;  let $\delta^{(t+1)}=\delta'$ with probability $p^{(t)}$, and let $\delta^{(t+1)}=\delta^{(t)}$  with probability $1-p^{(t)}$.
\end{itemize}
\end{algorithm}

\begin{algorithm}[H]
\caption{The Coordinate-Ascent Algorithm for Neuronized prior (CAAN)}\label{alg:Opt}
\begin{tabbing}
\enspace $\bullet\:\:$Initialize the parameters $\balpha$, $\alpha_0$, $\bw$, $\sigma^2$.\\
\enspace $\bullet\:\:$Set a candidate set of temperature, $\{t_{(1)},\dots,t_{(2L+1)}\}$, where $t_{(l)}>t_{(l+1)}$ and $t_{(2L+1)}=1$.\\
\enspace For $l = 1,\dots,2L+1$\\
\enspace\quad $\bullet\:\:$Set $t = t_{(l)}$.\\
\enspace\quad $\bullet\:\:$Set $\mathbf{r} = \by - X\btheta(\balpha,\bw)$. \\
\enspace\quad {For $M$ iterations}\\
   \enspace\quad\quad For $j = 1,\dots,p$ \\                                             
   \enspace\enspace\enspace\quad\quad $\bullet\:\:$Update $\mathbf{r} = \mathbf{r} + X_jT(\alpha_j-\alpha_0) w_j$.  \\
   \enspace\enspace\enspace\quad\quad $\bullet\:\:$Update $\alpha_j$ by optimizing   the logarithm of the marginalized posterior 
   \\ \enspace\enspace\enspace\quad\quad\quad  
   density function $-\log (v_j )/2 - \alpha_j^2/2+ v_jm_j^2/2$ with respect to $\alpha_j$,  \\
   \enspace\enspace\enspace\quad\quad\quad 
   where $v_j = X_j^\T X_j T^2(\alpha_j - \alpha_0) + \sigma^2/\tau_w^2$ and $m_j = \mathbf{r}^\T X_j T(\alpha_j-\alpha_0)/v_j$. \; \; --- ($\Diamond$) 
   \\
   \enspace\enspace\enspace\quad\quad$\bullet\:\:$Update $w_j$ by $m_j$. \\
   \enspace\enspace\enspace\quad\quad $\bullet\:\:$Update $\mathbf{r} = \mathbf{r} - X_j T(\alpha_j-\alpha_0) w_j$.   \\
     \enspace\quad\quad End. \\
    \enspace\quad\quad Every $N$ iterations,\\
    \enspace\enspace\enspace\quad\quad $\bullet$ Update $\sigma^2 = (\norm{\by-X\btheta(\balpha,\bw)}_2^2/t+2b_1)/\{n+2a_1+2
    \}$. \\
    \enspace\enspace\enspace\quad\quad\quad 
    If $l < L$ \\
    \enspace\enspace\enspace\enspace\enspace\quad\quad\quad $\bullet$ Set $\sigma^2 = \sigma^2 + z$, where $z\sim \exp(1)$. \\
    \enspace\enspace\enspace\quad\quad $\bullet$ Update $\alpha_0 = \{\sum_{j=1}^p\mathbb{I}(\alpha_j>\alpha_0)+a_0-1\}/(p+b_0-2)$. \\
 \enspace\quad End. \\
\enspace End.
\end{tabbing}
\end{algorithm}

\newpage

\noindent
{\bf\Large Tables}

\vspace{6mm}

\begin{table}[H]
\centering
\begin{tabular}{|c|c|}
  \hline
   Activation function $T(t)$ & Target Prior \\
   \hline
  $\max\{0,t\}$ (ReLU)  &  Discrete SpSL  with Laplace slab \\
   $t$ (linear) &  Bayesian Lasso \\
 $\exp\{0.5\mbox{sign}(t)t^2+0.733t\}$& Horseshoe \\
   $T(t)=\exp\{0.5t^2-1.27t+0.29\}$ & Cauchy   \\
   \hline
\end{tabular}
 \caption{The choice of $T$ for neuronized priors and the corresponding existing Bayesian priors. The default value of $\tau_w^2$ is set to be one, if it is not specified. }\label{tab:neuro}
\end{table}

\vspace{0.4in}
\begin{table}[H]
\centering
\resizebox{15cm}{!}{%
\centering
\begin{tabular}{| l | ccccr | ccccr |}
\hline
               
& \multicolumn{ 5 }{ c |}{ $(n=200,p=50)$ } & \multicolumn{ 5 }{  c | }{ $(n=400,p=100)$ }\\
Method          & MSE   & Cos      & MCC      & FP    & ESS     & MSE   & Cos & MCC  & FP    & ESS    \\ \hline
Oracle      & 0.069(0.008)  & 0.870   &                         &    &     & 0.069(0.013)  & 0.929 &    &    &     \\
SpSL-G(HCG) & 0.167(0.009) & 0.558 & 0.48(0.02) & 0.01 & 15242.5 & 0.295(0.014) & 0.581 & 0.49(0.01) & 0.03 & 2594.8 \\ 
N-SpSL-L(Exact) & 0.150(0.009) & 0.592 & 0.53(0.04) & 0.03 & 5826.2 & 0.261(0.014) & 0.630 & 0.53(0.01) & 0.07 & 1089.6 \\ 
SpSL-C(HCG) & 0.159(0.009) & 0.582 & 0.51(0.03) & 0.02 & 1023.6 & 0.275(0.014) & 0.613 & 0.51(0.01) & 0.05 & 277.7 \\ 
  N-SpSL-C(RW) & 0.168(0.009) & 0.554 & 0.48(0.02) & 0.01 & 1747.8 & 0.299(0.014) & 0.574 & 0.48(0.01) & 0.03 & 422.2 \\ 
HS & 0.142(0.008) & 0.594 & {\bf 0.55}(0.03) & 0.04 & 610.1 & 0.240(0.012) & 0.658 & 0.56(0.01) & 0.04 & 91.1 \\ 
 N-HS(RW) & 0.143(0.008) & 0.594 & {\bf 0.55}(0.03) & 0.03 & 1357.0 & 0.243(0.012) & 0.653 & 0.55(0.01) & 0.04 & 217.3 \\
BL & 0.198(0.011) & 0.569 & 0.51(0.01) & 2.94 & 2798.2 & 0.273(0.010) & 0.669 & 0.60(0.02) & 3.25 & 397.5 \\ 
  N-BL(RW) & 0.157(0.008) & 0.601 & 0.53(0.01) & 1.49 & 1152.8 & {\bf 0.218}(0.009) & {\bf 0.698} & {\bf 0.62}(0.02) & 0.99 & 361.2 \\ 

SkG              & 0.159(0.008) & 0.573      & 0.50(0.01)     & 0.02  & 9961.8 & 0.276(0.010) & 0.614 & 0.51(0.01) & 0.06  & 1189.6 \\

SpSL(MM)       & 0.193(0.012) & 0.481      & 0.41(0.04)     & 1.27  &         & 0.310(0.012) & 0.582 & 0.48(0.02) & 2.77  &        \\

  EMVS            & 0.225(0.010) & 0.436      & 0.45(0.02)     & 0.00  &         & 0.412(0.013) & 0.419 & 0.40(0.02) & 0.00  &        \\
SSLasso         & 0.208(0.009) & 0.449      & 0.41(0.02)     & 0.80  &         & 0.355(0.010) & 0.510 & 0.49(0.01) & 1.26  &        \\
N-SpSL(MAP) & 0.222(0.010) & 0.537 & 0.48(0.02) & 1.03 &  & 0.310(0.013) & 0.624 & 0.57(0.02) & 1.12 &  \\ 
N-BL(MAP)       & 0.152(0.009) & {\bf 0.616}      & 0.54(0.02)     & 1.70  &         & 0.226(0.011) & 0.684 & 0.61(0.01) & 1.86  &    \\
Lasso(CV)       & {\bf 0.134}(0.009) & 0.608      & 0.48(0.02)     & 5.34  &         & 0.228(0.011) & 0.668 & 0.45(0.01) & 13.61 &        \\
SCAD(CV)        & 0.222(0.009) & 0.528      & 0.39(0.02)     & 3.72  &         & 0.339(0.011) & 0.572 & 0.38(0.02) & 8.86  &        \\
Lasso(BIC)      & 0.174(0.010) & 0.465      & 0.48(0.02)     & 0.44  &         & 0.307(0.013) & 0.516 & 0.57(0.01) & 0.56  &        \\
SCAD(BIC)       & 0.187(0.010) & 0.465      & 0.45(0.02)     & 0.69  &         & 0.361(0.014) & 0.475 & 0.50(0.02) & 1.14  &   \\ 
\hline
\end{tabular}
}
\caption{Results for the low-dimensional setting with dependent covariates. SpSL, HS, and BL indicate the procedure based on the discrete SpSL, the horseshoe, and  Bayesian Lasso priors, respectively. The sign ``N'' stands for the neuronized version of the corresponding prior.}\label{tab:sim_low_dep}
\end{table}

\begin{table}[H]
\centering
\resizebox{15cm}{!}{%
\begin{tabular}{| l | ccccr | ccccr |}
\hline
                & \multicolumn{5}{c|}{$(n=100,p=300)$}  & \multicolumn{5}{c|}{$(n=150,p=1000)$}                         \\
Method          & MSE   & Cos & MCC  & FP    & ESS    & MSE   & Cos                       & MCC   & FP     & ESS   \\ \hline
Oracle      & 0.141(0.045)  &  0.956  &                         &    &     & 0.084  & 0.977 &    &    &\\    
 SpSL-G(HCG) & 0.872(0.059) & 0.630 & 0.56(0.04) & 0.11 & 3123.3 & 0.759(0.052) & 0.675 & 0.58(0.02) & 0.09 & 709.9 \\ 
N-SpSL-L(Exact) & 0.824(0.057) & 0.641 & 0.56(0.02) & 0.26 & 837.8 & 0.709(0.051) & 0.693 & 0.62(0.02) & 0.24 & 185.3 \\ SpSL-C(HCG) & 0.829(0.057) & {\bf 0.647} & {\bf0.58}(0.02) & 0.23 & 113.8 & {\bf 0.699}(0.046) & {\bf 0.705} & 0.63(0.01) & 0.16 & 30.9 \\ 
  N-SpSL-C(RW) & 0.882(0.059) & 0.625 & 0.56(0.04) & 0.06 & 261.3 & 0.759(0.055) & 0.673 & 0.58(0.02) & 0.10 & 66.4 \\ 
HS & 0.820(0.054) & 0.629 & 0.57(0.02) & 0.88 & 15.5 & 0.765(0.049) & 0.655 & 0.56(0.01) & 3.66 & 3.5 \\ 
 N-HS(RW) & 0.813(0.054) & 0.636 & {\bf 0.58}(0.02) & 0.82 & 122.8 & 0.738(0.049) & 0.670 & 0.57(0.01) & 3.70 & 8.1   \\ BL & 1.055(0.134) & 0.455 & 0.53(0.11) & 8.84 & 78.6 & 0.984(0.058) & 0.255 & 0.57(0.18) & 0.43 & 19.1 \\ 
  N-BL(RW) & 0.902(0.065) & 0.536 & {\bf 0.58}(0.08) & 5.74 & 124.2 & 0.967(0.061) & 0.451 & 0.59(0.17) & 0.09 & 15.4 \\ SkG              & 0.939(0.063) & 0.585 & 0.54(0.02) & 0.00  & 1950.7 & 0.924(0.050) & 0.587                       & 0.53(0.03)  & 0.01   & 530.7 \\
SpSL(MM)       & 1.022(0.069) & 0.519 & 0.45(0.12) & 1.05  &        & 1.282(0.078) & 0.380                       & 0.32(0.16)  & 2.77   &       \\
  EMVS            & 1.283(0.073) & 0.385 & 0.48(0.10) & 0.00  &        & 1.327(0.083) & 0.339                       & 0.49(0.12)  & 0.00   &       \\
SSLasso         & 0.965(0.057) & 0.587 & 0.56(0.03) & 0.00  &        & 0.752(0.047) & 0.672                       & {\bf 0.67}(0.02)  & 0.00   &       \\
N-SpSL(MAP)     & 1.057(0.064) & 0.538 & 0.51(0.05) & 0.32  &        & 0.999(0.062) & 0.554                       & 0.55(0.02)  & 0.00   &       \\
N-BL(MAP)       & 0.786(0.053) & 0.619 & 0.29(0.19) & 19.63 &        & 0.727(0.052) & 0.636                       & 0.25(0.23)  & 35.48  &       \\
Lasso(CV)       & {\bf0.782}(0.051) & 0.636 & 0.43(0.04) & 10.81 &        & 0.717(0.047)   & 0.664 & 0.39(0.08)  & 16.52  &       \\
SCAD(CV)        & 1.027(0.070) & 0.575 & 0.34(0.08) & 9.80  &        & 1.000(0.065) & 0.587                       & 0.34(0.10)  & 13.69  &       \\
Lasso(EBIC)      & 1.186(0.078) & 0.476 & 0.50(0.06) & 0.04  &        & 1.165(0.076) & 0.497                       & 0.55(0.03)  & 0.04   &       \\
SCAD(EBIC)       & 1.183(0.079) & 0.476 & 0.49(0.05) & 0.06  &        & 1.178(0.074) & 0.493                       & 0.54(0.03)  & 0.05   &       \\

 \hline
   \end{tabular}
   }
\caption{Results for the high-dimensional setting with dependent covariates.}\label{tab:sim_high_dep}
\end{table}

\begin{table}[H]
\centering
\resizebox{11cm}{!}{%
\begin{tabular}{| l | ccc | ccc |}
\hline
  & \multicolumn{ 3}{ c |}{ Boston housing } & \multicolumn{ 3 }{  c | }{ Bardet-Biedl }\\
  \hline
Method & MSPE & Cos(Angle) & MS & MSPE  & Cos(Angle) & MS  \\
   \hline
SpSL-G(HCG) & 25.246(0.914) & 0.841 & 6.82 & 0.425(0.026) & 0.697  & 2.64
\\ 
N-SpSL-L(Exact) & 25.288(0.903) & 0.841 & 6.98 & 0.421(0.024) &0.701& 2.28 \\ 
SpSL-C(HCG) &  25.252(0.907) & 0.841 & 6.18  & 0.421(0.026) & 0.689& 2.36   \\ 
N-SpSL-C(RW) & 25.203(0.892) & 0.841 & 6.08 & 0.452(0.038) & 0.696 & 2.28 \\ 
HS & 25.479(0.927) & 0.839 & 5.56  &  0.375(0.020) &0.697 &8.56 \\
  N-HS(RW) & 25.461(0.924) & 0.840 & 5.60 & 0.378(0.020) &0.696 &8.12\\
  BL & 25.448(0.938) & 0.829  & 6.10 & 0.357(0.021) & 0.642 & 80.97 \\ 
  N-BL(RW) & 25.411(0.903) & 0.829  & 6.10 & 0.364(0.015) & 0.661 & 94.53  \\ 
SkG &  25.332(0.891) & 0.841 & 8.00  & 0.766(0.047)  & 0.653  & 0.00   \\ 
  SpSL(MM) & 27.341(1.115) & 0.826 &  4.81 & 0.502(0.035) & 0.669 & 5.63 \\ 
  EMVS & 25.385(0.923) & 0.840  &  6.00 & 0.697(0.040)  & 0.685 &
  0.00 \\ 
  SSLasso & 25.058(0.897) & 0.842  &  6.00 & 0.491(0.038) & 0.648 & 2.90 \\ 
    N-SpSL(MAP) & {\bf24.043}(0.871) & {\bf0.848} & 6.90   &  {\bf 0.355}(0.017) & 0.689  & 2.76 \\ 
  N-BL(MAP) & 25.192(0.890) & 0.842  &  6.59 & 0.432(0.029) & 0.705 & 12.00 \\ 
  Lasso(CV) & 25.196(0.886) & 0.842 & 8.60 & 0.424(0.031) &{\bf  0.707} & 22.59 \\ 
  SCAD(CV) & 25.111(0.894) & 0.842 & 7.31 & 0.491(0.037)  & 0.694 & 9.77 \\ 
  Lasso(BIC) & 26.833(0.954) & 0.833 & 7.09 & 1.176(0.047) & 0.665 & 2.07 \\ 
  SCAD(BIC) & 25.515(0.926)  & 0.839 & 6.34 & 1.157(0.052) & 0.655 & 2.25 \\
   \hline   
 \end{tabular}
 }
\caption{Results for the real data sets.
MSPE and MS stand for the  mean squared prediction error (out-of-sample) and the model size (number of selected  variables), respectively.}\label{tab:real1}
\end{table}

\newpage

\noindent
{\bf\Large Figures}

\vspace{6mm}

\newpage

\appendix \label{appendix}

\centerline{\huge\bf Supplementary Materials}

\section{Proofs of Main Results}

\noindent{\em Proof of Proposition \ref{prop:a0_prior}}.
It is clear that $P(\theta_j\neq0\mid \alpha_0) = P(\alpha_j > \alpha_0\mid \alpha_0) = \Phi(-\alpha_0)$. The term $\Phi(-\alpha_0)$ in the neuronized prior controls the sparsity level, and it corresponds to the hyper-parameter $\eta$ in \eqref{eq:model_prior} for the standard SpSL priors. Then, after applying a change of variable as $\Phi(-\alpha_0)=\eta$, where $\eta\sim Beta(a_0, b_0)$, we obtain the transformed density function of $\alpha_0$ as $\Phi(-\alpha_0)^{a_0-1}(1-\Phi(-\alpha_0))^{b_0-1}\phi(\alpha_0)$.
\qed\\

\noindent{\em Proof of Lemma \ref{lem:BL}}. 
Let $\theta=\alpha w$ and $z = w$.  With a change of variable, we obtain the Jacobian term is $z^{-1}$. A simple plug-in of $\alpha=\theta/z$ and $w=z$  completes the proof. 
 \qed\\

\noindent{\em Proof of Proposition \ref{prop:BL}}. We first show that the lower bound holds. By the change of variable $u = z^2$, for any $0<\epsilon<1$, we have
\begin{eqnarray*}
\pi_L(\theta) &=& \int_0^\infty z^{-1}\exp\{ - \theta^2/(2\tau_w^2z^2) - z^2/2 \} dz\\
&=&(2\tau_w^2)^{-1}\int_0^\infty u^{-1}\exp\{ - \theta^2/(2\tau_w^2u) - u/2 \} du\\ 
&=&(2\tau_w^2)^{-1}\int_0^\infty u^{-1/2}\exp\{\epsilon u/2\}u^{-1/2}\exp\{ - \theta^2/(2\tau_w^2u) - (1/2+\epsilon/2)u \}du\\
&\geq& (2\tau_w^2)^{-1}\epsilon^{1/2}\exp\{1\}\int_0^\infty u^{-1/2}\exp\{ - \theta^2/(2\tau_w^2u) - (1/2+\epsilon/2)u \}du\\ 
&=& (2\tau_w^2)^{-1}\epsilon^{1/2}\exp\{1\}(\pi/(1/2+\epsilon/2))^{1/2}\exp\{-(1+\epsilon)^{1/2}|\theta|/\tau_w\}.
\end{eqnarray*}
Second, we show that the upper bound holds.
\begin{eqnarray*}
\pi_L(\theta) &=& \int_0^\infty z^{-1}\exp\{ - \theta^2/(2\tau_w^2z^2) - z^2/2 \} dz\\
&\leq& \int_0^\infty \exp\{ - (1-\epsilon)\theta^2/(2\tau_w^2z^2)- z^2/2 \} dz\\
&\propto&\exp\{ -(1-\epsilon)^{1/2}|\theta|/\tau_w\}. 
\end{eqnarray*}
\qed

\noindent{\em Proof of Proposition \ref{prop:HS}}. Without loss of generality, we assume $\alpha_0=0$ and $\tau_w^2=1$. Because the tail behavior of $\theta$ is governed by the positive region of $\alpha$, we assume that $\alpha>0$.  Then, letting $\theta = T(\alpha)w$ and $z=w$, it follows that
\begin{eqnarray*}
\exp\left\{-\frac{\alpha^2}{2} - \frac{w^2}{2}\right\}d\alpha dw = J(\theta,z)\exp\left\{ -\frac{\{T^{-1}(\theta/z)\}^2}{2} -\frac{z^2}{2} \right\} d\theta dz,
\end{eqnarray*}
where $J(\theta,z)$ is the determinant of the Jacobian term, and one can show that $J(\theta,w)=\left[ 2\theta\lambda_1^{1/2}\{\log(\theta/z)\}^{1/2}  \right]^{-1}$ when $T(t) = \exp\{\lambda_1 \mbox{sign}(t)t^2\}$. As a result, the marginal density of $\theta$ given $z$ is proportional to
\begin{eqnarray*}
\pi_E(\theta)\propto\int^{\infty}_{0}\lambda_1^{-1/2}\{\log(\theta/z)\}^{-1/2}\theta^{-1/(2\lambda_1)-1}z^{1/(2\lambda_1)}\exp\left\{ -\frac{z^2}{2} \right\}dz
\end{eqnarray*}
By the {\em dominated convergence theorem}, the proof is completed.    \qed \\


\noindent{\em Proof of Proposition \ref{prop:cond_alpha}}. 
We note that the conditional distribution of $\alpha_j$ given the others is 
\[
\pi(\alpha_j\mid \alpha_{(-j)}, w, \sigma^2, \by) \propto (2\pi)^{-1/2} \exp\left\{ -\frac{\norm{r_j - X_jT(\alpha_j -\alpha_0)w_j}^2_2}{2\sigma^2} -\frac{\alpha_j^2}{2}  \right\}.
\]
Since the activation function is the ReLU function, it follows that
\begin{eqnarray*}
\pi(\alpha_j\mid \alpha_{(-j)}, w, \sigma^2, \by) \propto \begin{cases}
(2\pi)^{-1/2} \exp\left\{ -\norm{r_j}^2_2/(2\sigma^2)  -\alpha_j^2/2 \right\},\mbox{ if $\alpha_j<\alpha_0$}\\
(2\pi)^{-1/2} \exp\left\{ - \norm{\widetilde r_j - X_j\alpha_j w_j}^2_2/(2\sigma^2)  -\alpha_j^2/2 \right\},\mbox{ if $\alpha_j\geq \alpha_0$},
\end{cases}
\end{eqnarray*}
where $\widetilde r_j = r_j + X_j\alpha_0 w_j$. By doing a simple calculation, we obtain that
\begin{eqnarray*}
(2\pi)^{-1/2} \exp\left\{ - \norm{\widetilde r_j - X_j\alpha_j w_j}^2_2/(2\sigma^2)  -\alpha_j^2/2 \right\} \\
= \widetilde \sigma_j \exp\{-\norm{\widetilde r_j}^2_2/(2\sigma^2)+ \widetilde \alpha_j^2/(2\widetilde\sigma_j^2) \} \phi(\alpha_j; \widetilde\alpha_j,\widetilde\sigma_j^2),
\end{eqnarray*}
where $\phi(\cdot;u,z)$ is the Gaussian density function with mean $u$ and variance $z$, and $\widetilde\alpha_j$ and $\widetilde\sigma_j^2$ are defined in the statement of the proposition. This completes the proof. 

\qed \\

\noindent{\em Proof of Theorem \ref{theo:post_SpSL}}. 
\cite{castillo2015bayesian} investigated asymptotic posterior behaviors for high-dimensional linear regression models. They suggested some sufficient conditions for a certain class of priors to achieve the model selection consistency and the optimal posterior contraction rate. We will show that the conditions on the neuronized SpSL prior satisfies the sufficient conditions proposed in  \cite{castillo2015bayesian} to achieve an optimal posterior contraction rate. The first condition is imposed on the model prior as 
\begin{eqnarray}\label{eq:post_model}
A_1p^{-A_3}\pi(|\gamma|-1) \leq \pi(|\gamma|)\leq A_2p^{-A_4}\pi(|\gamma|+1),
\end{eqnarray}
where $\gamma=\{\gamma_1,\dots,\gamma_p\}^\T$ for some positive constants $A_1$, $A_2$, $A_3$, and $A_4$, and $|\gamma|$ indicates the number of non-zero $\gamma_j$'s. It was  shown that the condition \eqref{eq:post_model} is met when a beta prior, $Beta(1,p^u)$ for some $u>1$, is imposed on $\eta$ in \eqref{eq:spike}. For the neuronized prior, this condition can be satisfied by imposing a hyper-prior of $\alpha_0$ proposed in Proposition \ref{prop:a0_prior} with $a_0 = 1$ and $b_0 = p^u$.

The other condition they considered is on the Laplace slab prior as follows:
\begin{eqnarray}\label{eq:post_laplace}
 \pi_1(\theta_j) =  2^{-1}\lambda_n\exp\{-\lambda_n|\theta_j|\}\mbox{ with } \norm{X}/p\leq \lambda_n \leq 4\norm{X}(\log p)^{1/2},
\end{eqnarray}
where $\norm{X} = \max_{1\leq j\leq p}\norm{X_j}_2$. 

As shown in Proposition \ref{prop:BL}, the tail behavior of the neuronized BL prior is decaying at a rate of $\exp\{-t/\tau_w\}$ when $t$ is large enough, so by plug-in $1/\tau_w$ in $\lambda_n$, its asymptotic property can be preserved by setting $(n\log p )^{-1}/16\leq\tau_w^2\leq n^{-1}p^2$ in the neuronized prior under (A2).

One important concept in \cite{castillo2015bayesian} is the \emph{compatibility condition} that is defined as below:
\begin{eqnarray*}\label{eq:compt}
\phi(\bk) = \inf_\theta\left\{ \frac{\norm{X\btheta}_2|\bk|^{1/2}}{\norm{X}\norm{\btheta_\bk}_1}: \norm{\btheta_{\bk^c}}_1\leq 7\norm{\btheta_{\bk}}_1   \right\}.
\end{eqnarray*}
   The other definitions used in \cite{castillo2015bayesian} follow
\begin{eqnarray}\label{eq:comp2}
\overline\phi(s) = \inf_{\theta_\bk, \bk}\left\{ \frac{\norm{X_\bk\btheta_\bk}_2}{\norm{X_\bk}\norm{\btheta_\bk}_1}: 0\neq |\bk|\leq s   \right\}, \:\:\: \widetilde\phi(s) = \inf_{\theta_\bk, \bk}\left\{ \frac{\norm{X_\bk\btheta_\bk}_2}{\norm{X_\bk}\norm{\btheta_\bk}_2}: 0\neq  |\bk| \leq s   \right\}
\end{eqnarray}
The first equation in \eqref{eq:comp2} is a stronger version of the compatibility condition, which uniformly controls the minimum eigenvalue of Gram matrices in a $l_1$ sense, and the second equation in \eqref{eq:comp2} is a restricted eigenvalue condition that is similar with (A3). Under these notations, one can show that 
$\overline\phi(s)\geq C_2^{-1}C_3^{1/2}s^{-1/2}$ by using (A2) and (A3).
Then, consider
\begin{eqnarray*}
\overline\psi(\bk) &=& \overline\phi\bigg( \Big(2+\frac{3}{A_4}+\frac{33\lambda_n}{2\phi(\bk)^2\norm{X}\sqrt{\log p}}  \Big)|\bk| \bigg)\\ 
\widetilde\psi(\bk) &=& \widetilde\phi\bigg( \Big(2+\frac{3}{A_4}+\frac{33\lambda_n}{2\phi(\bk)^2\norm{X}\sqrt{\log p}} \Big)|\bk| \bigg),
\end{eqnarray*}
where $A_4$ is defined in \eqref{eq:post_model} and $\lambda_n$ appears in \eqref{eq:post_laplace}. 


 Theorem 1 in \cite{castillo2015bayesian} states that $\sup_{\theta_0}\mathbb{E}_{\theta_0}\pi\big(|\bk| > |\bt|+M(1+32/\phi(\bt)^2)|\bt|/A_4 \mid \by \big) \to 0$, and the condition (A3) (restricted eigen value condition) implies a compatibility condition, i.e. $\phi(\bk)>0$ for $|\bk|\leq |\bt|\log n$, as shown in \cite{van2009conditions}. It thus follows that $\sup_{\theta_0}\mathbb{E}_{\theta_0}\pi\big(|\bk| > |\bt|\log n  \mid \by \big) \to 0$, since the term $M(1+32/\phi(\bt)^2)|\bt|/A_4$ is bounded when $\phi(\bt)>0$.  Now we can restrict our focus on models such that $\{\bk: |\bk| \leq |\bt|\log n\}$.

Using the aforementioned results, Theorem 2 in \cite{castillo2015bayesian} shows the following results:
\begin{eqnarray*}
&&\sup_{\theta_0}\mathbb{E}_{\theta_0}\pi\Big(\norm{\btheta-\btheta_0}_2 > \frac{M}{\widetilde{\psi}(\bt)^2}\frac{\sqrt{|\bt|\log p}}{\norm{X}\phi(\bt)}  \:\big\rvert \: \by \Big) \to 0\\
&&\sup_{\theta_0}\mathbb{E}_{\theta_0}\pi\Big(\norm{\btheta-\btheta_0}_1 > \frac{M}{\overline{\psi}(\bt)^2}\frac{|\bt|\sqrt{\log p}}{\norm{X}\phi(\bt)^2} \:\big\rvert \: \by \Big) \to 0,
\end{eqnarray*}
for a large enough constant $M>0$. Since the restricted eigenvalue condition (A3) implies that $\phi(\bk)>0$ for $|\bk|<|\bt|\log n$, by using condition (A2) and  (A3)), it follows that
\begin{eqnarray*}
&&\sup_{\theta_0}\mathbb{E}_{\theta_0}\pi\Big(\norm{\btheta-\btheta_0}_2 > M'C_2^{2}C_3^{-1}\sqrt{|\bt|\log p/n}  \:\big\rvert \: \by \Big) \to 0\\
&&\sup_{\theta_0}\mathbb{E}_{\theta_0}\pi\Big(\norm{\btheta-\btheta_0}_1 > M''C_2^{2}C_3^{-1}|\bt|\sqrt{\log p/n} \:\big\rvert \: \by \Big) \to 0,
\end{eqnarray*}
for some constant $M'$ and $M''$ that are larger than $M$.
\qed \\

\noindent{\em Proof of Theorem \ref{theo:post_conti}}. 
We will show that our proposed conditions on the continuous neuronized prior satisfy the sufficient conditions introduced in \cite{song2017nearly}, and as a result, the optimal contraction rate for the standard shrinkage prior also can be applied to its neuronized counterpart.

We first list the regularity conditions in \cite{song2017nearly} as follows:

\noindent{$B_1(1)$} :  All covariates are uniformly bounded. \\
\noindent{$B_1(2)$} : The dimensionality is high $p\succeq n$.  \\
\noindent{$B_1(3)$} : There exist some integer $\bar p$ and fixed constant $\lambda_0$ such that 
\[\bar p \succ |\bt|, \ \ \mbox{and } \inf_{\bk: |\bk| < \bar p}\lambda_{\min}(X_\bk^\T X_\bk)\geq n\lambda_0.\]   

\noindent{$B_2(1)$} : $|\bt|\log p \prec n$.   \\
\noindent{$B_2(2)$} : $\max_{1\leq j \leq p}|\theta_{0,j}/\sigma^2_0|\leq \gamma_3 E_n$ for some fixed $\gamma\in(0,1)$ and $E_n$ is a non-decreasing sequence.  \\

It is clear that our condition (A2) guarantees $B_1(1)$, and our (A1) and (A3) imply $B_1(1)$, and $B_2(1)$. We further assume that $\bar p=|\bt|\log n$ to assure that (A3) leads to $B_1(3)$. Also, (A4) leads to $B_2(2)$. Thus, our conditions (A1) -- (A4) satisfy these regularity conditions.   

In Corollary 3.1 in \cite{song2017nearly}, under $B_1$ and $B_2$, they proposed some conditions on the shrinkage prior to achieve the optimal posterior contraction rate for standard continuous shrinkage priors. Consider a continuous prior with $r$ degree of polynomial tails, e.g. a Cauchy attains $r=2$, and the prior has a scale parameter $\lambda_n$.  Then, their conditions on the global shrinkage parameter follows:
\begin{eqnarray*}
\tau_w \leq a_n p^{-(u+1)/(r-1)+1},\:\:\: -\log \tau_w = O(\log p),
\end{eqnarray*}
for some $u>0$ and $a_n\asymp (|\bt|\log p /n)^{1/2}/p$.


By Proposition \ref{prop:HS}, setting $T(t) = \exp\{t^2/\{2(r-1)\}\}$ guarantees that the resulting marginal density of the coefficient decays at a polynomial rate with $r\geq2$. Also, we set $-\log \tau_w = O(\log p)$ and $\tau_w = O(p^{-(u+1)/(r-1)}\sqrt{|\bt|\log p /n})$ for some $u>0$. This completes the proof.
\qed \\

\noindent{\em Proof of Theorem \ref{theo:ge_N}}. Without loss of generality, we assume that $\sigma^2=1$ and $\alpha_0=0$. Since $n^{-1/2}X$ is orthogonal, it follows that
\begin{eqnarray*}
\pi(\alpha_j\mid \by) &=& \int\pi(\alpha_j, w_j|\by)dw_j \\ 
&\propto& \{nT^2(\alpha_j)+1/\tau_w^2\}^{-{1\over 2}}\exp[(X_j^\T
\by)^2/\{2(nT^2(\alpha_j)+1/\tau_w^2)\} - \alpha_j^2/2]
\end{eqnarray*}
and  $\pi(\balpha\mid\by)=\prod_{j=1}^p\pi(\alpha_j\mid\by)$. Then, it follows that 
\begin{eqnarray}\label{eq:ge_decomp}
\norm{P^t(\balpha^{(0)}, \cdot) - \pi_\by(\cdot) }_{TV} \leq \max_{1\leq j\leq p}\Vert{P_j^t(\alpha_j^{(0)}, \cdot) - \pi_{\by,j}(\cdot) \Vert}_{TV},
\end{eqnarray}
where $\pi_{\by,j}(\alpha_j) = \pi(\alpha_j\mid \by)$, $\pi_\by(\balpha) = \pi(\balpha\mid\by)=\prod_{j=1}^p\pi(\alpha_j\mid \by)$, and $P_j^t$ is a Markov transition kernel of the Metropolis algorithm for $\alpha_j$ at iteration $t$. Since the conditional posterior distribution of $\bw$ given $\balpha$ is explicitly represented, which is a product of independent Gaussians with mean $X_j^\T\by/(nT^2(\alpha_j)+1/\tau_w^2)$ and variance $(nT^2(\alpha_j)+1/\tau_w^2)^{-1}$, the convergence behavior of Algorithm 1 is solely determined by the convergence rate of $\max_{1\leq j\leq p}\Vert{P_j^t(\alpha_j^{(0)}, \cdot) - \pi_{\by,j}(\cdot) \Vert}_{TV}$, so it is sufficient to show that $P_j^t$ results in a geometrical ergodicity for any $j\in\{1,\dots,p\}$.  

To simplify the description, we first introduce some  concepts regarding a distribution. We consider a distribution with a density function $\pi$, and define
\begin{eqnarray}\label{eq:ge_tails}
V=\limsup_{|x|\to\infty}\frac{x}{|x|}\nabla \log \pi(x).
\end{eqnarray}
The distribution is called \emph{super-exponentially light} if  $V=-\infty$ in \eqref{eq:ge_tails} ;  \emph{exponentially light} if $V$ is a negative constant; and \emph{sub-exponentially light} if $V=0$
\citep{johnson2012variable, mengersen1996rates, roberts1996geometric}. Using these definitions, Theorem 4.3 in \cite{jarner2000geometric} considers a Metropolis transition kernel induced by a proposal density  that contains strictly positive amount of density around zero. Since we are using a Gaussian kernel in Algorithm \ref{alg:MCMClin}, our case satisfies this condition. Then, their theorem implies that the resulting random-walk Metropolis algorithm targeting $\pi$ is geometrically ergodic, if $\pi$ is super-exponentially light and satisfies 
\begin{eqnarray}\label{eq:ge_cond}
\limsup_{|x|\to\infty}\frac{x}{|x|} \frac{ \nabla\pi(x)}{|\nabla\pi(x)|} < 0.
\end{eqnarray}
However, in one-dimensional cases, equation \eqref{eq:ge_tails} implies \eqref{eq:ge_cond}. Thus, the proof will be completed if we show that $\pi_{\by,j}$ is  super-exponentially light.


Note that
\begin{equation}\label{eq:super}
\frac{x}{|x|}\nabla \log \pi_{\by,j}(x) = sgn(x)\left\{ -\frac{nT(x)T'(x)}{nT^2(x)+1/\tau_w^2}  - \frac{n T(x) T'(x)(X_j^\T\by)^2}{(nT^2(x)+1/\tau_w^2)^2} - x\right\},
\end{equation}
where $sgn$ is a sign function. Since the activation function $T$ has stable tails, i.e.,  $\exists \ C_1, C_2, C_3> 0$
such that (a) when $ x<-C_3$, either $|T'(x)|\leq C_1$ or $|T'(x)|\geq C_2$ and the sign of $T'(x)$ does not change;
and (b) when $ x>C_3$, either $|T'(x)|\leq C_1$ or $|T'(x)|\geq C_2$ and the sign of $T'(x)$ does not change.
It is clear that for either tail, if $|T'(x)|$ is bounded from above, then the RHS of \eqref{eq:super} is dominated by $-|x|$ and hence   diverges to $-\infty$ as either $x \rightarrow \infty$ or $x\rightarrow -\infty$.
If $|T'(x)|$ is bounded from below and $T'(x)$ does not change sign after $x>C_3$, then, as $x\rightarrow\infty$, either $T'(x)\geq C_2$, which implies that $T(x)$ will become positive eventually and thus $\lim_{x\rightarrow\infty} T(t)T'(t)\geq 0$; or  $T'(x)\leq -C_2$, which means that $T(x)$ will become negative eventually and also  $\lim_{x\rightarrow\infty} T(t)T'(t)\geq 0$. Thus, all the three terms inside the parenthesis of the RHS of \eqref{eq:super} are of the same sign and, hence, the RHS diverges to $-\infty$. 
As $x\rightarrow -\infty$,  we see by the same argument as above that, if $|T'(x)|\geq C_2$ and $T'(x)$ does not change sign after $x<-C_3$, $\lim_{x\rightarrow-\infty} T(x)T'(x)<0$. Thus, all terms inside the parenthesis of the RHS of 
 \eqref{eq:super} are of the same sign and hence \eqref{eq:super} diverges to $-\infty$.

As a result, there exist 
$C_j$ and $\rho_j\in(0,1)$ such that
\[
  \Vert{P_j^t(\alpha_j^{(0)}, \cdot) - \pi_{\by,j}(\cdot) \Vert}_{TV}\leq C_j(\alpha_j) \rho_j^t,
\]
for $j=1,\dots,p$. By plugging this to \eqref{eq:ge_decomp}, it follows that
\begin{eqnarray*}
\norm{P^t(\balpha^{(0)}, \cdot) - \pi_\by(\cdot) }_{TV} \leq \max_{1\leq j\leq p}\{C_j(\alpha_j)\}\max_{1\leq j\leq p}\{\rho_j\}^t.
\end{eqnarray*}
\qed 
\\

\noindent{\em Proof of Theorem \ref{theo:ge_HS}}.
 We first note that when there exists no moment generating function of a target density of the Metropolis-Hastings algorithm, the resulting MH algorithm cannot achieve the geometric ergodicity \citep{mengersen1996rates}. Moreover, it is well-known that if any single conditional density in a Metropolis-Hastings-within-Gibbs sampler is not geometrically ergodic, neither the full MCMC is \citep{roberts2001optimal,diaconis2008gibbs,robert1995convergence}. So, it is sufficient to show  that the moment generating function of $\pi(\tau_j\mid\beta_j)$ does not exist regardless of the value of $\beta_j$. 
 
 Consider the following conditional posterior density of $\tau_j$ for some $j\in\{1,\dots,p\}$:
 \[
 \log \pi(\tau_j\mid\beta_j) = -(1/2)\log (\tau_j^2) - \beta_j^2/(2\tau_j^2) - c\tau_j^\kappa + C,
 \]
 where $C$ is some constant.  Because $0<\kappa<1$, it is clear that for any $t>0$ and $\beta_j\in\mathbb{R}$, $\tau_j t + \log\pi(\tau_j\mid \beta_j)$ diverges to infinity as $\tau_j$ increases, which concludes that this conditional posterior density cannot have a proper moment generating function.  \qed
 
 \section{Updating  Matrix Inversion and Determinant}\label{appendix:algebra}
 In this section, under a discrete SpSL Gaussian-conjugate prior, we provide an instruction on how to efficiently evaluate some linear algebra calculations that are required to implement the fully-collapsed Gibbs sampler for the Bayesian linear model selection. When implementing the  collapsed Gibbs sampler, one needs to compute the inversion and determinant of a modified sample covariance matrix at each iteration. To improve computational efficiencies, we can use the following linear algebra techniques.

Let $A$ be a $m\times m$ symmetric matrix and $B=\begin{pmatrix} A & b\\  b^T & c  \end{pmatrix}$, where $b$ is an $m\times 1$ vector.  Then, 
\begin{equation}\label{matrix_inverse_plus}
    B^{-1}\equiv \begin{pmatrix} Q_{11} & q_{12} \\ q_{21} & q_{22} \\ \end{pmatrix}= \begin{pmatrix} A^{-1}+\frac{1}{k}A^{-1}bb^TA^{-1}& -\frac{1}{k}A^{-1}b \\ 
-\frac{1}{k}b^TA^{-1} & \frac{1}{k}\\
   \end{pmatrix},
\end{equation}
   where $ k=c-b^TA^{-1}b$,   and 
\begin{equation}\label{determinant_plus}
    \text{det}(B)\equiv \text{det}\begin{pmatrix} A & b\\
b^T & c\\
 \end{pmatrix} = \text{det}(A) \times (c-b^TA^{-1}b).
\end{equation} 
Conversely, if we want to update from $B$ to $A$, we have 
 \[A^{-1}=Q_{11} -q_{12}\times q_{21}/q_{22},\]
 and 
 \[ \det(A)=\det(B)/(c-b^TA^{-1}b).\]

To apply the above updating formulas to the fully-collapsed Gibbs sampler, we let the current model be $\boldsymbol{\gamma}$,  randomly select one index $j\in\{1,\dots,p\}.$ If $\gamma_j=0$, we propose a candidate model by adding $X_j$ to the current model, and the  binary representation of the proposed model is  $\boldsymbol{\gamma}'=\{\gamma'_1,\dots,\gamma'_p\}$, where $$\gamma_h'=\begin{cases} 
1\:\:\:\:\text{if $\gamma_h=1$ or $k=j$ },\\
0\:\:\:\:\text{otherwise}
\end{cases},$$
for $h=1,\dots,p$.
Let  $x = X_j$ and $A = X_{\gamma}^TX_\gamma +\frac{\sigma^2}{\tau_2^2}I$,  and assume that for the current model, the inverse and the determinant of $X_{\gamma}^TX_{\gamma} + (\sigma^2/\tau_w^2)I$ are known. We can obtain
 the inverse matrix and the determinant of $X_{\gamma'}^TX_{\gamma'} + (\sigma^2/\tau_w^2)I$ economically using formulas \eqref{matrix_inverse_plus} and \eqref{determinant_plus}:
\begin{eqnarray*}
&&\left(X_{\gamma'}^TX_{\gamma'}+\frac{\sigma^2}{\tau_w^2}I\right)^{-1}
   = \begin{pmatrix} A^{-1}+\frac{1}{k}A^{-1}X_\gamma^Txx^TX_\gamma A^{-1}& -\frac{1}{k}A^{-1}X_\gamma^Tx \\ 
-\frac{1}{k}x^TX_\gamma A^{-1} & \frac{1}{k}\\
   \end{pmatrix},
\end{eqnarray*}
where $k = x^Tx+\sigma^2/\tau - x^TX_\gamma A^{-1}X_\gamma^Tx$, and
$$\text{det}\left(X_{\gamma'}^TX_{\gamma'}+\frac{\sigma^2}{\tau_w^2}I\right) = \text{det}\left(A\right)\times (x^Tx + \sigma^2/\tau_w^2- x^TX_\gamma A^{-1} X_\gamma^Tx).$$
   
  If $\bgamma_j=1$,  the candidate model is the same as the current model but with $X_j$ excluded, i.e., $\bgamma'$ is $$
  \gamma_h'=\begin{cases} 
0\:\:\:\:\text{if $\gamma_h=0$ or $h=j$ },\\
1\:\:\:\:\text{otherwise},
\end{cases}
  $$
  for $h=1,\dots,p$. 
 Then, it follows that
 \begin{eqnarray}\label{eq:inverse}
 \left(X_{\gamma'}^TX_{\gamma'} + \frac{\sigma^2}{\tau_w^2}I\right)^{-1}  =Q_{11}-q_{12}q_{21}/q_{22},\end{eqnarray} and  $$
 \text{det}\left(X_{\gamma'}^TX_{\gamma'} + \frac{\sigma^2}{\tau_w^2}I\right) = \frac{\text{det}(X_{\gamma}^TX_{\gamma} + \frac{\sigma^2}{\tau_w^2}I)}{c - b^TD^{-1}b}, 
 $$ where $Q_{11}$, $q_{12}$, $q_{21}$, and $q_{22}$ are block components of 
 $$
 \left(X_\gamma^T X_\gamma + \frac{\sigma^2}{\tau_w^2}I \right)^{-1} = \begin{pmatrix} Q_{11} & q_{12}\\ 
q_{21} & q_{22}\\
   \end{pmatrix},
 $$
 and the second block corresponds to $X_j$. Also, $c$, $b$, and $D$ are block components of $X_\gamma^T X_\gamma + \frac{\sigma^2}{\tau_w^2}I$; i.e.,
 $$
X_\gamma^T X_\gamma + \frac{\sigma^2}{\tau_w^2}I = \begin{pmatrix} D & b\\ 
b^T & c\\
   \end{pmatrix} = \begin{pmatrix} X_{\gamma'}^T X_{\gamma'} + \frac{\sigma^2}{\tau_w^2}I & X_{\gamma'}^TX_j\\ 
X_j^TX_{\gamma'} & X_j^TX_j + (\sigma^2/\tau_w^2)\\
   \end{pmatrix},
 $$
 where $\bgamma\setminus{j}$ is the model where $X_j$ is discarded from $\bgamma$, and $D^{-1}$ can be evaluated from \eqref{eq:inverse}. 
 
Once these inverse matrix and determinant are evaluated, the Metropolis acceptance probability can be defined as $\min\left\{1, \frac{\pi(\bgamma'\mid \by, \eta,\sigma^2)}{\pi(\bgamma\mid \by, \eta,\sigma^2)} \right\}$, where 
$$
\pi(\bgamma\mid\by,\eta,\sigma^2) \propto |X_\gamma^TX_\gamma + (\sigma^2/\tau_w^2)I|^{-1/2}\exp\left\{\by^T\widetilde P_\gamma \by/2\right\}\eta^{|\bgamma|+a_0-1}(1-\eta)^{p-|\bgamma|+b_0-1},
$$
 and $\widetilde P_\gamma = X_\gamma(X_\gamma^TX_\gamma + \sigma^2/\tau_w^2I)^{-1}X_\gamma^T$. We note that this posterior probability is based on a prior setting with $\theta_\gamma\sim N(0,\tau_w^2I)$ and $\pi(\sigma^2)\propto 1/\sigma^2$.
 
 The computational complexity of this linear algebra calculation, given the inverse matrix and determinant for the current model, is $O(|\bgamma|n)+O(|\bgamma|^2)$. This updating rule is more efficient than a naive evaluations without the guidance of the previous result, which requires $O(|\bgamma|^2n)+O(|\bgamma|^3)$. However, the computational gain would be slightly diluted in overall, because after evaluating the inverse matrix and the determinant, evaluating the marginal likelihood takes an additional complexity $O(|\bgamma|n)$ that is equally applied to both procedures. 
 
 In contrast, the half-collapsed Gibbs sampler and  N-SpSL(Exact) do not require the evaluation of the determinant nor the inverse matrix, and their computational complexity for a single sampling $\gamma_j$ is  lower than that required for the fully-collapsed Gibbs, $O(n)$. The HCG and the neuronized SpSL procedure thus  appear to be more efficient, in terms of ESS per second, than the FCG at least for our limited examples.    
 


\section{Some Auxiliary Results}
\subsection{Additional optimization paths for CAAN}

\begin{figure}[H]
    \centering
        \begin{subfigure}[b]{0.24\textwidth}
        \includegraphics[width=\textwidth, height=3cm]{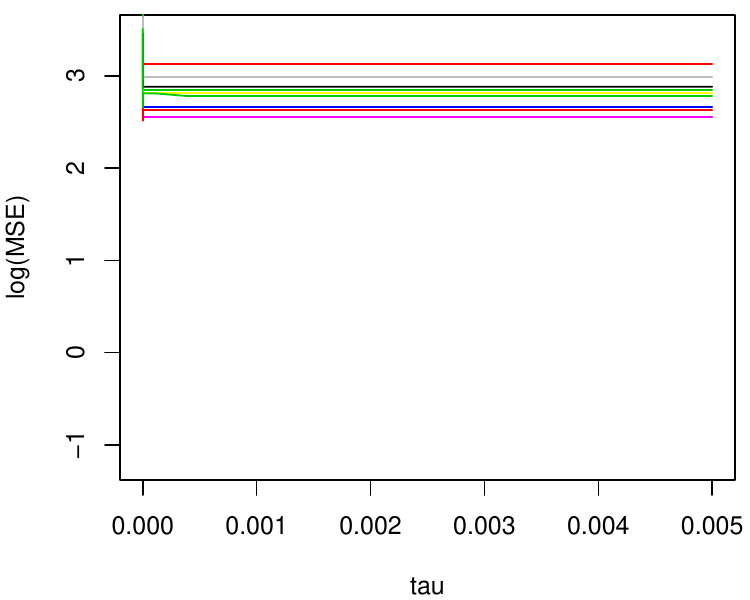}
\end{subfigure}                
\begin{subfigure}[b]{0.24\linewidth}
        \includegraphics[width=\textwidth, height=3cm]{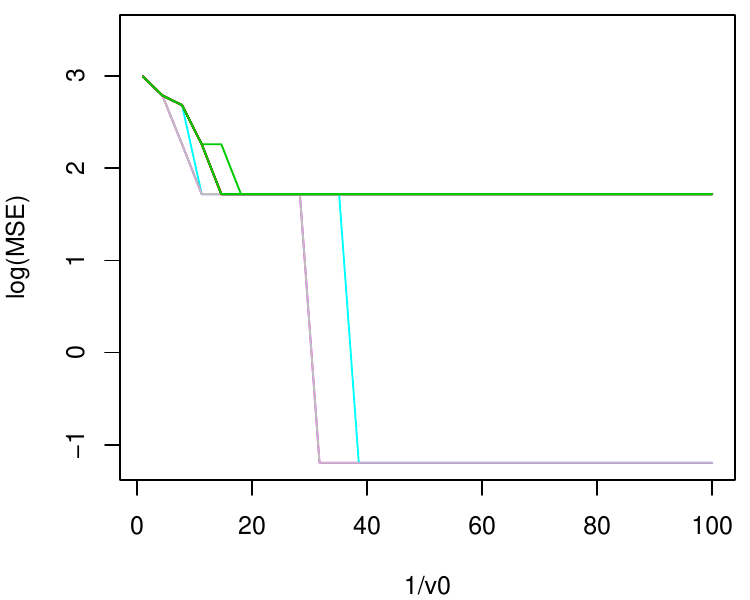}
\end{subfigure}
 \begin{subfigure}[b]{0.24\linewidth}
        \includegraphics[width=\textwidth, height=3cm]{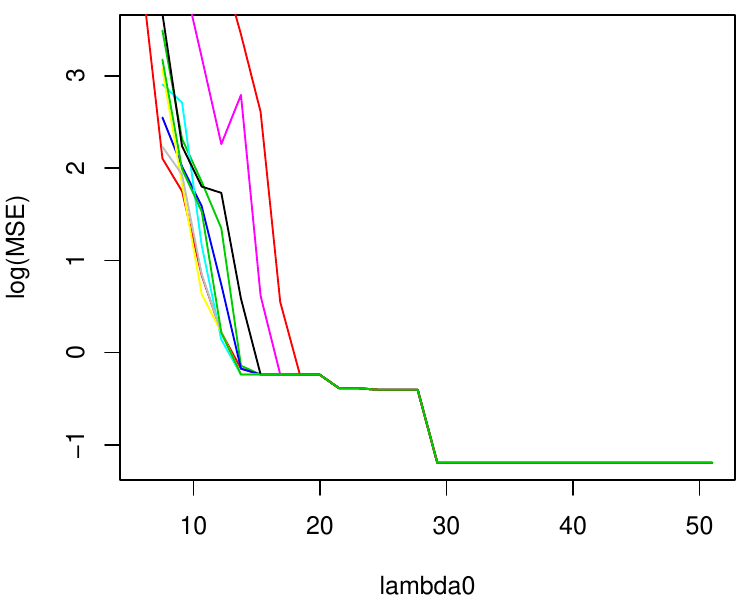}
\end{subfigure} 
 \begin{subfigure}[b]{0.24\linewidth}
        \includegraphics[width=\textwidth, height=3cm]{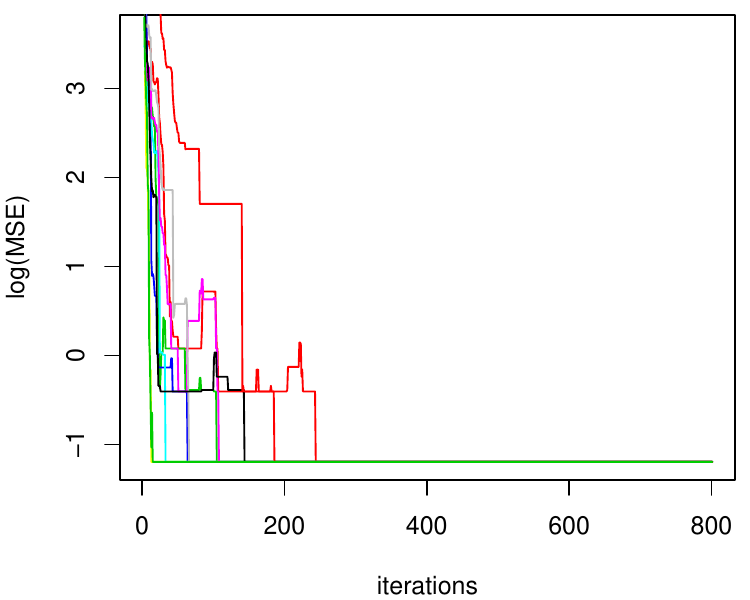}
\end{subfigure}
        \begin{subfigure}[b]{0.24\textwidth}
        \includegraphics[width=\textwidth, height=3cm]{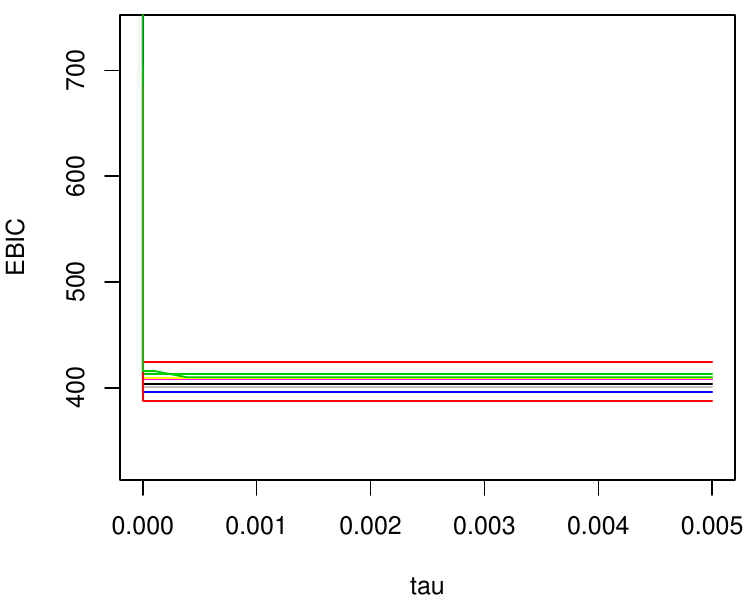}\caption{MM}
\end{subfigure}                
\begin{subfigure}[b]{0.24\linewidth}
        \includegraphics[width=\textwidth, height=3cm]{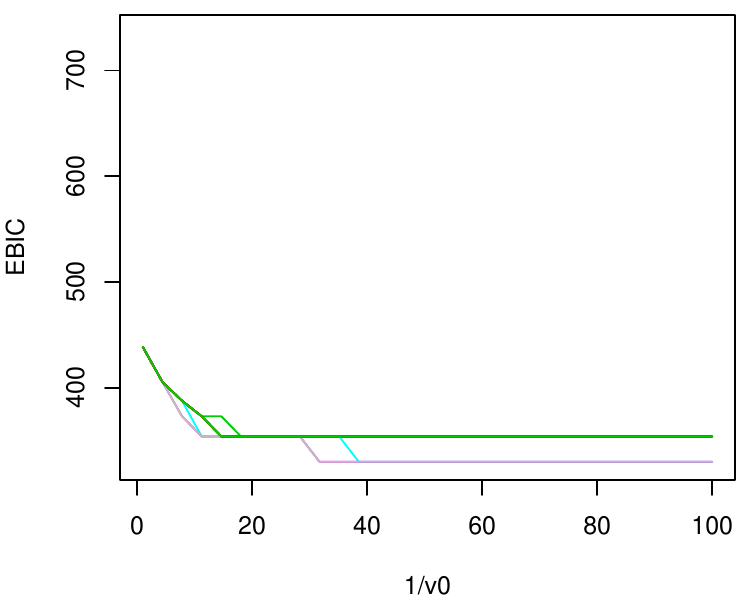}\caption{EMVS}
\end{subfigure}
 \begin{subfigure}[b]{0.24\linewidth}
        \includegraphics[width=\textwidth, height=3cm]{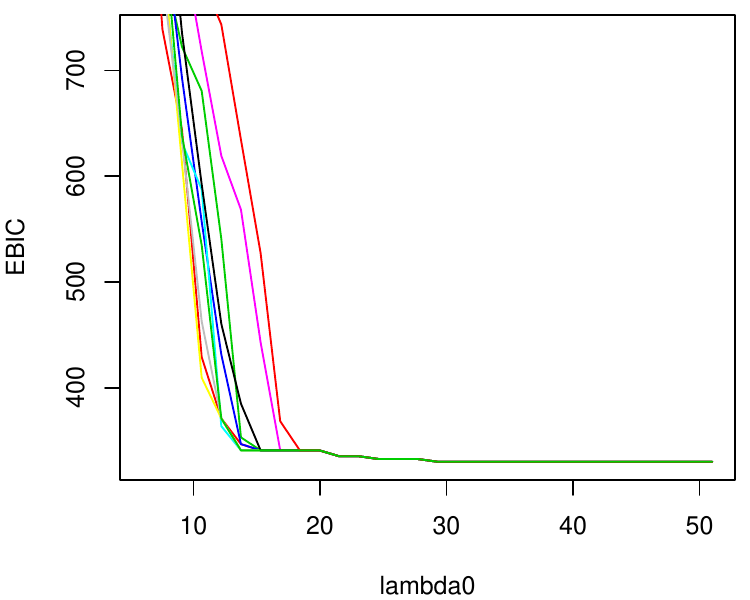}\caption{SSLasso}
\end{subfigure} 
 \begin{subfigure}[b]{0.24\linewidth}
        \includegraphics[width=\textwidth, height=3cm]{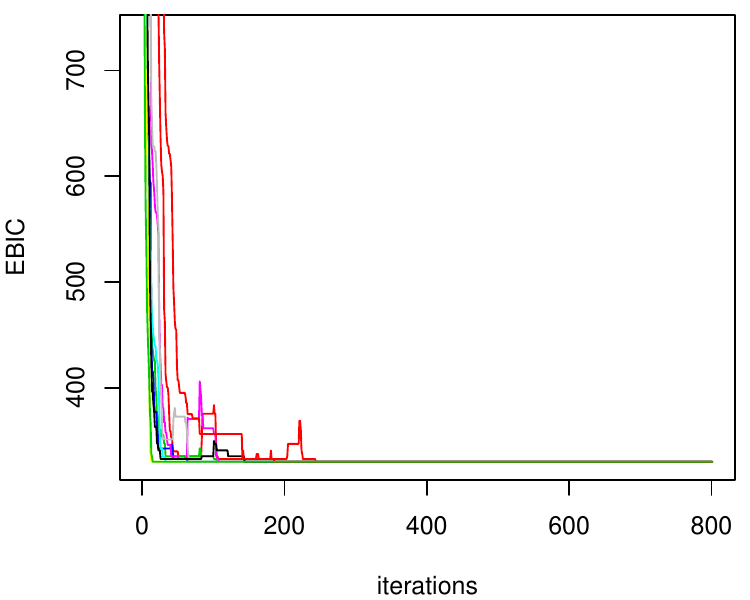}\caption{CAAN}
\end{subfigure}
\caption{Trace plots of the log-MSE (top row) and EBIC (bottom row) paths from 10 different initial points for the four optimization algorithms, based on a synthetic data set generated from the Bardet-Biedl dataset ($n= 120$ and $p=  200$) with the true model size 5.  The MM procedure used $\tau_3=10^{-2}$.}
\label{fig:p5}
\end{figure}

As a supplement of the synthetic example in Section \ref{comp_optim}, we examine a scenario where the true model size is five (the other settings are equivalent to the example in the main text). Figure \ref{fig:p5} show that the CAAN and the SSLasso procedures consistently chose the same  model via EBIC across all ten random initial values, while the MM and the EMVS fail to achieve the consistency.

\subsection{Comparisons Between Different MCMC Algorithms} 
{In this section, we consider extra simulation studies. We first compare the ESS (per second) of ``N-SpSL-L(Exact)'' and ``N-SpSL-L(RW)'', and the results are shown in Table \ref{tab:ESS}. The column ``Ind'' and ``Dep'' indicates scenarios where the covariates are generated from iid standard Gaussian and from the Toeplitz design considered in Section \ref{sec:sim}, respectively. The other settings are exactly the same with these in the simulation studies in the main paper. The results show that ``N-SpSL-L(Exact)'' is at least two times more efficient in terms of ESS.

\begin{table}[H]
\centering
\renewcommand{\arraystretch}{0.6}
\resizebox{13cm}{!}{%
\begin{tabular}{| l | rr|rr| rr| rr|}
\hline
  &\multicolumn{ 8 }{ c |}{ Low-dimension  }\\
  \hline
Sample size  & \multicolumn{ 4 }{ c |}{ $(n=200,p=50)$ } &  \multicolumn{ 4 }{  c | }{ $(n=400,p=100)$ }\\
 \hline
Signal strength & \multicolumn{ 2 }{ c |}{ Weak } &\multicolumn{ 2 }{ c |}{ Strong} &\multicolumn{ 2 }{ c |}{ Weak } & \multicolumn{ 2 }{ c |}{ Strong}   \\
\hline
Covariate  &  Ind  &  Dep  &  Ind  &  Dep  &Ind  &  Dep  &  Ind  &  Dep \\
\hline
\hline
N-SpSL-L(Exact) & 7625.6 & 5949.1 & 8255.1 & 4551.7 & 2793.6 & 1123.9 & 3479.3 &    515.0 \\
N-SpSL-L(RW) & 2238.5 & 1666.8 & 2582.7 & 1397.9 & 1000.5 &370.7  & 889.3 & 210.5     \\
\hline 
\hline
  &\multicolumn{ 8 }{ c |}{ High-dimension  }\\
  \hline
Sample size  & \multicolumn{ 4 }{ c |}{ $(n=100,p=300)$ } &  \multicolumn{ 4 }{  c | }{ $(n=150,p=1000)$ }\\
 \hline
Signal strength & \multicolumn{ 2 }{ c |}{ Weak } &\multicolumn{ 2 }{ c |}{ Strong} &\multicolumn{ 2 }{ c |}{ Weak } & \multicolumn{ 2 }{ c |}{ Strong}   \\
\hline
Covariate  &  Ind  &  Dep  &  Ind  &  Dep  &Ind  &  Dep  &  Ind  &  Dep \\
\hline
\hline
N-SpSL-L(Exact) & 919.7 & 874.5 & 1271.1 & 561.6 & 217.2 & 131.9 & 262.2 & 114.5   \\
N-SpSL-L(RW) & 221.9 & 203.6 & 294.7 & 136.6 & 69.4 & 43.7  & 80.3 & 36.6      \\
\hline 

\end{tabular}
}
\caption{A comparison of ESS per second between different neuronized SpSL procedures.}\label{tab:ESS}
\end{table}
Table \ref{tab:ESS2} compares two different MCMC algorithms: the half-collapsed Gibbs sampler used in the main manuscript (SpSL-G(HCG)) vs. the fully-collapsed Gibbs sampler (SpSL-G(FCG)). Briefly, by taking advantages of Gaussian conjugacy, SpSL-G(FCG) marginalizes out all the continuous coefficients to obtain the target distribution $\pi(\bgamma\mid \by)$ and considers as a proposal to flip a randomly selected indicator from $\gamma_j$ to $1-\gamma_j$. It is well-known that ``SpSL-G(FCG)'' is highly inefficient \citep{ji2013adaptive}, and this finding is also confirmed again in Table \ref{tab:ESS2}. The ESS of ``SpSL-G(FCG)'' is significantly smaller than that from ``SpSL-G(HCG)''. In particular, under high-dimensional settings, its ESS is less than 10, while ``SpSL-G(HCG)'' attains at least hundreds of ESS per second.  

\begin{table}[H]
\centering
\renewcommand{\arraystretch}{0.6}
\resizebox{13cm}{!}{%
\begin{tabular}{| l | rr|rr| rr| rr|}
\hline
  &\multicolumn{ 8 }{ c |}{ Low-dimension  }\\
  \hline
Sample size  & \multicolumn{ 4 }{ c |}{ $(n=200,p=50)$ } &  \multicolumn{ 4 }{  c | }{ $(n=400,p=100)$ }\\
 \hline
Signal strength & \multicolumn{ 2 }{ c |}{ Weak } &\multicolumn{ 2 }{ c |}{ Strong} &\multicolumn{ 2 }{ c |}{ Weak } & \multicolumn{ 2 }{ c |}{ Strong}   \\
\hline
Covariate  &  Ind  &  Dep  &  Ind  &  Dep  &Ind  &  Dep  &  Ind  &  Dep \\
\hline
\hline
SpSL-G(HCG) & 15781.0 & 15422.8  & 20366.7 & 9752.9  & 6175.9  & 2521.9 & 8939.6  & 1205.3  \\
SpSL-G(FCG) &  82.2   &  48.9  & 487.5   & 54.0    & 184.6    & 21.6   & 540.7   & 38.26       \\
\hline 
\hline
  &\multicolumn{ 8 }{ c |}{ High-dimension  }\\
  \hline
Sample size  & \multicolumn{ 4 }{ c |}{ $(n=100,p=300)$ } &  \multicolumn{ 4 }{  c | }{ $(n=150,p=1000)$ }\\
 \hline
Signal strength & \multicolumn{ 2 }{ c |}{ Weak } &\multicolumn{ 2 }{ c |}{ Strong} &\multicolumn{ 2 }{ c |}{ Weak } & \multicolumn{ 2 }{ c |}{ Strong}   \\
\hline
Covariate  &  Ind  &  Dep  &  Ind  &  Dep  &Ind  &  Dep  &  Ind  &  Dep \\
\hline
\hline
SpSL-G(HCG) &  2773.3  &  3015.4 & 3960.2   & 1896.2    & 744.1   & 506.3  &    819.5& 385.6     \\
SpSL-G(FCG) &    1.5&6.8    & 4.8   &    6.5& 7.0   &    4.1 &    7.3 & 8.8       \\
\hline 
\end{tabular}
}
\caption{A comparison of ESS per second between different SpSL procedures.}\label{tab:ESS2}
\end{table}

\subsection{Additional simulation studies of sparse regression algorithms} 

We provide the results of more simulation studies for independent covariate cases with different signal strengths in Table \ref{tab:sim_low_ind}  and \ref{tab:sim_high_indep}, and Table \ref{tab:sim_low_dep_SM} and \ref{tab:sim_high_dep_SM} show simulations results for strong signals. The first five true regression coefficients are non-zero, and the non-zero coefficients of the low-dimensional and high-dimensional settings are set to be  $\pm s$ and $s\times\{\pm 0.4,\pm 0.45,\pm 0.5,\pm 0.55,\pm 0.6\}$, respectively.}

\begin{table}[H]
\renewcommand{\arraystretch}{0.5}
\centering
\resizebox{15cm}{!}{%
\begin{tabular}{| l | ccccr | ccccr |}
\hline
                & \multicolumn{10}{c|}{Strong Signal ($s=0.3$)}                                  \\ \hline
  & \multicolumn{ 5 }{ c |}{ $(n=200,p=50)$ } & \multicolumn{ 5 }{  c | }{ $(n=400,p=100)$ }\\
Method          & MSE   & Cos(Angle) & MCC  & FP    & ESS     & MSE   & Cos(Angle) & MCC  & FP    & ESS    \\ \hline
Oracle      & 0.025  & 0.979   &                         &    &     & 0.028  & 0.987 &    &    &     \\
SpSL-G(HCG) & 0.085 & 0.909 & 0.89 & 0.17 & 20877.1 & {\bf 0.036} & {\bf 0.981} & 0.99 & 0.14 & 6414.5 \\ 
 N-SpSL-L(Exact) & {\bf 0.072} & 0.924 & {\bf0.92} & 0.40 & 7957.3 & 0.042 & 0.978 & 0.98 & 0.47 & 4100.8 \\ 
SpSL-C(HCG) & 0.073 & 0.923 & {\bf0.92} & 0.28 & 1608.1 & 0.038 & {\bf 0.981} & 0.98 & 0.27 & 795.7 \\ N-SpSL-C(RW) & 0.091 & 0.901 & 0.89 & 0.10 & 2423.0 & {\bf 0.036} & {\bf 0.981} & 0.99 & 0.13 & 1307.5 \\ 
HS & 0.087 & 0.906 & 0.90 & 0.15 & 815.6 & 0.054 & 0.972 & 0.99 & 0.14 & 718.5 \\ 
  N-HS(RW) & 0.088 & 0.906 & 0.90 & 0.14 & 1754.4 & 0.052 & 0.973 & 0.99 & 0.13 & 619.5 \\ 
  BL & 0.154 & 0.844 & 0.79 & 4.14 & 3374.9 & 0.134 & 0.918 & 0.92 & 1.69 & 846.0 \\ 
  N-BL(RW) & 0.122 & 0.866 & 0.82 & 2.17 & 1822.3 & 0.114 & 0.936 & 0.97 & 0.67 & 575.5 \\         
SkG              & 0.074 & 0.922 & 0.91 & 0.26  & 9238.6 & 0.038 & 0.980 & 0.98 & 0.29  & 4712.8 \\
SpSL(MM)       & 0.099 & 0.905 & 0.76 & 3.16  &         & 0.112 & 0.939 & 0.78 & 5.31  &        \\
 N-SpSL-L(MAP) & 0.078 & {\bf0.928} & 0.88 & 1.07 & & 0.058 & 0.970 & 0.93 & 1.49 &  \\ 
EMVS            & 0.238 & 0.718 & 0.71 & 0.03  &         & 0.089 & 0.954  & 0.96 & 0.00  &        \\
SSLasso         & 0.096  & 0.894 & 0.88 & 1.09  &         & 0.037 & {\bf0.981} & 0.93 & 1.38  &        \\
Lasso(CV)       & 0.091 & 0.906 & 0.52 & 10.70 &         & 0.095 & 0.958 & 0.53 & 19.72 &        \\
SCAD(CV)        & 0.080 & 0.920 & 0.55 & 8.94  &         & 0.049 & 0.974 & 0.63 & 12.47 &        \\
Lasso(BIC)      & 0.216 & 0.852 & 0.90 & 0.94  &         & 0.339 & 0.904 & 0.95 & 1.10  &        \\
SCAD(BIC)       & 0.211 & 0.847 & 0.89 & 0.99  &         & 0.306 & 0.896 & 0.94 & 1.21  &        \\ 
N-BL(MAP)       & 0.107 & 0.881 & 0.78 & 2.94  &         & 0.106 & 0.942 & 0.94 & 1.34  &        \\

\hline
\end{tabular}
}
\caption{Results for the low-dimensional setting with independent covariates. SpSL, HS, and BL indicate the procedure based on the discrete SpSL, the horseshoe, and  Bayesian Lasso priors, respectively. The sign ``N'' stands for the neuronized version of the corresponding prior. }\label{tab:sim_low_ind}
\end{table}

\begin{table}[H]
\centering
\renewcommand{\arraystretch}{0.5}
\resizebox{15cm}{!}{%
\begin{tabular}{| l | ccccr | ccccr |}
\hline
                & \multicolumn{10}{c|}{Strong Signal ($s=1.5$)}                                     \\ \hline
                & \multicolumn{5}{c|}{$(n=100,p=300)$}     & \multicolumn{5}{c|}{$(n=150,p=1000)$}   \\
Method          & MSE   & Cos(Angle)  & MCC  & FP     & ESS     & MSE    & Cos(Angle)  & MCC  & FP    & ESS   \\ \hline
Oracle      & 0.055  & 0.992   &                         &    &     & 0.037  & 0.995 &    &    &\\    
SpSL-G(HCG) & 0.095 & 0.985 & 0.98 & 0.23 & 4409.6 & 0.054 & 0.992 & 0.99 & 0.15 & 1168.0 \\ 
 N-SpSL-L(Exact) & 0.139 & 0.977 & 0.94 & 0.75 & 1317.1 & 0.084 & 0.987 & 0.96 & 0.49 & 389.2 \\ 
 SpSL-C(HCG) & 0.120 & 0.981 & 0.96 & 0.48 & 157.9 & 0.068 & 0.989 & 0.97 & 0.34 & 55.7 \\ 
 N-SpSL-C(RW) &{\bf 0.090} & 0.986 & 0.98 & 0.17 & 421.2 & 0.052 & 0.992 & 0.99 & 0.12 & 131.2 \\ 
 HS & 0.164 & 0.973 & 0.86 & 1.98 & 56.6 & 0.191 & 0.968 & 0.68 & 6.20 & 7.3 \\ 
  N-HS(RW) & 0.155 & 0.975 & 0.87 & 1.83 & 182.1 & 0.190 & 0.968 & 0.68 & 6.36 & 12.3 \\ BL              & 1.015 & 0.808  & 0.41 & 24.20  & 42.1    & 1.512  & 0.699  & 0.65 & 5.32  & 12.6  \\
N-BL(RW)        & 0.864 & 0.826  & 0.38 & 29.32  & 50.2    & 1.439  & 0.736  & 0.67 & 5.93  & 11.9  \\
SkG              & 0.097 & 0.985  & 0.99 & 0.01  & 2827.0 & 0.054  & 0.992  & {\bf1.00} & 0.00  & 949.7 \\
SpSL(MM)       & 0.489 & 0.909  & 0.82 & 1.58   &         & 1.932  & 0.613  & 0.43 & 8.35  &       \\
  N-SpSL-L(MAP)     & 0.109 & 0.982  & 0.98 & 0.03   &         & {\bf0.041}  & {\bf0.994}  & {\bf1.00} & 0.03  &       \\
  EMVS            & 0.483 & 0.910  & 0.88 & 0.01   &         & 1.215  & 0.743  & 0.71 & 0.00  &       \\
SSLasso         & {\bf 0.090} & {\bf0.986}  & {\bf0.99} & 0.02   &         & 0.042  & 0.994  & {\bf 1.00} & 0.04  &       \\
Lasso(CV)       & 0.412 & 0.947  & 0.44 & 24.25  &         & 0.332  & 0.965  & 0.40 & 33.72 &       \\
SCAD(CV)        & 0.153 & 0.975  & 0.53 & 13.72  &         & 0.095  & 0.985  & 0.50 & 18.09 &       \\
Lasso(EBIC)      & 1.740 & 0.821  & 0.96 & 0.08   &         & 1.577  & 0.861  & 0.99 & 0.05  &       \\
SCAD(EBIC)       & 1.690 & 0.825  & 0.96 & 0.08   &         & 1.568  & 0.859  & 0.99 & 0.05  &       \\ 
N-BL(MAP)       & 0.394 & 0.941  & 0.36 & 30.40  &         & 0.332  & 0.955  & 0.30 & 48.70 &       \\

\hline
   \end{tabular}
   } 
\caption{Results for the high-dimensional setting with independent covariates.}\label{tab:sim_high_indep}
\end{table}

\begin{table}[H]
\centering
\renewcommand{\arraystretch}{0.5}
\resizebox{15cm}{!}{%
\centering
\begin{tabular}{| l | ccccr | ccccr |}
\hline
 & \multicolumn{10}{c|}{Strong Signal ($s=0.3$)}                                  \\ \hline
  & \multicolumn{ 5 }{ c |}{ $(n=200,p=50)$ } & \multicolumn{ 5 }{  c | }{ $(n=400,p=100)$ }\\
Method          & MSE   & Cos(Angle) & MCC  & FP    & ESS     & MSE   & Cos(Angle) & MCC  & FP    & ESS    \\ 
\hline
Oracle      & 0.071  & 0.939   &                         &    &     & 0.071  & 0.966 &    &    &\\    
SpSL-G(HCG) & 0.305 & 0.645 & 0.56 & 0.05 & 12566.5 & 0.411 & 0.757 & 0.70 & 0.04 & 1243.8 \\ 
N-SpSL-L(Exact) & 0.269 & 0.683 & 0.59 & 0.14 & 4260.9 & 0.326 & 0.807 & 0.77 & 0.15 & 578.3 \\ 
SpSL-C(HCG) & 0.280 & 0.677 & 0.58 & 0.11 & 840.2 & 0.346 & 0.799 & 0.76 & 0.11 & 153.5 \\ 
 N-SpSL-C(RW) & 0.311 & 0.637 & 0.56 & 0.05 & 1601.8 & 0.431 & 0.743 & 0.69 & 0.03 & 204.3 \\ 
 HS & 0.278 & 0.669 & 0.61 & 0.09 & 584.0 & 0.362 & 0.782 & 0.76 & 0.02 & 101.1 \\ 
  N-HS(RW) & 0.279 & 0.667 & 0.61 & 0.08 & 1224.2 & 0.369 & 0.778 & 0.76 & 0.02 & 168.7 \\ 
 BL & 0.247 & 0.702 & 0.62 & 4.11 & 3124.1 & {\bf 0.265} & 0.814 & 0.79 & 6.41 & 548.0 \\ 
  N-BL(RW) & {\bf 0.241} & {\bf0.720} & {\bf0.64} & 2.03 & 1396.5 & 0.285 & 0.834 & 0.81 & 5.18 & 302.3 \\ 
SkG              & 0.289 & 0.662 & 0.57 & 0.10  & 5027.2 & 0.357 & 0.791 & 0.75 & 0.10  & 459.8 \\
SpSL(MM)       & 0.328 & 0.573 & 0.49 & 1.23  &         & 0.464 & 0.711 & 0.61 & 2.66  &        \\
N-SpSL-L(MAP) & 0.310 & 0.671 & 0.62 & 0.86 &  & 0.268 & {\bf 0.860} & {\bf0.82} & 1.49 &  \\ 
EMVS            & 0.454 & 0.478 & 0.49 & 0.01  &         & 0.729 & 0.554 & 0.55 & 0.00  &        \\
SSLasso         & 0.363 & 0.572 & 0.53 & 0.81  &         & 0.499 & 0.714 & 0.68 & 1.19  &        \\

Lasso(CV)       & 0.244 & 0.686 & 0.46 & 6.96  &         & 0.316 & 0.815 & 0.47 & 20.19 &        \\
SCAD(CV)        & 0.357 & 0.614 & 0.41 & 4.89  &         & 0.367 & 0.800 & 0.50 & 12.74 &        \\
Lasso(BIC)      & 0.356 & 0.541 & 0.56 & 0.60  &         & 0.568 & 0.635 & 0.67 & 1.74  &        \\
SCAD(BIC)       & 0.385 & 0.524 & 0.52 & 0.89  &         & 0.628 & 0.602 & 0.59 & 2.97  &        \\ 
  N-BL(MAP)       & 0.247 & 0.694 & 0.59 & 2.19  &         & 0.280 & 0.835 & 0.80 & 2.53  &        \\
\hline
\end{tabular}
}
\caption{Results for the low-dimensional setting with dependent covariates. SpSL, HS, and BL indicate the procedure based on the discrete SpSL, the horseshoe, and  Bayesian Lasso priors, respectively. The sign ``N'' stands for the neuronized version of the corresponding prior.}\label{tab:sim_low_dep_SM}
\end{table}

\begin{table}[H]
\centering
\renewcommand{\arraystretch}{0.5}
\resizebox{15cm}{!}{%
\begin{tabular}{| l | ccccr | ccccr |}
\hline
& \multicolumn{10}{c|}{Strong Signal ($s=1.5$)}                                  \\ \hline
                & \multicolumn{5}{c|}{$(n=100,p=300)$}  & \multicolumn{5}{c|}{$(n=150,p=1000)$} \\
Method          & MSE   & Cos(Angle) & MCC  & FP    & ESS    & MSE   & Cos(Angle) & MCC  & FP    & ESS   \\ \hline
Oracle      & 0.150  &  0.980  &                         &    &     & 0.080  & 0.989 &    &    &\\    
SpSL-G(HCG) & 0.980 & 0.823 & 0.76 & 0.11 & 2149.6 & 0.607 & 0.890 & 0.84 & 0.11 & 592.5 \\ 
  N-SpSL-L(Exact) & 0.948 & 0.827 & 0.76 & 0.28 & 557.4 & 0.610 & 0.890 & 0.84 & 0.33 & 162.6 \\ 
 SpSL-C(HCG) & {\bf 0.858} & {\bf 0.848} & 0.79 & 0.22 & 68.8 & 0.509 & 0.910 & 0.87 & 0.27 & 26.7 \\ 
 N-SpSL-C(RW) & 1.046 & 0.809 & 0.75 & 0.11 & 175.9 & 0.628 & 0.886 & 0.84 & 0.10 & 58.1 \\ 
HS & 1.059 & 0.805 & 0.79 & 0.80 & 20.2 & 0.769 & 0.859 & 0.69 & 4.13 & 4.2 \\ 
  N-HS(RW) & 1.019 & 0.814 & 0.79 & 0.84 & 95.2 & 0.719 & 0.868 & 0.71 & 4.03 & 7.9 \\ 
 BL & 1.631 & 0.503 & 0.41 & 17.68 & 102.1 & 2.030 & 0.515 & 0.74 & 1.21 & 24.2 \\ 
  N-BL(RW) & 1.494 & 0.682 & 0.62 & 7.53 & 130.0 & 1.715 & 0.622 & 0.74 & 0.96 & 34.0 \\ 
  SkG              & 1.402 & 0.736 & 0.65 & 0.02  & 1068.0 & 1.304 & 0.751 & 0.66 & 0.00  & 317.3 \\
SpSL(MM)       & 2.106 & 0.577 & 0.50 & 1.36  &        & 2.143 & 0.544 & 0.44 & 2.24  &       \\
 N-SpSL-L(MAP)     & 1.640 & 0.678 & 0.65 & 0.06  &        & 1.105 & 0.782 & 0.76 & 0.03  &       \\

EMVS            & 2.317 & 0.538 & 0.56 & 0.00  &        & 2.553 & 0.451 & 0.51 & 0.00  &       \\
SSLasso         & 0.993 & 0.821 & {\bf0.81} & 0.07  &        & {\bf 0.491} & {\bf0.910} & {\bf0.90} & 0.10  &       \\
Lasso(CV)       & 1.240 & 0.759 & 0.46 & 14.56 &        & 1.258 & 0.762 & 0.38 & 26.23 &       \\
SCAD(CV)        & 1.484 & 0.731 & 0.39 & 12.40 &        & 1.338 & 0.752 & 0.36 & 21.14 &       \\
Lasso(EBIC)      & 2.422 & 0.544 & 0.62 & 0.00  &        & 2.311 & 0.562 & 0.65 & 0.08  &       \\
SCAD(EBIC)       & 2.467 & 0.537 & 0.60 & 0.07  &        & 2.383 & 0.547 & 0.61 & 0.14  &  \\
N-BL(MAP)       & 1.204 & 0.758 & 0.35 & 21.03 &        & 1.223 & 0.760 & 0.28 & 37.99 &       \\
\hline
   \end{tabular}
}
\caption{Results for the high-dimensional setting with dependent covariates.}\label{tab:sim_high_dep_SM}
\end{table}

\subsection{Numerical Approximation Errors for Horseshoe Prior}
In the simulation and real data studies examined in Section \ref{sec:sim} and \ref{sec:real}, it was shown that the horseshoe prior and its neuronized counterpart produced slightly different numerical results, even though they should have resulted in exactly the same posterior distribution  for the coefficients. We here investigate a high-dimensional example with a much larger number of MCMC iterations and show that the observed differences are due to numerical approximation errors of MCMC. 

\begin{figure}
    \centering
    \begin{subfigure}[b]{0.47\textwidth}
        \includegraphics[width=\textwidth]{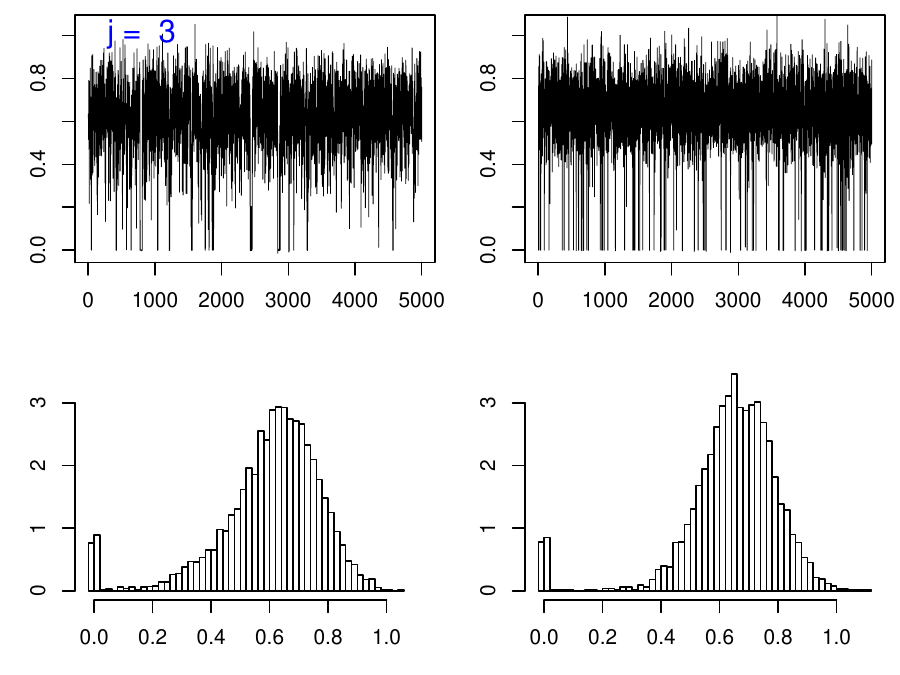}
        \caption{}
    \end{subfigure}
    ~ 
    \begin{subfigure}[b]{0.47\textwidth}
        \includegraphics[width=\textwidth]{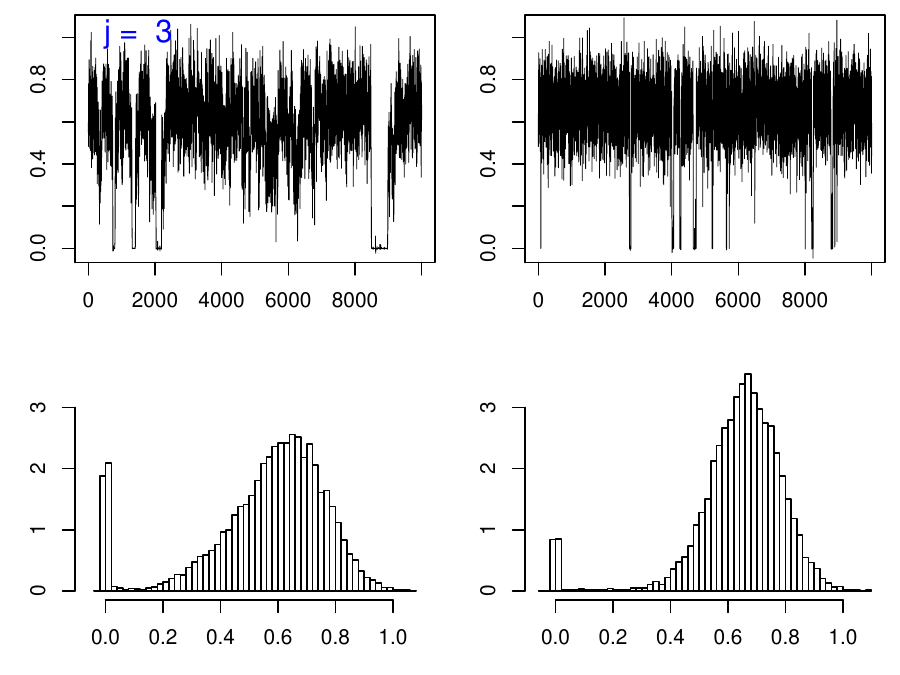}
        \caption{}
    \end{subfigure}
         \begin{subfigure}[b]{0.47\textwidth}
        \includegraphics[width=\textwidth]{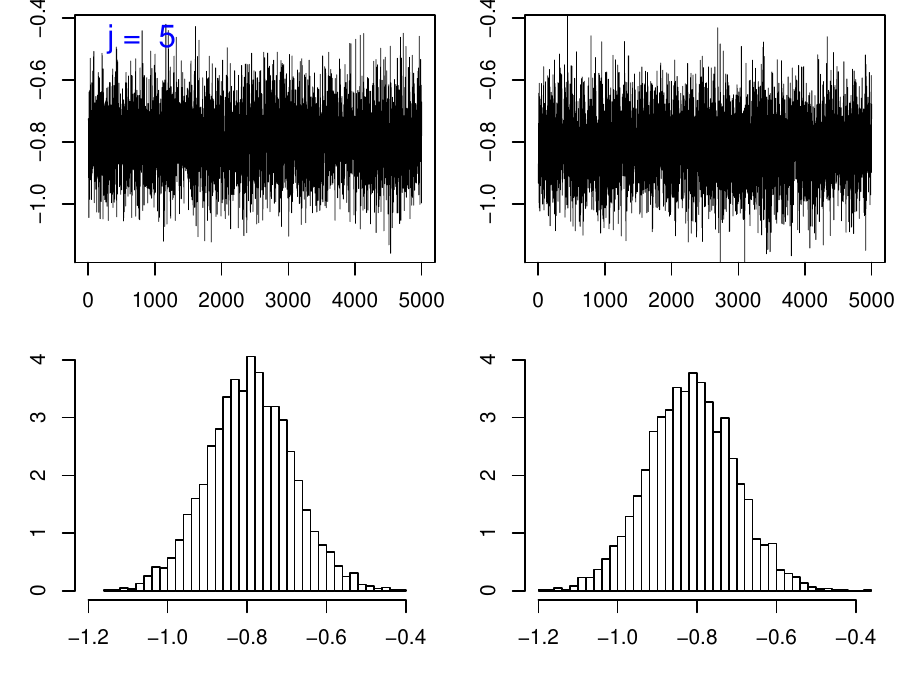}
        \caption{}
    \end{subfigure}
    ~ 
    \begin{subfigure}[b]{0.47\textwidth}
        \includegraphics[width=\textwidth]{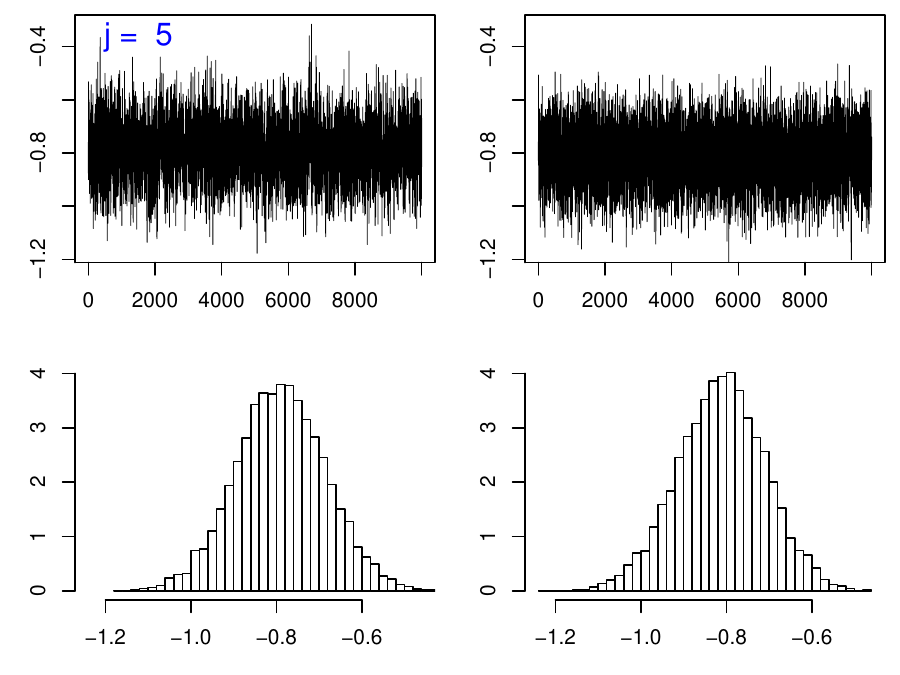}
        \caption{}
    \end{subfigure}
          \begin{subfigure}[b]{0.47\textwidth}
        \includegraphics[width=\textwidth]{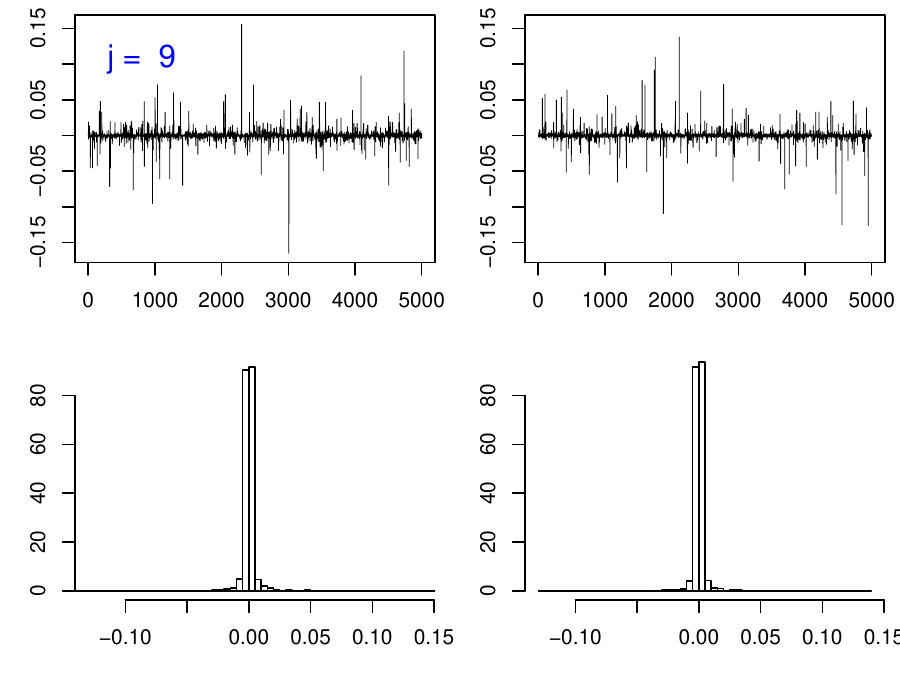}
        \caption{}
    \end{subfigure}
    ~ 
    \begin{subfigure}[b]{0.47\textwidth}
        \includegraphics[width=\textwidth]{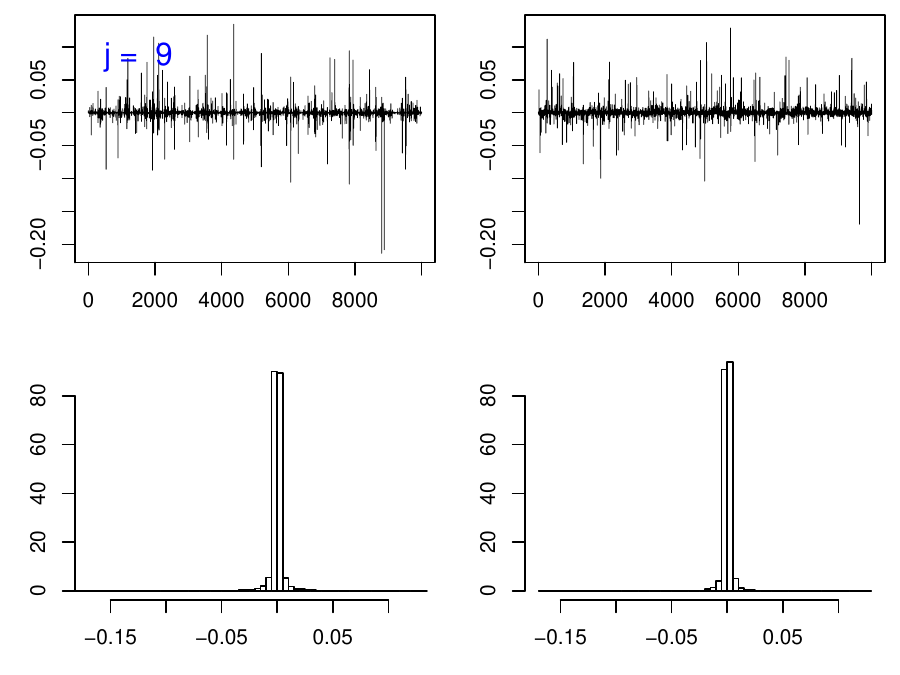}
        \caption{}
    \end{subfigure}
    \caption{The first two columns indicates the cases with a long chain; the other columns shows the results with a short chain. (a) and (b) illustrate the posterior distribution of $\theta_3$; (c) and (d) are for the posterior distributions of  $\theta_5$; (e) and (f) are for $\theta_9$. The left  and  rights panel within each sub-figure represent the standard horseshoe prior and the neuronized horseshoe prior, respectively. }\label{fig:HS_long}
\end{figure}

We generate a synthesized  data set by following the same high-dimensional setting used in Section \ref{sec:sim}, with a strong signal, $n=150$ and $p=1000$. We consider  100,000 iterations after 10,000 burn-in (20 thinning size). The resulting approximated posterior distributions for several coefficients are illustrated in the first two columns of Figure \ref{fig:HS_long}. A short chain with  10,000 iterations and 2,000 burn-in steps is also presented on the other columns.    

Figure \ref{fig:HS_long} shows that when the length of the chain is large enough, the standard horseshoe prior and its neuronized counterpart lead to nearly identical posterior distributions for $\theta_3$, $\theta_5$, and $\theta_9$. For short MCMC chains, the both standard and neuronized  procedures successfully approximate the posterior distributions of $\theta_5$ and $\theta_9$.  However,  the
shorter chain did not provide a good mixing for the posterior distribution of $\theta_3$ under the standard horseshoe prior (the left panel of (b)), with the chain stuck around the origin for a long time, leading to 
an over-estimation of the posterior probability around zero. Comparing with the result from the longer chain, we observe that the algorithm with the neuronized HS prior appears to have done a much better job mixing for the shorter chain.


\end{document}